\documentclass[aps,prx,reprint,eqsecnum]{revtex4-1}
\usepackage{graphicx}
\usepackage{amssymb}
\usepackage{amsmath}
\usepackage{bm}
\usepackage{verbatim}
\usepackage[bookmarks=false,colorlinks=true,urlcolor=blue,citecolor=blue,linkcolor=blue]{hyperref}
\usepackage[normalem]{ulem}

\begin{document}

\title{
Configuration-Controlled Many-Body Localization and the Mobility Emulsion}

\author{Michael Schecter}\thanks{These authors contributed equally to this work.}
\affiliation{Condensed Matter Theory Center and Joint Quantum Institute, Department of Physics, University of Maryland, College Park, Maryland 20742, USA}

\author{Thomas Iadecola}\thanks{These authors contributed equally to this work.}
\affiliation{Condensed Matter Theory Center and Joint Quantum Institute, Department of Physics, University of Maryland, College Park, Maryland 20742, USA}

\author{Sankar Das~Sarma}
\affiliation{Condensed Matter Theory Center and Joint Quantum Institute, Department of Physics, University of Maryland, College Park, Maryland 20742, USA}

\date{\today}

\begin{abstract}

We uncover a new non-ergodic phase, distinct from the many-body localized (MBL) phase, in a disordered two-leg ladder of interacting hardcore bosons. The dynamics of this emergent phase, which has no single-particle analog and exists only for strong disorder and finite interaction, is determined by the many-body configuration of the initial state. Remarkably, this phase features the \emph{coexistence} of localized and extended many-body states at fixed energy density and thus does not exhibit a many-body mobility edge, nor does it reduce to a model with a single-particle mobility edge in the noninteracting limit.  We show that eigenstates in this phase can be described in terms of interacting emergent Ising spin degrees of freedom (``singlons") suspended in a mixture with inert charge degrees of freedom (``doublons" and ``holons"), and thus dub it a \emph{mobility emulsion} (ME). We argue that grouping eigenstates by their doublon/holon density reveals a transition between localized and extended states that is invisible as a function of energy density.  We further demonstrate that the dynamics of the system following a quench may exhibit either thermalizing or localized behavior depending on the doublon/holon density of the initial product state. Intriguingly, the ergodicity of the ME is thus tuned by the initial state of the many-body system. These results establish a new paradigm for using many-body configurations as a tool to study and control the MBL transition. The ME phase may be observable in suitably prepared cold atom optical lattices.

\end{abstract}

\maketitle

\section{Introduction}\label{sec:intro}

Non-ergodic quantum systems have attracted much attention in recent years due to rapid progress in a growing number of experimental systems including Rydberg atoms~\cite{Bernien17,Guardado-Sanchez17,Lienhard17} and interacting disordered systems such as cold atoms~\cite{Schreiber15,Kondov15,Bordia16,Kaufman16,Choi16,Bordia17,Luschen17,Luschen18,Lukin18} and trapped ions~\cite{Smith16,Zhang17} (see also \cite{Choi17}), making possible a systematic study of their nontrivial dynamical behavior. They are also of deep conceptual interest in understanding the applicability of quantum statistical mechanics to isolated systems. Non-ergodic systems are exceptional in that they are fundamentally incompatible with the laws of statistical mechanics and generally do not relax to thermal equilibrium as described by the Gibbs ensemble. It is well-known that non-ergodicity arises in exactly solvable quantum integrable models (similar to what happens in classical integrable systems, e.g., the Fermi-Pasta-Ulam model ~\cite{Fermi55,Zabusky65,Gallavotti08}) due to their extensive number of integrals of motion (see~\cite{Caux11} for a recent discussion). However, it is frequently believed that quantum integrability in a many-body system is generally unstable to the addition of weak integrability-breaking perturbations due to the absence of a quantum analog of the KAM theorem (see also~\cite{Brandino15}). Thus, such explicitly integrable quantum systems are not generic, although the associated lack of thermalization has been studied in carefully prepared laboratory experiments~\cite{Kinoshita06}.

An alternative route to robust non-ergodicity has recently emerged in the context of many-body localization (MBL), which occurs in (presumably generic) interacting quantum systems subject to quenched disorder~\cite{Gornyi05,Basko06,Oganesyan07,Znidaric08,Vosk10,Pal10,Berkelbach10,Bardarson12,Serbyn13,Huse14,Barlev14,Serbyn14,Serbyn15,Chandran15a,Luitz15,Potter15,Agarwal15,Ros15,Gopalakrishnan15,Znidaric16,Rademaker16,Imbrie16a,Imbrie16b,OBrien16,Khemani17,Yang17,Li15,Li17,Gornyi17}. It is well-established that the MBL phase possesses a robust and \emph{emergent} integrability associated with an extensive number of local integrals of motion (LIOMs)~\cite{Serbyn13,Huse14,Ros15,Imbrie16a,Imbrie16b,Chandran15a,OBrien16,Rademaker16}. This manifestation of non-ergodicity represents a spectacular departure from the laws of statistical mechanics (i.e. isolated interacting quantum systems may not be thermal generically) as it allows the possibility of spontaneous symmetry-breaking and topological phase transitions even at infinite temperature, i.e. for eigenstates with \emph{arbitrary} energy density~\cite{Huse13,Bauer13,Pekker14,Kjall14,Chandran14,Slagle15,Bahri15,Potter16,Vasseur16,Friedman17,Parameswaran17,Chandran17,Prakash17}. Despite the richness of the MBL phase itself, most previous studies have focused on systems that exhibit only two types of dynamical behavior: the fully MBL phase with complete emergent integrability (i.e. the number of LIOMs matches the number of degrees of freedom) or the usual thermal ergodic phase that satisfies the eigenstate thermalization hypothesis (ETH)~\cite{Deutsch91,Page93,Srednicki94,D'Alessio16,Mori17} and obeys the laws of statistical mechanics. The present work introduces the possibility of an intriguing intermediate phase between the MBL and ETH phases, which occurs at finite interaction and large disorder, and is truly \emph{emergent} in the sense that it has no single-particle analog.  The intermediate phase we find is qualitatively different from other intermediate phases which have recently been discussed in the MBL literature, as it is not rooted in any mobility-edge physics (either single-particle or many-body) or Griffiths physics of rare regions.

Intermediate phases that possess a single-particle~\cite{Modak15,Li15,Li16,Li17} or many-body~\cite{Basko06, Luitz15,Gornyi17} mobility edge represent possible non-ergodic phases with incomplete integrability, but the existence of a many-body mobility edge is still controversial~\cite{DeRoeck16}.  However, studying this physics experimentally is challenging; for example, probing the critical energy or energy density of such an intermediate phase via quench dynamics requires the ability to prepare the initial state in an energy-resolved manner.  Other models whose disorder respects continuous $SU(2)$ symmetries~\cite{BarLev16,Prelovsek16,Protopopov17,Kozarzewski18} have been argued to exhibit non-ergodicity with incomplete integrability for small system sizes, but it is likely that such systems ultimately undergo thermalization at long times in the thermodynamic limit due to the fundamental incompatibility of MBL and the non-Abelian symmetry~\cite{Protopopov17,Potter16,Vasseur16,Friedman17,Prakash17}.

%%%%%%%%%%%%%%%%%%
%%%%%%%%%%%%%%%%%%
\begin{figure}[t!]
\includegraphics[width=\columnwidth]{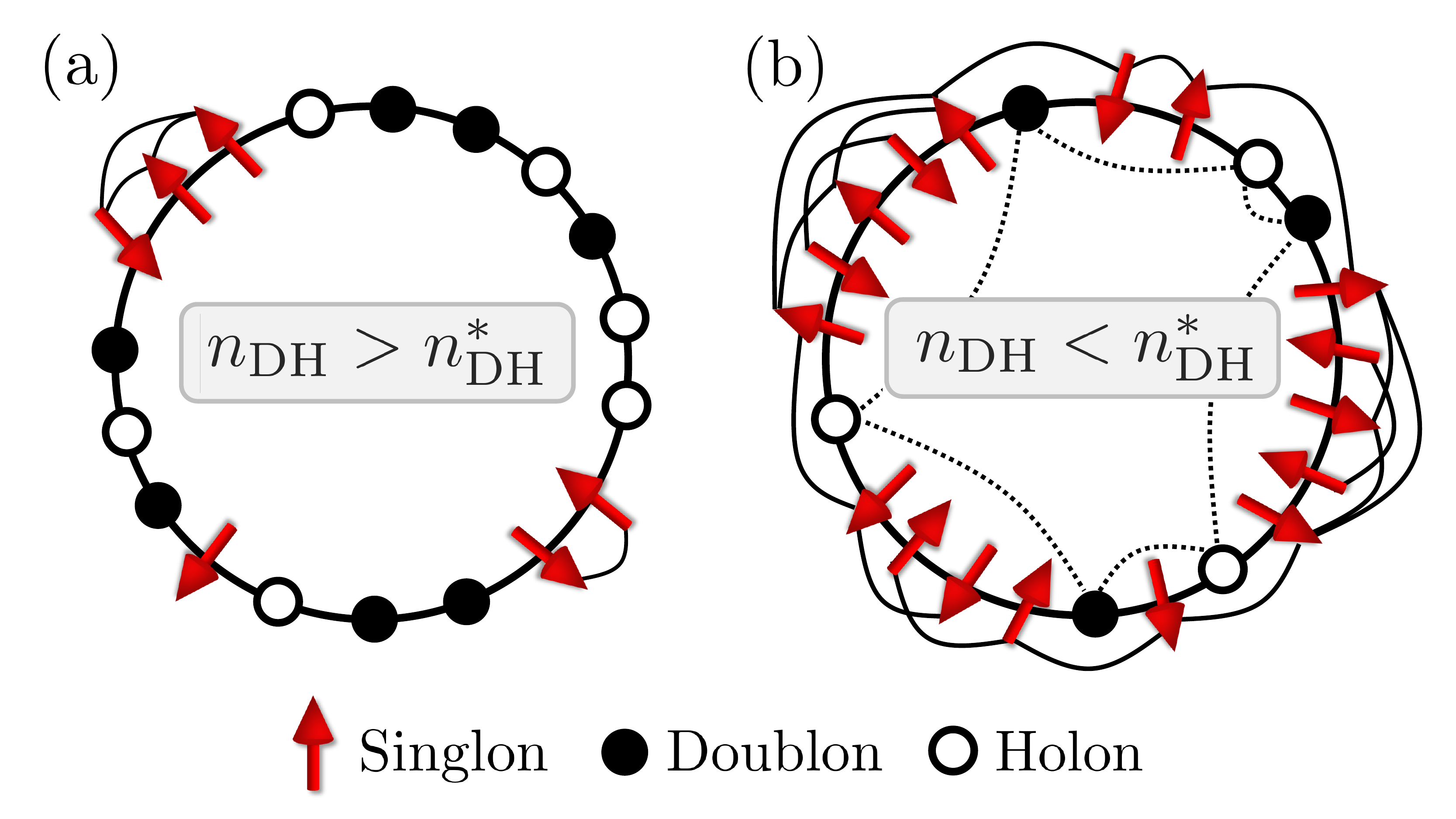}
\caption{Schematic depiction of the mobility emulsion.  The solid black bonds represent entanglement between emergent Ising spins (singlons) c.f.~Fig.~\ref{fig:2}. (a) Above the critical doublon/holon density, a typical eigenstate consists of a sparse network of singlons suspended in a background of localized doublons and holons.  The system supports clusters of interacting singlons, but these clusters are typically much farther apart than their typical size, so that the network of singlons is many-body localized.  Processes admixing these clusters with doublons and holons are far off resonance at strong disorder, and the ``emulsion" is stable. (b) Below the critical doublon/holon density, a typical eigenstate consists of a sparse set of doublons and holons suspended in a thermalizing bath of singlons.  Doublons exchange energy with this bath and undergo variable-range hopping (dashed bonds), thereby
mediating the transport of charge.}
\label{fig:1}
\end{figure}
%%%%%%%%%%%%%%%%%%
%%%%%%%%%%%%%%%%%%

In this paper we report on the existence of a non-ergodic intermediate phase that neither satisfies ETH nor is fully MBL, but at the same time is not connected with any many-body (or single-particle) mobility edge. Rather, this phase is defined by the \emph{coexistence} of localized and extended many-body states at fixed energy density. We show that eigenstates of the system can be described in terms of a \emph{mobility emulsion} (ME), wherein emergent interacting Ising spin degrees of freedom (denoted ``singlons")  become suspended in a mixture with inert charge degrees of freedom (denoted ``doublons" and ``holons"), see Fig.~\ref{fig:1}. The disorder in this case acts as a ``surfactant" that stabilizes the singlon/doublon emulsion provided that the emergent doublon/holon density, $n_{\rm DH}$, is sufficiently large. We argue that grouping eigenstates by $n_{\rm DH}$ reveals a transition, at a critical value $n^*_{\rm DH}$, between localized and extended states that is absent as a function of energy density. Thus, the ME phase, in addition to requiring finite interaction and strong disorder, is configurationally controlled through the relative singlon and doublon/holon densities in the initial state.    We emphasize that the standard ETH and MBL phases also exist in the system for small and large disorder, respectively.

In eigenstates with sufficiently large doublon/holon density, $n_{\rm DH}>n_{\rm DH}^*$, the sparse singlons interact weakly and are frozen into paramagnetic configurations, remaining MBL. In eigenstates with $n_{\rm DH}<n_{\rm DH}^*$, the enhanced density of singlons strengthens their mutual interactions, allowing the singlons to thermalize among themselves. The thermal bath of singlons then mediates variable-range hopping of the remaining dilute doublons and holons, which interact and thermalize in turn.

%%%%%%%%%%%%%%%%%%
%%%%%%%%%%%%%%%%%%
\begin{figure}[t!]
\includegraphics[width=\columnwidth]{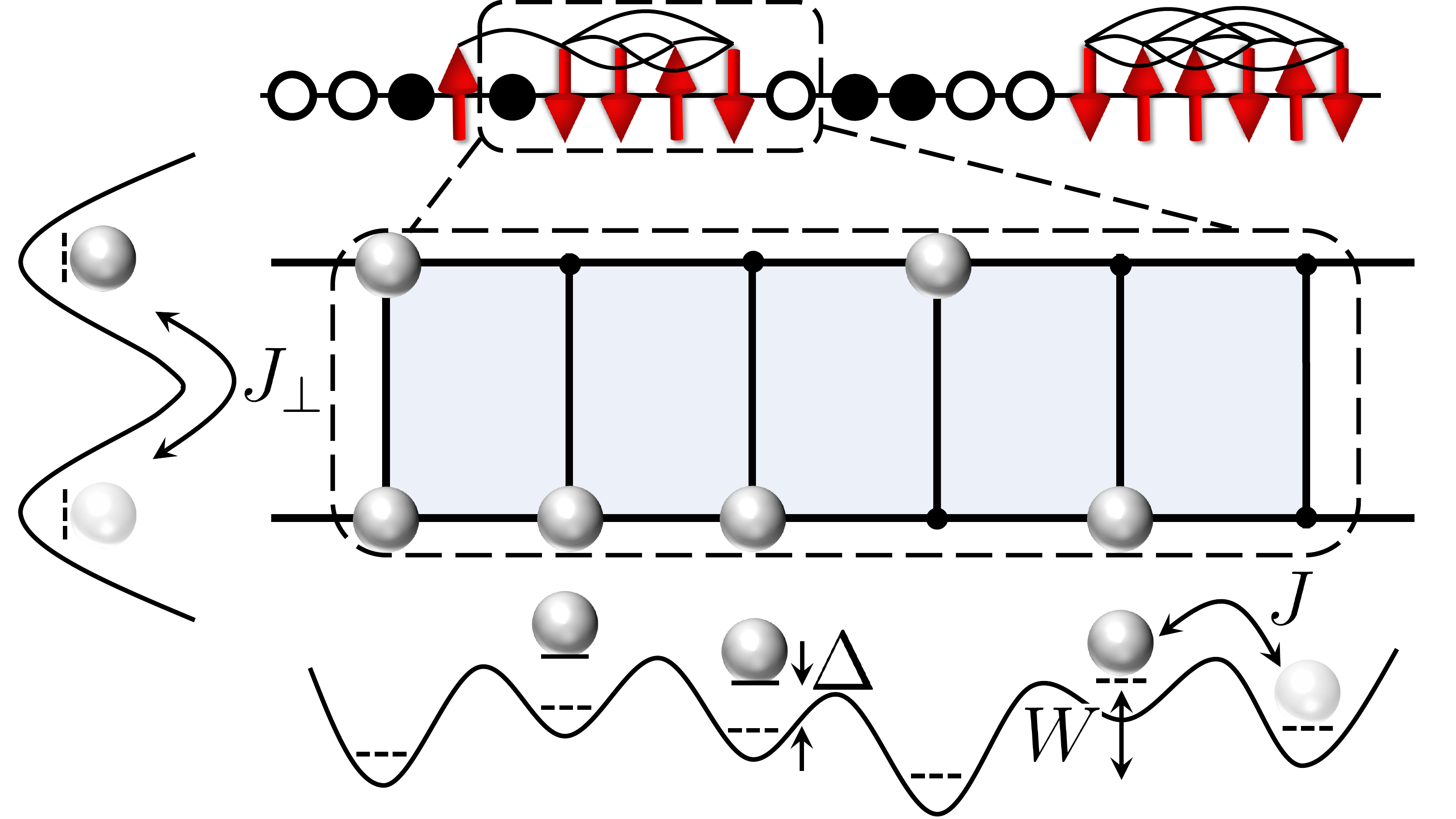}
\caption{Two-leg ladder with identical (mirror-symmetric) disorder potentials on the two legs. For strong disorder, mirror-related sites that share one boson form emergent Ising spin degrees of freedom (red arrows) suspended in a mixture of doublons (doubly occupied rungs) and holons (empty rungs) denoted by full and open circles, respectively. }
\label{fig:2}
\end{figure}
%%%%%%%%%%%%%%%%%%
%%%%%%%%%%%%%%%%%%

These two cases can be distinguished sharply in an experimental setting by observing the dynamics of the system following quenches from initial product states with different values of $n_{\rm DH}$.  Initial states with $n_{\rm DH}>n_{\rm DH}^*$ retain memory of the initial state at long times, while initial states with $n_{\rm DH}<n_{\rm DH}^*$ lose this memory. Remarkably, this distinction can be made even in a fixed disorder realization and for fixed Hamiltonian parameters; one need only tune the doublon/holon density of the initial state. This configuration-controlled localization in an intermediate phase between ETH and MBL leads
 to a new paradigm in the study and manipulation of non-ergodic phases of matter.

To exemplify this paradigm, we focus on interacting hardcore bosons in a two-leg ladder whose legs are subject to identical disorder potentials, see Fig.~\ref{fig:2}. This system can equivalently be viewed as a coupled pair of identical random-field XXZ spin chains. Here, the singlon and doublon/holon states correspond to rungs of the ladder that either host a single particle or are full/vacant, respectively (see Fig.~\ref{fig:2}). Because the two legs have the same disorder potential, the system possesses a $\mathbb Z_2$ mirror symmetry that exchanges the legs of the ladder. As we show below, in the MBL phase this symmetry can break spontaneously, giving rise to ``mirror-glass" order in a nonzero fraction of states at nonzero energy density. The mechanism behind this symmetry breaking was studied recently~\cite{Iadecola18} in a case involving a single chain with mirror-symmetric disorder. In the current two-leg ladder system, the mirror symmetry arises naturally by virtue of the disorder being the same in the two individual chains. Remarkably, the long-range mirror-glass order that arises in the two-leg ladder (and in Ref.~\cite{Iadecola18}) can occur \emph{only} in states with a nonzero energy density above (below) the ground (ceiling) states, which remain symmetric throughout the MBL phase. This ``inverse freezing" effect is due to the fact that the 
ground state essentially contains only doublons and holons, which transform trivially under the mirror symmetry. Singlons, on the other hand, carry nontrivial representations of the mirror symmetry, but only become \emph{activated} in excited states~\cite{Iadecola18}. When a nonzero density of singlons are activated, their mutual interactions drive the mirror-symmetry-breaking transition. Unlike the case of disrete non-Abelian symmetries~\cite{Potter16,Vasseur16,Friedman17,Prakash17}, in the ladder model studied here there is no obstruction to having mirror-symmetric MBL eigenstates at finite energy density in which the singlons form a paramagnetic state. Indeed, as one tunes the inter-leg hopping amplitude $J_\perp$ at large disorder, we find that the MBL mirror-glass phase melts directly into the symmetric non-ergodic ME phase as shown in the phase diagram of Fig.~\ref{fig:3}. Thus, the ME phase requires finite interaction, large disorder, and strong interchain hopping for its existence in the two-leg ladder system.  We mention here that, in addition to not having a single-particle analog (e.g. no single-particle mobility edge), the ME phase also has no single-chain analog as it is driven explicitly by tuning the interchain hopping to produce the appropriate singlon-holon-doublon dynamics necessary for its existence.

The remaining part of the paper is organized as follows. In Sec.~\ref{sec:model} we introduce the model of the two-leg ladder studied throughout. We present its infinite-temperature phase diagram and discuss the numerical diagnostics used to construct it. In Sec.~\ref{sec:singlon} we derive an effective singlon Hamiltonian using a Schrieffer-Wolff transformation to systematically eliminate the longitudinal hopping at large disorder. This allows us to characterize and develop intuition for both the MBL and ME phases, and to estimate the critical point that separates them.  In Sec.~\ref{sec:mixed}, we present further numerical results characterizing the ME phase, showing in particular the lack of a mobility edge as a function of many-body energy density and the correlation between the degree of  entanglement in an eigenstate and its doublon/holon density. In Sec.~\ref{sec:dynamics} we consider the dynamics of local observables following quantum quenches from various local-density product states, which are readily preparable experimentally. We find substantial qualitative differences in the late-time behavior of local observables depending on the initial value of the doublon/holon density. Discussion and conclusions are presented in Sec.~\ref{sec:conclusion}.

\section{Model and phase diagram}\label{sec:model}

We study hardcore bosons hopping on a disordered two-leg ladder with a $\mathbb Z_2$ leg-permutation (mirror) symmetry. The Hamiltonian is given by
\begin{subequations}
\label{eq:H}
\begin{equation}
\label{eq:Ha}
H=H_1+H_2+H_\perp,
\end{equation}
where
\begin{align}
H_\alpha
&\!=\!
\!\sum^{L}_{i=1}\!
\left[\frac{J}{2}\!\left(\!b^\dagger_{\alpha,i} b_{\alpha,i+1}\!+\!{\rm H.c.}\!\right)\!+\!\Delta\!\left(n_{\alpha,i}\!-\!\frac{1}{2}\right)\!\!\left(n_{\alpha,i+1}\!-\!\frac{1}{2}\right)\!\right]\nonumber\\
&\qquad+2\sum^L_{i=1} h_{\alpha,i}\! \left(n_{\alpha,i}-\frac{1}{2}\right) ,\label{eq:Hb}
\end{align}
and
\begin{equation}\label{eq:Hc}
H_\perp=\frac{J_\perp}{2} \sum^{L}_{i=1} \left(b^\dagger_{1,i}b_{2,i}+{\rm h.c.}\right).
\end{equation}
Here, $b^\dagger_{\alpha,i}/b_{\alpha,i}$ are boson creation/annihilation operators on rung $i$ and leg $\alpha=1,2$, $n_{\alpha,i}=b^\dagger_{\alpha,i} b_{\alpha,i}$ is the local boson density, and $L$ is the system length.  We assume the bosons interact strongly onsite and thus satisfy the hardcore constraint $n_{\alpha,i}(n_{\alpha,i}-1)=0$ with commutation relations $[b_{\alpha,i},b^\dagger_{\beta,j}]=\delta_{ij}\delta_{\alpha\beta}(1-2n_{\alpha,i})$, which allows one to map the problem onto a pair of coupled XXZ chains. The Hamiltonian $H$ possesses a global $U(1)$ symmetry associated with conservation of total particle number and a $\mathbb Z_2$ mirror symmetry $M$, which interchanges the leg indices, $1\leftrightarrow 2$, and implies that the disorder in the two legs is identical, 
\begin{align}
h_{1,i}=h_{2,i}\equiv h_{i}.
\label{eq:Hd}
\end{align}
\end{subequations}
 We focus on the case of half filling (equivalently, on the zero-magnetization sector of the XXZ ladder) with periodic boundary conditions and consider independent random onsite potentials $h_i$ drawn from a normal distribution with mean zero and standard deviation $W/2$ for $i=1,\dots,L$. The mirror symmetry can be realized experimentally in a number of ways, but perhaps the simplest is to use a two-dimensional optical lattice subject to an additional longitudinal disorder potential and a transverse confining potential supporting two minima, see Fig.~\ref{fig:2}.

We performed an exact diagonalization study of the model~\eqref{eq:H} at small system sizes.  The numerically determined infinite-temperature phase diagram of $H$ as a function of $W$ and $J_\perp$ is shown in Fig.~\ref{fig:3}. We devote the remainder of this section to understanding this phase diagram and describing how it is obtained.  

\subsection{Qualitative understanding of the phase diagram}\label{subsec:qualitative}

We begin by considering the limit of decoupled legs, $J_\perp=0$, where the phase diagram is easiest to understand.  In this case, the problem reduces to that of two independent copies of the random-field XXZ chain, where it is numerically established~\cite{Oganesyan07,Znidaric08,Pal10,Berkelbach10,Bardarson12,Serbyn13,Barlev14,Serbyn14,Serbyn15,Chandran15a,Luitz15,Agarwal15,Gopalakrishnan15,Znidaric16,Rademaker16,OBrien16,Khemani17,Yang17} that the system undergoes a phase transition from an ergodic phase to an MBL phase at a critical disorder strength. At finite $J_{\perp}$, it is reasonable to expect that the ergodic phase persists so long as the disorder strength $W$ is sufficiently weak. Eigenstates at nonzero energy density remain mirror-symmetric in the ergodic phase, since long-range order at nonzero temperature is thermodynamically forbidden in one dimension~\cite{LandauStatMPhys1980,Mermin66,Hohenberg67}.

%%%%%%%%%%%%%%%%%%
%%%%%%%%%%%%%%%%%%
\begin{figure}[t!]
\includegraphics[width=\columnwidth]{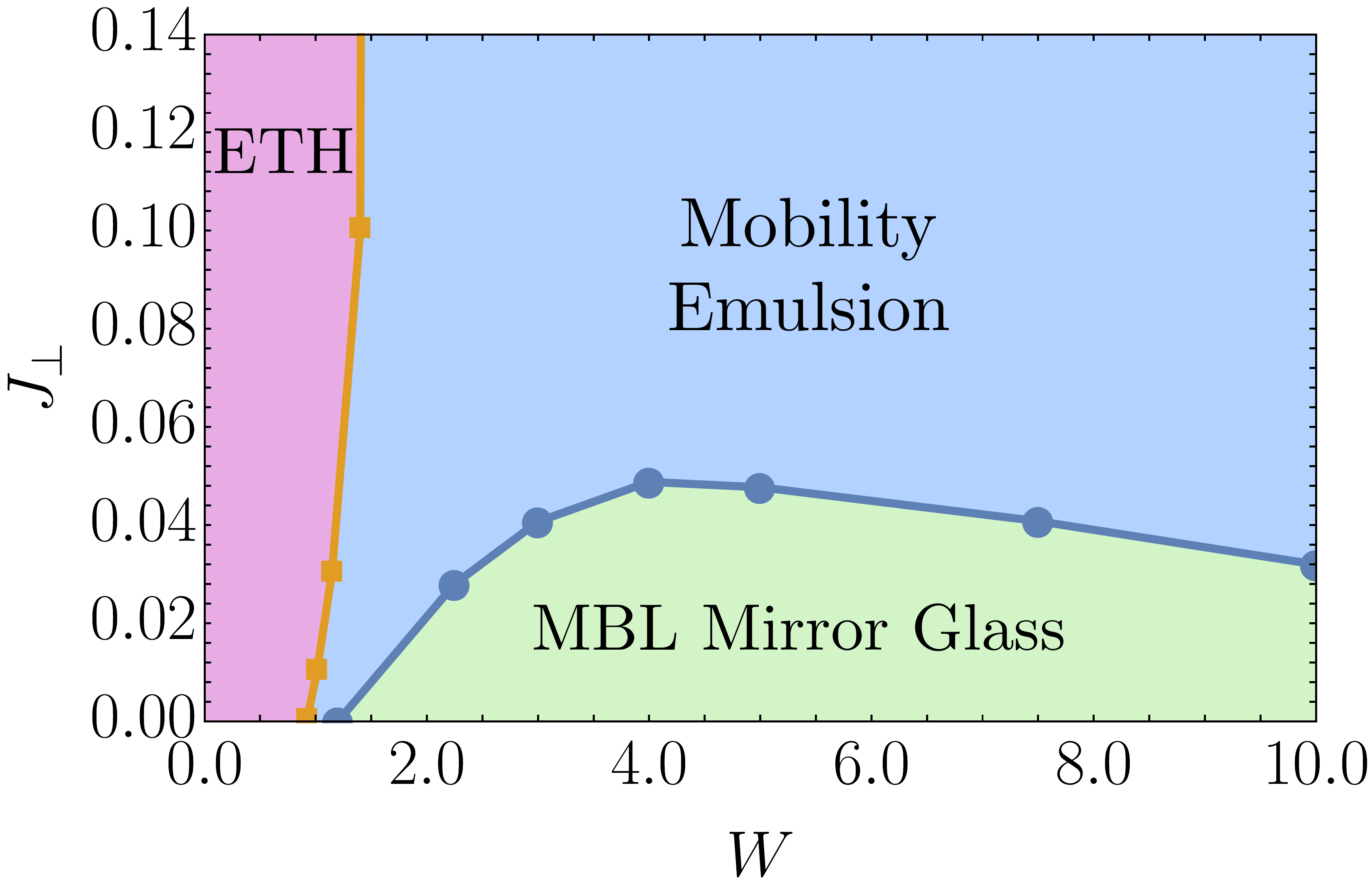}
\caption{The dynamical phase diagram of Eq.~\eqref{eq:H} at $\Delta=0.5$. Here and in the remainder of the paper, we work in units such that the intra-leg hopping amplitude $J=1$.  In addition to the ETH and MBL phases, we find a new phase---the mobility emulsion (ME)---that predominates at strong disorder. We determine transitions using the mirror-glass order parameter $q_n$ [Eq.~\eqref{eq:q}] for the MBL/ME phase boundary line and the doublon correlator $p_n$ [Eq.~\eqref{eq:p}] for the ETH/ME phase boundary line. See Sec.~\ref{sec:transition_indicators} for representative examples of the exact-diagonalization data used to obtain the points on the above phase diagram.}
\label{fig:3}
\end{figure}
%%%%%%%%%%%%%%%%%%
%%%%%%%%%%%%%%%%%%

We now turn to the MBL phase at $J_\perp=0$. In this case, the eigenstates on each leg can be labeled by suitably dressed occupation factors, which constitute the eigenvalues $0$ or $1$ of the LIOMs $\tilde n_{\alpha,i}$. When the intra-leg hopping $J=0$, the local state on rung $i$ can be written as $|n_{1,i},n_{2,i}\rangle\equiv|n_{1},n_{2}\rangle_i$, with $n_{1,2}=0,1$; this yields \emph{four} states per rung. Of these four states, only two transform nontrivially under the mirror symmetry $M$, namely the ``singlon" states
\begin{subequations}
\label{eq:singlon doublon/holon states}
\begin{align}
|1,0\rangle_i\equiv|\!\uparrow\rangle_{i}
\indent\text{and}\indent
|0,1\rangle_i\equiv|\!\downarrow\rangle_{i},
\label{eq:singlon states}
\end{align}
which are degenerate because the disorder potential respects $M$. Indeed, from Eqs.~\eqref{eq:Hb} and \eqref{eq:Hd}, one sees that the singlon states do not couple directly to the disorder potential. In contrast, the mirror-symmetric ``doublon" and ``holon" states,
\begin{align}
|1,1\rangle_i\equiv |\bullet\rangle_i 
\indent\text{and}\indent
|0,0\rangle_i\equiv |\circ\rangle_{i},
\label{eq:doublon/holon states}
\end{align}
respectively, are split in energy by an amount of order $W$ due to the disorder potential. Doublon and/or holon states on distinct rungs $i$ and $j$ are thus far off-resonance at large $W$, whereas nearby singlon states are split comparatively weakly by interactions $\Delta\ll W$.  We emphasize that singlons, doublons, and holons are only well-defined (i.e., long-lived) degrees of freedom at strong disorder.  In the ETH phase, there is no meaningful distinction among these degrees of freedom, as the local boson density on each rung is not approximately conserved, as it is at strong disorder.
\end{subequations}

At strong disorder, the singlon states~\eqref{eq:singlon states} can thus be viewed as local states of an effective spin-$1/2$ chain in which the mirror symmetry $M$ becomes an onsite $\mathbb Z_2$ symmetry (the doublons and holons, on the other hand, are essentially inert)~\cite{Iadecola18}.  Crucially, this effective spin-$1/2$ chain is \textit{disorder-free} at $J=0$, since the disorder potential does not couple directly to the singlons. At small but finite intra-leg hopping $J\ll W$, an effective interaction is generated between these effective spin configurations, due to the repulsion $\Delta$ between neighboring particles, and is randomly renormalized by the dressing of the occupation factors, $\Delta n_{\alpha,i}n_{\alpha,j}\to \tilde\Delta_{ij}\tilde n_{\alpha,i}\tilde n_{\alpha,j}$. In Sec.~\ref{sec:singlon}, we calculate the random corrections to the interaction energy in the limit of strong disorder, where perturbation theory in $J/W$ is controlled (up to rare-region effects). These renormalized random interactions among singlons lead to spontaneous breaking of the $\mathbb Z_2$ symmetry of the effective singlon spin chain in eigenstates for which a finite fraction of the rungs occupy singlon states~\cite{Iadecola18}.  (As mentioned earlier, these eigenstates are necessarily at finite energy density owing to the fact that the ground and ceiling states consist overwhelmingly of doublon and holon states on each rung.)  The MBL phase at $J_{\perp}=0$ thus breaks the mirror symmetry $M$ by default, yielding long-range mirror-glass order in typical eigenstates at finite energy density.

Next we consider the fate of the decoupled-chain MBL phase upon adding a finite interchain coupling $J_\perp$. In Sec.~\ref{sec:singlon}, we show that this coupling induces a ``transverse field" in the singlon spin model, thereby enhancing the quantum fluctuations of the singlons.  If the transverse field is much weaker than the random interactions induced by the finite intra-leg hopping $J$, then we expect that the system remains MBL and, furthermore, that $M$ remains spontaneously broken.  However, if the quantum fluctuations induced by the transverse field dominate over the effective bond randomness in the interactions, localization is no longer guaranteed. Thus, as $J_\perp$ increases we expect MBL eigenstates with broken mirror symmetry to give way to thermalizing eigenstates that necessarily preserve the mirror symmetry.  It is the thermalization of singlons due to the interchain coupling that drives the strong-disorder transition between the MBL mirror glass phased at weak $J_\perp$ and the ME phase at larger $J_\perp$. Furthermore, because the singlon bond randomness is induced by virtual transitions between singlon and doublon states (see Sec.~\ref{sec:singlon}), it becomes \textit{weaker} asymptotically as the direct disorder strength $W$ is increased; thus, we expect the critical value of $J_\perp$ at which the transition takes place to \textit{decrease} with increasing $W$, as observed in Fig.~\ref{fig:3}. At strong disorder, the phase diagram thus becomes dominated by the ME phase at nonzero $J_\perp$.  The phase diagram for different finite values of interaction ($\Delta$) and intra-leg hopping ($J$) is similar to Fig.~\ref{fig:3}.

We stress that the ME phase is not simply a reentrance of the ergodic phase, where ETH holds in \textit{all} eigenstates at finite energy density; on the contrary, there is a sharp transition between them, as we show below.  Indeed, although the singlons tend to delocalize as $W$ is increased at finite $J_{\perp}$, the doublons and holons tend to localize more strongly as they become further off-resonance with the singlons and with each other.  Thus, while eigenstates in which singlons predominate tend to thermalize, eigenstates in which doublons and holons predominate tend to localize more strongly.  This is perhaps suggestive of a many-body mobility edge that separates the thermalizing states dominated by singlons from the localized states dominated by doublons and holons. However, the term ``many-body mobility edge" presupposes the existence of a critical many-body energy density that separates localized and extended states; this is not the case here.  Eigenstates in which all rungs of the ladder are occupied by singlons generically arise in the middle of the many-body spectrum, but eigenstates in which all rungs are occupied by doublons or holons can arise at \textit{any} energy density.  This suggests that there is no discernible transition between MBL and ETH eigenstates as a function of energy density. (We will present vivid numerical proof of this fact in Sec.~\ref{sec:mixed}.) Rather, we will argue in Sec.~\ref{sec:singlon} that there is a \textit{finite doublon/holon density} $n^*_{\rm DH}$ at which a transition between thermalizing and localizing behavior occurs.  Generic eigenstates of the two-leg ladder consist of a mixture of ``hot" singlons and ``cold" doublons and holons, and their interconversion is heavily suppressed by disorder. The singlons, doublons, and holons are thus suspended in a mixture with one another, and configurational properties of this mixture  determine whether or not an eigenstate is thermal or MBL.  This is the essence of the mobility emulsion.

%%%%%%%%%%%%%%%%%%%%%%%
\subsection{Quantitative understanding of the phase diagram}\label{sec:transition_indicators}
%%%%%%%%%%%%%%%%%%%%%%%

We now discuss the quantitative indicators used to calculate the phase diagram shown in Fig.~\ref{fig:3}.  Given the discussion in the previous section, it is necessary to keep track of the singlon and doublon degrees of freedom that become stable excitations at strong disorder.  In the MBL mirror glass phase, both singlons and doublons/holons are localized in space as the onsite boson density is approximately conserved.  Moreover, the singlon ``spin" degree of freedom is frozen into a pattern of ``magnetization" that spontaneously breaks the $\mathbb Z_2$ mirror symmetry $M$.  The ``spin state" ($|\!\uparrow\rangle_i$ or $|\!\downarrow\rangle_i$) of each singlon can be measured using the local polarization
\begin{subequations}
\label{eq:sigma and d}
\begin{align}
\sigma_i=n_{1,i}-n_{2,i},
\label{eq:sigma}
\end{align}
which gives $\pm 1$ when acting on the state $|\!\uparrow\rangle_i$ or $|\!\downarrow\rangle_i$, respectively, and $0$ when acting on the doublon ($|\bullet\rangle_i$) or holon ($|\circ\rangle_i$) state.  Note that the polarization $\sigma_i$ is odd under $M$.  In the ME phase, the singlons delocalize in the manner discussed in the previous section; however, a finite fraction of all eigenstates contain localized doublons and holons.  One can keep track of whether rung $i$ hosts a doublon or holon using the local density
\begin{align}
d_i=n_{1,i}+n_{2,i}-1,
\label{eq:d}
\end{align}
which gives $\pm 1$ when acting on the doublon and holon states $|\bullet\rangle_i$ and $|\circ\rangle_i$, respectively, and $0$ when acting on the singlon states $|\!\uparrow\rangle_i$ and $|\!\downarrow\rangle_i$.  Note that the number of doublons must equal the number of holons in the system at half filling.  We will use these quantities to define two indicators that distinguish the three phases in Fig.~\ref{fig:3}.
\end{subequations}

To keep track of the freezing of singlons in the MBL mirror glass phase, we make use of the mirror-glass order parameter defined in Ref.~\cite{Iadecola18}.  This order parameter is defined at the level of individual eigenstates $|E_n\rangle$ with many-body energy $E_n$:
\begin{equation}\label{eq:q}
q_n=\frac{1}{L^2}\sum_{i,j=1}^{L}\langle E_n| \sigma_i\sigma_j|E_n\rangle^2.
\end{equation}
The definition of $q_n$ is motivated as follows. At infinite temperature, a generic eigenstate in the mirror-glass phase has nonzero local polarization, $\langle\sigma_i\rangle\neq0$, but the sign of this polarization is generically random, so that $\sum_i\langle\sigma_i\rangle=0$. However, the squares of these expectation values add coherently to yield $q_n>0$ for such states.  In this sense, $q_n$ is a faithful detector of spontaneous mirror-symmetry breaking in a many-body eigenstate.  Such spontaneous symmetry breaking can only occur at infinite temperature if a finite fraction of the eigenstates are MBL. [However, the converse of this statement does not hold since singlons can form symmetric (paramagnetic) MBL states when their typical separation is large and their mutual interactions are weak.]  Thus, we also take the presence of a nonvanishing infinite-temperature expectation value of the mirror-glass order parameter as evidence of MBL.  Moreover, $q_n=0$ generically in the ergodic phase, owing to the no-go theorems mentioned in the previous section.

\begin{figure}[t!]
\includegraphics[width=.95\columnwidth]{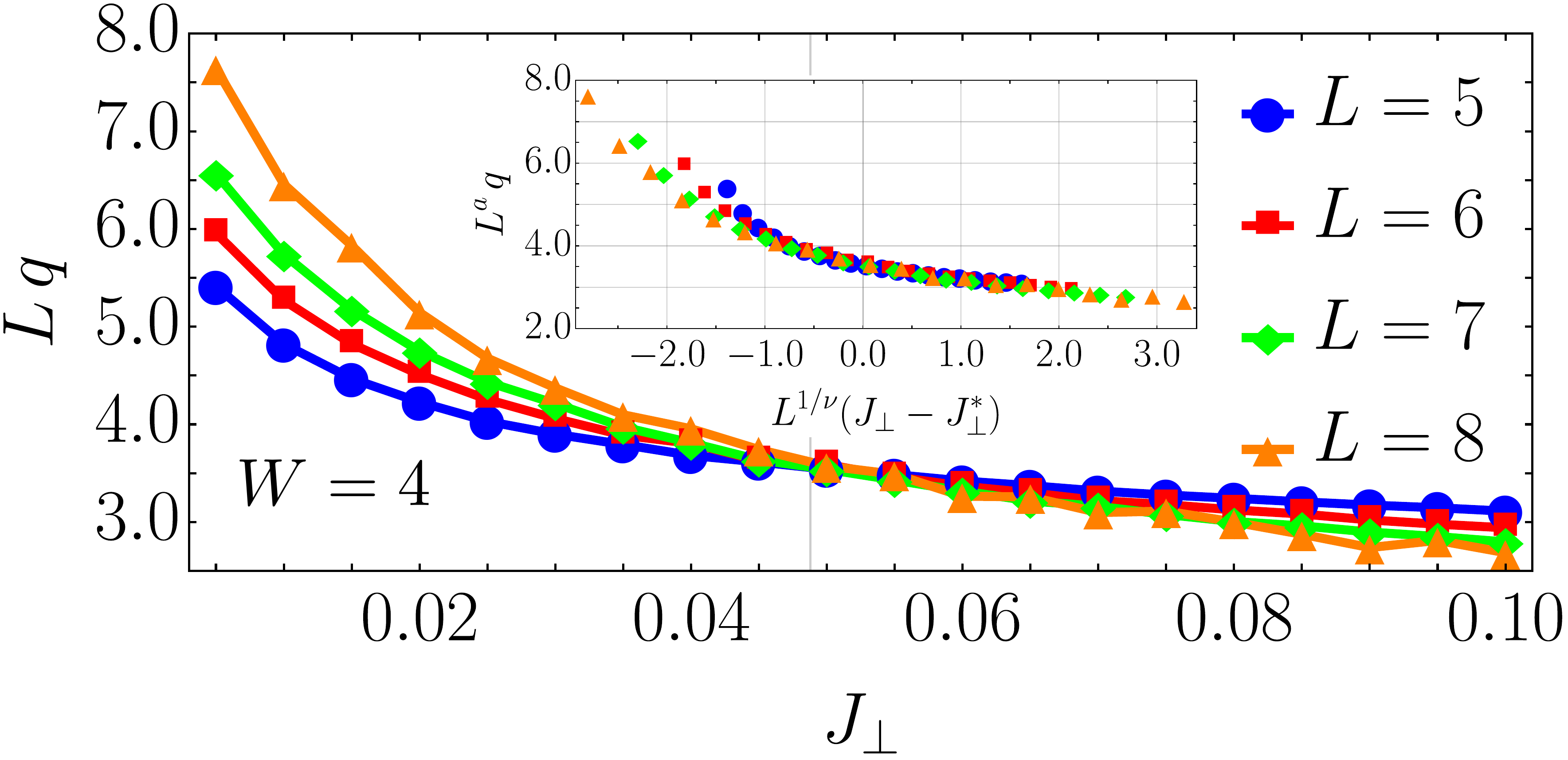}
\includegraphics[width=.95\columnwidth]{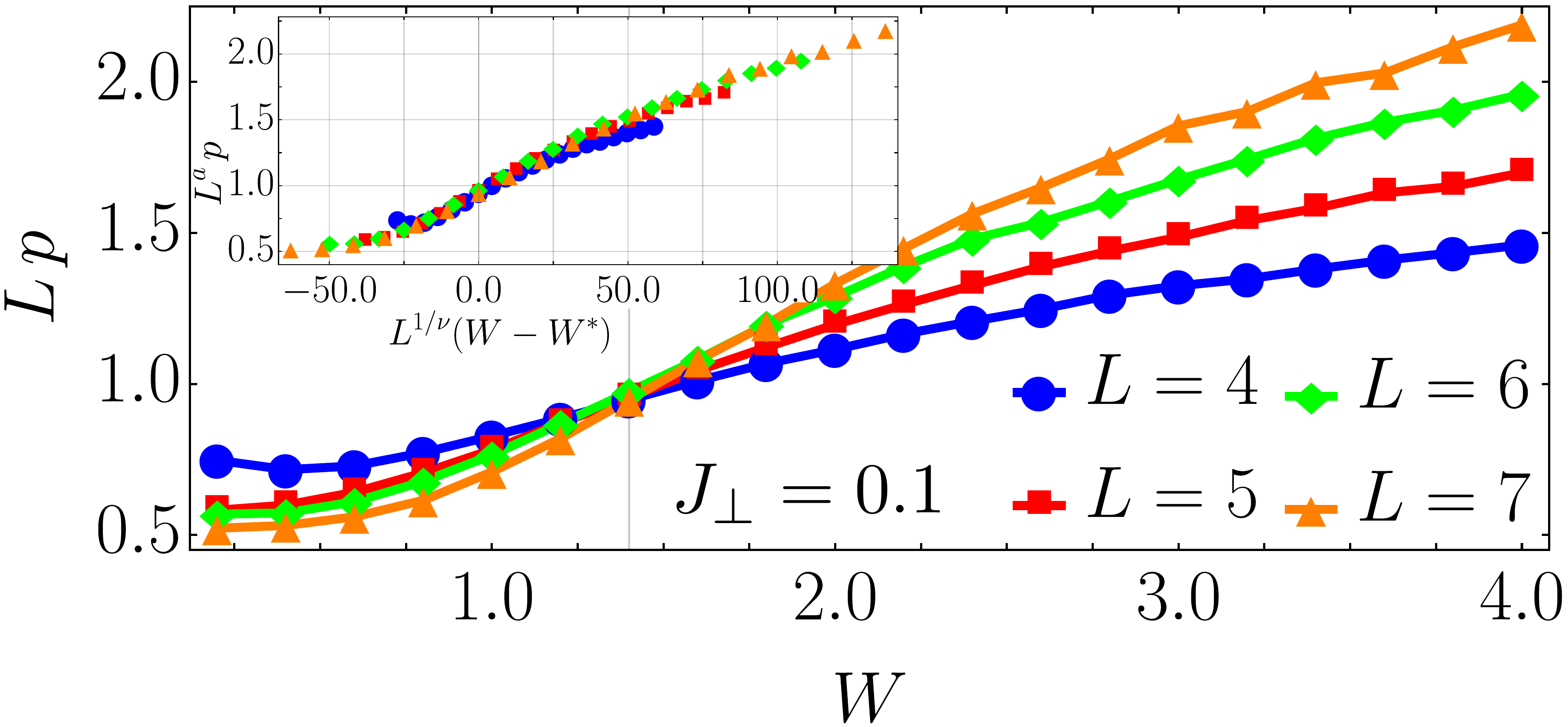}
\caption{
Representative plots of the mirror-glass order parameter \eqref{eq:q} and the doublon correlator \eqref{eq:p} used to determine the horizontal and vertical phase boundary lines, respectively, in Fig.~\ref{fig:3}.  Top: The energy- and disorder-averaged mirror-glass order parameter \eqref{eq:q} in the zero doublon/holon sector, rescaled by the system size $L$, at $W=4$. For all data shown we set $\Delta=0.5$ and $J=1$.  The inset shows the scaling collapse of the data near the transition point $J^*_\perp$, for $a\approx1$, $\nu\approx2/3$, and $J^*_\perp\approx 0.049$. The approximate location of the transition is shown as a grey vertical line in the main plot.  Bottom: The infinite-temperature disorder-averaged doublon correlator \eqref{eq:p}, rescaled by the system size $L$, at $J_\perp = 0.1$.  The inset shows the scaling collapse of the data near the transition point $W^*$, for $a\approx1$, $\nu\approx2/3$, and $W^*\approx 1.4$. 
}
\label{fig:4}
\end{figure}

We plot the infinite temperature average of $q_n$ in Fig.~\ref{fig:4} across the MBL-ME phase boundary (the horizontal blue curve in Fig.~\ref{fig:3}) as a function of the inter-leg coupling $J_\perp$ for a representative choice of disorder strength $W$.  To compute it, we performed shift-invert exact diagonalization to target states in the middle of the many-body spectrum at system sizes up to $L=8$ (i.e., for systems containing as many as $2L=16$ sites).  Results are averaged over 50 energy eigenstates and at least 1250 disorder realizations.  In order to obtain clearer finite-size scaling, we calculate $q_n$ for eigenstates in the zero-doublon/holon sector. Eigenstates in this sector have the strongest tendency to thermalize, as we will see in the next section.  We use the following definition of the doublon/holon density,
\begin{align}
n_{\mathrm{DH},n}=\frac{1}{L}\sum^L_{i=1}\langle E_n| d_i |E_n\rangle^2,
\label{eq:n_DH}
\end{align}
which evaluates to zero in an eigenstate containing only singlons, and which evaluates to one in an eigenstate containing only doublons and holons.  Once this postselection has been made, a clear finite size scaling collapse of the quantity $L\, q$ is observed near the transition (see inset of top panel of Fig.~\ref{fig:4}). The correlation-length exponent $\nu$ obtained from this scaling collapse is approximately $2/3$. The critical inter-leg coupling $J^*_\perp$ is estimated from the crossing point of the finite-size curves: to the left of the crossing, the quantity $L\, q$ scales to a finite value, while to the right of the crossing, it scales to zero.  

To probe the localization of doublons and holons at strong disorder, which is characteristic of both the MBL and ME phases, we define the doublon/holon correlator,
\begin{equation}\label{eq:p}
p_n=\frac{1}{L^2}\sum_{ij}\langle E_n|d_i d_j|E_n\rangle^2.
\end{equation}
This correlator operates on a similar principle to the mirror-glass order parameter \eqref{eq:q}, although it is defined in terms of the operators $d_i$, which are even under the mirror symmetry $M$.  In particular, when finite, it indicates that the eigenstate $|E_n\rangle$ contains a frozen pattern of doublons and holons.
In the ETH phase, the infinite temperature average of $p_n$, namely $\sum_n p_n/\mathcal D$, where $\mathcal D$ is the Hilbert space dimension, is zero because doublons are not generically stable at nonzero energy densities (with or without interactions). We show this behavior in the bottom panel of Fig.~\ref{fig:4}, where we show the infinite temperature average of $p_n$ for system sizes $L$ up to $7$ (i.e. for systems with up to $14$ sites) as a function of the disorder strength $W$ for a representative value of $J_\perp$.  We obtain these data using full diagonalization and average the results over at least 1250 disorder realizations. We again obtain finite-size scaling collapse with a correlation-length exponent $\nu\approx 2/3$ (see inset of bottom panel of Fig.~\ref{fig:4}). In the ETH phase, we see that the quantity $L\, p$ scales to zero with increasing $L$, while it grows with $L$ in the MBL phase and scales to a nonzero value. The crossing point of the finite-size curves can again be used to estimate the critical disorder strength $W^*$ for the transition out of the ETH phase. We expect a weak dependence of the critical parameters $J_\perp^*$ and $W^*$ (see Fig.~\ref{fig:4}) on the interaction ($\Delta$) and intraleg coupling ($J$) strength. The exponent $\nu$ is likely to be universal although its precise value may necessitate more numerical studies beyond the scope of the current work.

%%%%%%%%%%%%%%%%%%
%%%%%%%%%%%%%%%%%%
%%%%%%%%%%%%%%%%%%
%%%%%%%%%%%%%%%%%%

Ultimately, we use the mirror glass order parameter $q$ and the doublon correlator $p$ to distinguish among the ETH, MBL, and ME phases in Fig.~\ref{fig:3} as follows.  In the ETH phase, neither $q$ nor $p$ scales to a finite value in the thermodynamic limit---the system is thermalizing, so neither singlons, doublons, nor holons are stable degrees of freedom.  In the MBL mirror glass phase, both $q$ and $p$ scale to finite values.  The singlons freeze and break $M$ spontaneously at infinite temperature, so that $q$ is finite, but the doublons and holons are also localized.  Indeed, the singlon polarization $\sigma_i$ and the doublon density $d_i$ can be used to reconstruct the local densities $n_{\alpha,i}$ on the legs $\alpha=1,2$, which are both approximately conserved quantities and have finite overlap with the LIOMs in the MBL phase.  Finally, the ME phase is characterized by $q$ scaling to zero (due to the thermalization or paramagnetism of singlons) while $p$ scales to a finite value due to the localization of doublons and holons. We emphasize that the characterization of the three dynamical phases---ETH, ME, and MBL---using the $q$ and $p$ parameters as described above and shown in Fig.~\ref{fig:4} is unique and computationally tractable.  A finite $p$ value (``localized doublons/holons") along with a vanishing $q$ (``thermalized or paramagnetic singlons") uniquely distinguishes the ME phase from both the ETH and MBL phases.
  We provide further numerical evidence for the thermalization of the singlons in Secs.~\ref{sec:singlon}, \ref{sec:mixed} and \ref{sec:dynamics}.

%%%%%%%%%%%
\section{Singlon thermalization and its breakdown in the mobility emulsion}\label{sec:singlon}
%%%%%%%%%%%

In this section, we derive an effective model that takes explicit advantage of the approximate conservation of the doublon/holon density~\eqref{eq:n_DH} in the limit of strong disorder $W\gg J,J_\perp,\Delta$. In this limit, all doublons and holons in the system are strongly confined to their rungs and can be considered completely frozen to first approximation.  Indeed, any eigenstate in which each rung of the ladder is occupied by a doublon or holon (i.e., $n_{\rm DH}=1$) is manifestly fully localized in this limit, since hopping between legs is forbidden by Pauli exclusion and each leg is assumed to be MBL in the decoupled limit $J_\perp=0$.

However, since the singlons do not couple directly to the disorder potential (due to the mirror symmetry), the eigenstates in the all-singlon sector (i.e., $n_{\rm DH}=0$) are highly nontrivial. To determine the fate of such states we derive an effective model for the singlon degrees of freedom using a Schrieffer-Wolff transformation.  The resulting model will give insight into the thermalization of eigenstates in the all-singlon sector in the ME phase.  Furthermore, we will show that this thermalization is stable to the addition of a finite density of doublons and holons.  However, we will also argue that this thermalization breaks down when the singlons are sufficiently dilute and occupy a small but finite fraction of the rungs of the ladder.  This will lead us to hypothesize that there is a \textit{critical doublon/holon density} $n^*_{\rm DH}$ above which the system remains localized, and below which the system thermalizes.

\subsection{Effective singlon model}\label{sec:effective}

%%%%%%%%%%%%%%%%%%
%%%%%%%%%%%%%%%%%%
\begin{figure*}[t!]
\includegraphics[width=0.75\textwidth]{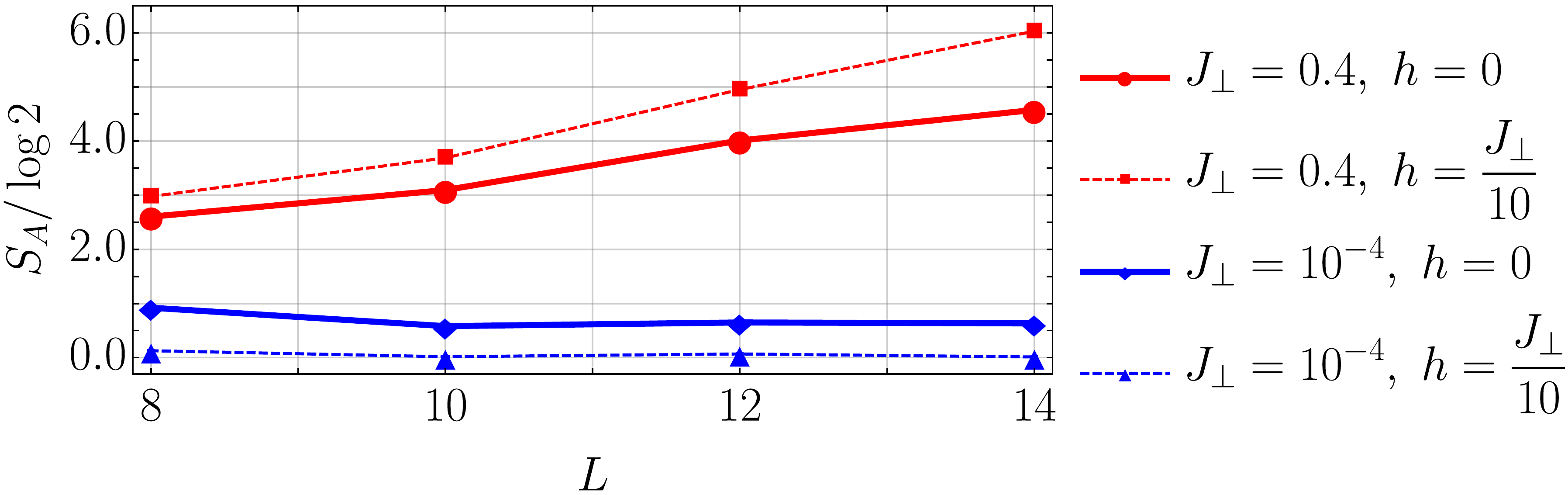}
\caption{
Bipartite entanglement entropy in the effective model \eqref{eq:Heff} describing states in the all-singlon sector $n_{\rm DH}=0$.  We fix $W=10$ and, as in Figs.~\ref{fig:3} and \ref{fig:4}, we work at $\Delta=0.5$ and $J=1$.
At small $J_\perp$ (blue curves) the entanglement entropy exhibits a very weak dependence on system size and appears to saturate to a value of order $\log 2$.  At larger $J_{\perp}$ (red curves), it increases monotonically with system size.  To obtain the dashed curves, a weak uniform ``longitudinal field" $h=J_\perp/10$, which breaks the mirror symmetry $M$, was added.  This field does not change the qualitative behavior, indicating that the ME phase is not protected by mirror symmetry.
}
\label{fig:5}
\end{figure*}
%%%%%%%%%%%%%%%%%%
%%%%%%%%%%%%%%%%%%

To derive an effective model for the singlons, we use a Schrieffer-Wolff transformation~\cite{Schrieffer66} to eliminate the intra-leg hopping piece of the full Hamiltonian $H$, which we denote as
\begin{align}
\hat{J}=\frac{J}{2}\sum_{\alpha,i}(b^\dagger_{\alpha,i}b_{\alpha,i+1}+{\rm H.c.}).
\end{align}
This is achieved with the unitary transformation
\begin{subequations}
\begin{align}\label{eq:HeffSW}
H_{\rm eff}
=
e^{S}H e^{-S}
\!\!=\!
H
\!+\!
[S,H]
\!+\!
\frac{1}{2}[S,[S,H]]+\dots,
\end{align}
with the generator $S$ chosen such that
\begin{align}\label{eq:S}
\hat J=[H_0,S],
\end{align}
where
\begin{align}
H_0=H-\hat{J}-H_\perp.
\end{align}
\end{subequations}
Working to second order in $J$, one then obtains 
\begin{align}\label{eq:Heff2}
H_{\rm eff}=H_0+H_\perp+\frac{1}{2}[S,\hat{J}]+\frac{1}{2}[S,[S,H_\perp]],
\end{align}
which can now be projected into a sector with a fixed configuration of doublons and holons.  After projection, the Hamiltonian for the singlons can be written in terms of the effective Ising spin operators
\begin{subequations}
\begin{align}
\sigma^x_i = |\!\uparrow\rangle_i\langle \downarrow\!|_i+ |\!\downarrow\rangle_i\langle \uparrow\!|_i,
\end{align}
and
\begin{align}
\sigma^z_i= |\!\uparrow\rangle_i\langle \uparrow\!|_i- |\!\downarrow\rangle_i\langle \downarrow\!|_i,
\end{align}
which are defined in terms of the local singlon states \eqref{eq:singlon states}.
\end{subequations}
To zeroth order in the intra-leg hopping $J$, the effective model is just the projection of $H_0+H_\perp$, which gives the clean transverse field Ising chain,
\begin{subequations}
\label{eq:Heff}
\begin{equation}\label{eq:Heff clean}
H_{\rm eff}=\frac{J_\perp}{2} \sum_i \sigma^x_i+\frac{\Delta}{2} \sum_i \sigma^z_i\sigma_{i+1}^z+\delta H_{\rm eff}.
\end{equation}
Higher-order virtual processes are contained in $\delta H_{\rm eff}$, which is obtained by projecting $\frac{1}{2}[S,\hat{J}]+\frac{1}{2}[S,[S,H_\perp]]+\cdots$ into a sector with a fixed configuration of doublons and holons. For the remainder of this section we will focus on the terms that arise in the zero-doublon sector; we will consider what occurs for singlons in a generic doublon-holon background in Sec.~\ref{sec:stability}.

In addition to the clean part of Eq.~\eqref{eq:Heff}, there are virtual processes that involve pairs of singlons transitioning to doublon-holon pairs and back. Such processes are generated by the Schrieffer-Wolff transformation discussed above and are the mechanism by which randomness enters the effective model, as the intermediate doublon-holon states depend explicitly on the disorder potential. Keeping corrections to second order in $J$ and first order in $\Delta,J_\perp$, we obtain
\begin{widetext}
\begin{equation}\label{eq:dHeff}
\delta H_{\rm eff}=\sum_i \frac{J^2 \Delta}{32\,\delta h_{ii+1}^2} \left( \sigma^z_{i-1} \sigma^z_{i}+\sigma^z_{i+1} \sigma^z_{i+2}-\sigma^z_i\sigma^z_{i+2}-\sigma^z_{i-1}\sigma^z_{i+1}\right)
- \frac{J^2 J_\perp}{16}\sum_i\left(\delta h_{ii-1}^{-2}+\delta h_{ii+1}^{-2}\right)\sigma^x_i,
\end{equation}
\end{widetext}
\end{subequations}
where $\delta h_{ij}=h_i-h_{j}\sim W$. We now see that as the disorder strength $W$ increases, the energy scales for the effective randomness in $\delta H_{\rm eff}$, namely $\Delta J^2/W^2$ and $J_\perp J^2 /W^2$, \emph{decrease} rapidly. This implies that the effective singlon Hamiltonian $H_{\rm eff}$ becomes \emph{cleaner} the stronger the potential disorder.

In the limit $J_\perp\ll\Delta$, where the phase diagram in Fig.~\ref{fig:3} resides, the effective model \eqref{eq:Heff} can be understood as a theory of weakly interacting domain walls in a weak disorder potential.  The transverse field constitutes a kinetic term for the domain walls, while the nearest-neighbor Ising interaction constitutes a potential energy term that is disordered on the scale $(J/W)^2$.  The next-nearest-neighbor Ising interaction can be viewed as a density-density interaction for the domain walls; hence, the effective singlon model cannot be mapped to one of Anderson-localized free particles.
Heuristically, one expects that when the transverse field is much smaller than the effective domain-wall disorder potential, i.e., $J_\perp\ll J^2\Delta/W^2\ll\Delta$, the system remains fully localized since the domain wall kinetic energy, of order $J_\perp$, is insufficient to overcome the relatively large effective disorder potential. However, in the range $J^2\Delta/W^2\ll J_\perp \ll \Delta$, the kinetic energy of the domain walls dominates and one expects ergodic behavior at nonzero energy densities as the domain walls begin to propagate and interact. As a result, we predict the singlons to exhibit a localization phase transition at strong disorder as a function of the inter-leg coupling $J_\perp$ at a critical value of order $J^2\Delta/W^2$.  Note that this behavior is reflected qualitatively in the phase diagram in Fig.~\ref{fig:3}, which was obtained by simulating the full model \eqref{eq:H}: at sufficiently large disorder strength $W$, the critical coupling $J^*_{\perp}$ at which the mirror-glass order is lost decreases as $W$ increases.

Interestingly, the primary transport that arises due to the thermalization of singlons is \emph{energy}, rather than charge, transport. This is because the singlons are actually quasidegenerate states of a single particle localized to a rung of the ladder.  Thus, charge transport is still heavily suppressed at short times despite the thermalization of the singlons. Ultimately, however, when the singlon density in the initial state is sufficiently large we predict slow charge transport to arise, as discussed in Sec.~\ref{sec:stability}.

To further substantiate the claim that the all-singlon sector indeed undergoes a localization transition as a function of $J_\perp$, we have performed an exact diagonalization study of the effective model \eqref{eq:Heff}.  This allows us to reach system sizes up to $L=14$, which is roughly twice as large as those used to construct Fig.~\ref{fig:3}.  We focus on our results for the bipartite entanglement entropy $S_A$, which is calculated by partitioning the system into subsystems $A$ and $B$ of size $L/2$:
\begin{align}\label{eq:sA}
S_A=-\rho_A\log \rho_A,
\end{align}
where $\rho_A=\text{tr}_{B}\rho$ and $\rho$ is the density matrix of $H_{\rm eff}$.  In Fig.~\ref{fig:5}, we show the infinite-temperature average of $S_A$ over the all-singlon sector as a function of system size $L$.  Our results are averaged over at least 20 disorder realizations, and error bars representing one standard error are smaller than the plot markers.  At small inter-leg coupling, $J_\perp=10^{-4}$, the entanglement entropy exhibits a very weak dependence on system size and appears to have saturated to a small value of order $\log 2$ by $L=14$.  This is consistent with the expected area-law entanglement scaling in the MBL phase.  At larger interleg coupling, $J_\perp=0.4$, the entanglement entropy increases monotonically with system size in an approximately linear fashion, consistent with the volume-law scaling expected for a thermalizing system.

We tested the stability of the area- and volume-law regimes by adding a weak uniform longitudinal field to Eqs.~\eqref{eq:Heff}, such that
\begin{align}
H_{\rm eff}\to H_{\rm eff}+h\sum_i\sigma^z_i
\end{align}
with $h=J_{\perp}/10$, which weakly breaks the $\mathbb Z_2$ symmetry of $H_{\rm eff}$. In the full model \eqref{eq:H}, this amounts to weakly breaking the mirror symmetry by adding a uniform bias between the two legs of the ladder (up to small random corrections to this bias that would result from the Schrieffer-Wolff transformation). The results for this case are plotted as dashed lines in Fig.~\ref{fig:5}.  Evidently, the area- and volume-law regimes are stable to breaking the underlying mirror symmetry of the problem; in fact, the discrepancy between the two regimes is enhanced, as may be expected due to the mixing of the symmetry sectors. This provides evidence that the ME phase is not a symmetry-protected phase and can exist even in the absence of mirror symmetry (unlike the mirror-glass, by definition).

\subsection{Stability of the localized and thermalizing limits}\label{sec:stability}

So far, we have demonstrated with numerical and analytical evidence that the ME phase is characterized by full MBL in the all-doublon/holon sector with $n_{\rm DH}=1$, and by thermalization in the all-singlon sector with $n_{\rm DH}=0$.  We will now argue that these two extreme limits are stable to the addition of a finite density of singlons and doublons or holons, respectively.  This will lead us to the hypothesis that the ME phase is characterized by the existence of a critical doublon/holon density $n^*_{\rm DH}$ below which eigenstates thermalize, and above which MBL sets in.

\subsubsection{Stability of the localized limit}\label{sec:localized}

\begin{figure}[t!]
\includegraphics[width=\columnwidth]{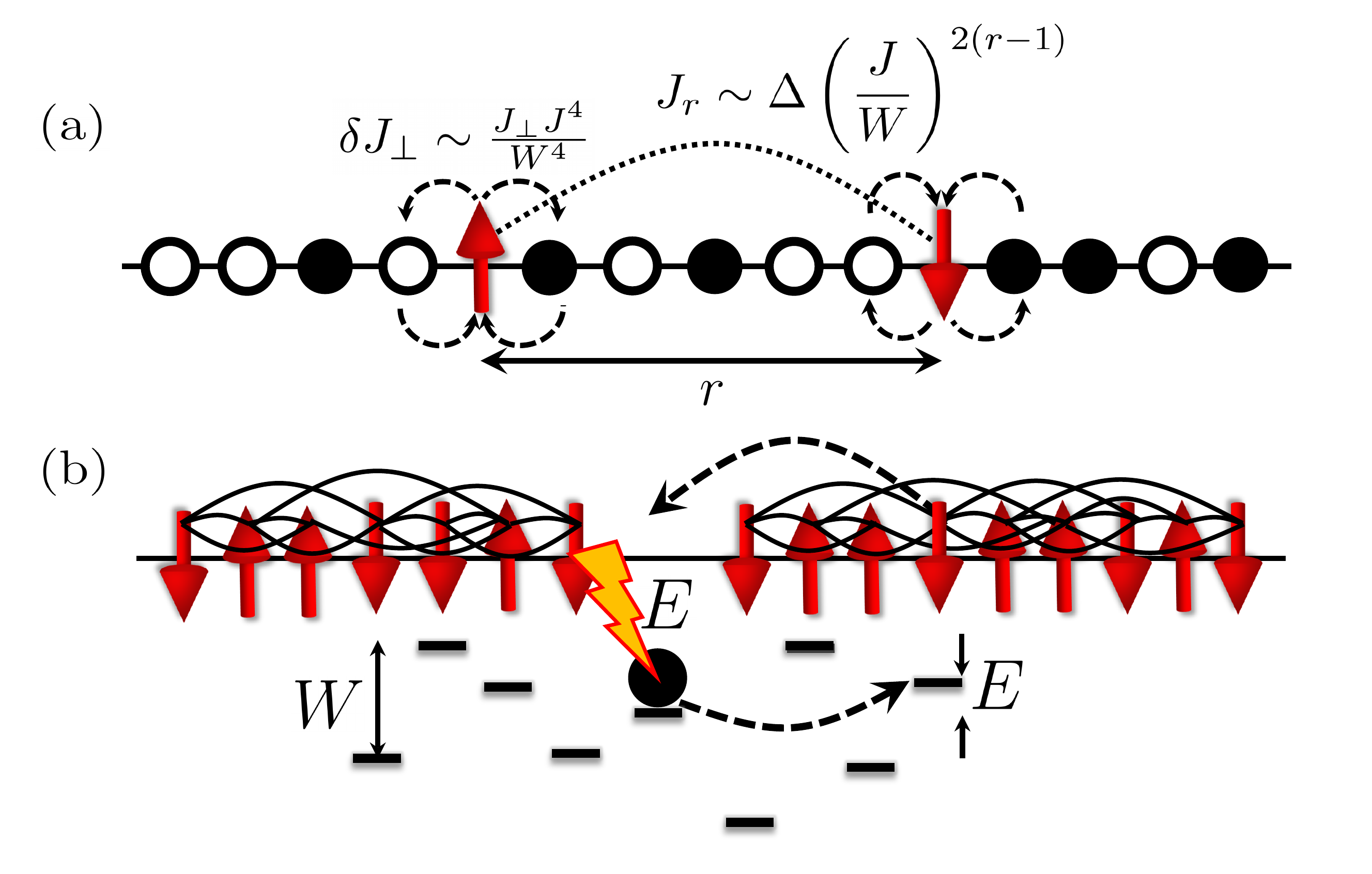}
\caption{
Schematic depiction of the localized and thermalizing limits discussed in Secs.~\ref{sec:localized} and \ref{sec:thermalizing}.  (a) Eigenstates containing a sparse distribution of singlons remain localized because the self-generated random field $\delta J_{\perp}$ is parametrically stronger at strong disorder than the inter-singlon interaction $J_{r}$, which arises at a higher order in perturbation theory when the inter-singlon distance $r$ is sufficiently large.  (b) Eigenstates containing a sparse distribution of doublons and holons thermalize because the doublons and holons undergo variable-range hopping mediated by the thermalizing bath of singlons.  A doublon absorbing energy $E$ from the bath can hop a distance that scales as $W/E$, but the amplitude for this process is exponentially suppressed by the decay of its wavefunction.  This leads to an optimal hopping distance and rate that is calculated in the Appendix.
}
\label{fig:6}
\end{figure}

Consider an eigenstate in which the vast majority of rungs occupy doublon/holon states, but that contains a very sparse random distribution of singlons [see Fig.~\ref{fig:6}(a)].
For an isolated singlon embedded in such a localized background, we generically find a random renormalization of the inter-leg hopping that depends on the local doublon-holon configuration. In particular, a special case arises when a singlon is surrounded by only doublons or by only holons. In this case the inter-leg hopping in fact does \emph{not} get renormalized, because in this sector there is only one boson (or one hole) in the entire system and the hardcore constraint becomes irrelevant. As a result, the Hamiltonian in the single-particle sector commutes with the inter-leg hopping term, Eq.~\eqref{eq:Hc}, and leads to the conservation of $H_\perp$, whose value is carried and preserved by the boson as it hops. Since the eigenvalues of $H_\perp$ cannot mix in the single-particle sector, the local transverse hopping $J_\perp$ does not get renormalized. 

However, in every sector with two or more particles (or holes) in the system, the term $H_\perp$ is no longer conserved because the transverse and longitudinal hopping terms do not commute for hardcore bosons. This is an interaction effect that leads to a weak renormalization of the inter-leg tunneling amplitude that scales as $\delta J_\perp\sim J_\perp (J/W)^2$ when a singlon neighbors another singlon, cf. Eq.~\eqref{eq:dHeff}. However, when a singlon is surrounded by a doublon-holon pair, e.g.~as depicted in Fig.~\ref{fig:6}, the renormalization is weaker and scales, to leading order in $1/W$, as
\begin{equation}\label{eq:Jperp2}
\delta J_{\perp}\propto J_\perp \left(\frac{J}{W}\right)^4.
\end{equation}
This renormalization can be calculated explicitly by using the generator \eqref{eq:S} to obtain the effective Schrieffer-Wolff Hamiltonian \eqref{eq:HeffSW} to fourth order in $J$, and then projecting into the space of states with singlons on the appropriate sites.  All lower-order corrections to $J_\perp$ vanish identically.
Intuitively, the extra factor of $(J/W)^2$ in Eq.~\eqref{eq:Jperp2} as compared to Eq.~\eqref{eq:dHeff} results from the price a doublon-holon pair must pay to fluctuate into and out of a virtual singlon pair with at least one of the virtual singlons neighboring the target singlon, which then generates a random correction as in Eq.~\eqref{eq:dHeff}.

%%%%%%%%%%%%%%%%%%
%%%%%%%%%%%%%%%%%%
%%%%%%%%%%%%%%%%%%
%%%%%%%%%%%%%%%%%%

The renormalized inter-leg hopping acts as a random local transverse field applied to the singlon [see Fig.~\ref{fig:6}(a)] that must compete with the mutual interactions between the singlons.
The interaction between the sparse singlons arises at higher order in perturbation theory and decays exponentially with their separation due to the strong localization of the doublons and holons that mediate the coupling.  Two singlons separated by a distance $r$ experience a mutual interaction of order
\begin{equation}\label{eq:Jr}
J_{r}\sim \Delta\left(\frac{J}{W}\right)^{2(r-1)},
\end{equation}
see Fig.~\ref{fig:6}(a).
Thus, when the average inter-singlon spacing is sufficiently large this effective interaction is parametrically weaker than the random part of the ``transverse field," Eq.~\eqref{eq:Jperp2}. As a result, singlons in states with a low (but finite) singlon density (corresponding to $n_{\rm DH}$ near, but not equal to, one) are localized and paramagnetic. This implies that the dynamics of an initial density product state with a sufficiently small number of singlons will show neither particle nor energy transport. Such states will also respect mirror-symmetry since the effective spins are polarized along the $\sigma^x$ direction, i.e., there is no mirror-glass order. This manifests itself dynamically as spin precession of singlons initialized along the $\sigma^z$ direction, which corresponds to uncorrelated single-particle hopping between the two legs of the ladder. Thus, the localized sector is stable to the addition of a small singlon density.

\subsubsection{Stability of the thermalizing limit}\label{sec:thermalizing}
We now turn to the opposite limit, considering the case of a single doublon embedded in a sea of thermalizing singlons [see Fig.~\ref{fig:6}(b)]. We shall see that the coupling between the doublon and the singlons leads to singlon-mediated hopping processes reminiscent of Mott variable-range hopping (VRH)~\cite{Mott69}. A key difference, however, is that we must consider states for which the singlon bath is at infinite temperature $T$. In the limit of strong disorder $W\gg J_\perp, \Delta$, the singlon single-particle bandwidth is narrow, so that individual excitations cannot mediate hopping with large energy transfers of order $\sim W$. Thus, the doublon must either hop to a faraway site that has small energy transfer $E\sim J_\perp,\Delta$, or it must absorb an $N$-particle excitation with higher energy $N J_\perp$ or $N\Delta\sim W$. The former (single-particle) process is limited by the decay of the localized wavefunction with distance from the initial site, while the latter ($N$-particle) process is limited by the smallness of the coupling between the doublon and singlons, which suppresses the amplitude of the multiparticle resonance. As we show in the Appendix, the competition between these processes leads to an optimal hopping distance and rate which controls the mobility of doublons. The optimal hopping rate is of the form
\begin{align}
J^*\sim J\, e^{-\alpha\sqrt{W/\Delta}},
\end{align}
where $\alpha$ depends only weakly on $J,W,J_\perp,$ and $\Delta$. At large $W$, the optimal VRH rate is parametrically smaller than $J_\perp$ and $\Delta$, which set the rate of energy transport by the excitations of the thermalizing effective singlon spin chain.
At sufficiently low doublon/holon density $n_{\rm DH}$, such singlon-mediated VRH leads to the motion and subsequent interaction of doublons and holons in the chain, thereby ultimately leading to slow diffusive charge transport. Thus, the thermalized sector is stable to the introduction of a small density of holons/doublons.

The fact that singlons carry energy and not charge indicates that charge transport in the ME phase is heavily suppressed at all timescales when $n_{\rm DH}$ is zero.  However, when the doublon/holon density is small but finite, charge transport is mediated via the VRH discussed above.  This suggests a parametrically large separation of timescales for charge and energy transport when the system is deep in the ME phase and the doublon/holon density is low.

\subsubsection{Critical doublon/holon density for localization}
\label{subsubsec: Critical doublon/holon density for localization}

As the density of singlons is increased, the strength of the interactions among them increases rapidly due to the exponential nature of their effective coupling. We focus now on the regime where the clean component of the transverse field $J_\perp$ is much larger than the induced randomness (in a zero-doublon background), $\Delta(J/W)^2$, so that the zero-doublon sector is strongly thermalizing. We want to estimate the critical doublon/holon density $n^*_{\rm DH}$ where the singlon delocalization transition takes place. Starting from $n_{\rm DH}\sim 1$, i.e., from a dilute set of singlons, one expects that when their interaction becomes comparable to the random component of the ``transverse field" (in a zero-singlon background), Eq.~\eqref{eq:Jperp2}, then the singlons will undergo the delocalization phase transition. Inspecting Eq.~\eqref{eq:Jr}, one sees that third-neighbor singlons separated by doublon/holon sites interact with a strength $\Delta(J/W)^4$, which is comparable to their random field when $J_\perp\sim\Delta$. If the singlons are separated by a further neighbor, their interaction is suppressed by an additional power of $J/W$ and becomes negligible compared to the random field. This suggests that the critical singlon spacing $r\sim 3$ sites, corresponding to a critical doublon/holon density $n^*_{\rm DH}\approx 2/3$ when the system is deep in the ME phase and $J_\perp\sim\Delta\ll W$.

This simple estimate of $n^*_{\rm DH}$ does not take into account what happens in eigenstates in which clusters of several singlons are separated by intervening regions of doublons and holons.  Any two such clusters interact to leading order via a coupling of the form~\eqref{eq:Jr}; the clusters effectively decouple when this interaction is smaller than the minimum level spacing of the two clusters.  Thus, the critical separation between clusters depends on the cluster size in a nontrivial way, and this dependence must be taken into account in order to precisely determine $n_{\rm DH}^*$.

Furthermore, as the system is tuned towards the ETH or mirror-glass phase transitions it is possible, if not highly likely, that $n_{\rm DH}^*$ deviates strongly from its value deep in the ME phase. Near the mirror-glass phase, for example, it is reasonable to expect $n_{\rm DH}^*\to0$ continuously  upon entering the MBL phase where all singlons are localized. This is because the thermalizing singlon states near the phase boundary are extremely fragile so that adding even a small fraction of doublons and holons would immediately lead to localization.   Near the ETH phase transition, however, the localized doublon/holon states become fragile so that adding any fraction of singlons leads to thermalization. This would imply $n_{\rm DH}^*\to1$ upon approaching the ETH phase boundary.  We should note that this speculation also does not take into account the effect of rare configurations at fixed $n_{\rm DH}$ in which singlons are anomalously close together. Further studies with larger systems will be required to reach more definitive conclusions regarding the precise value of the critical density discussed above, which we have argued is generically finite and nontrivial $0<n^*_{\rm DH}<1$ in the ME phase. A precise determination of $n^*_{\rm DH}$ is an important future problem of interest, but our work establishes that such an $n^*_{\rm DH}$ exists in the ME phase.

It is interesting to note that the arguments above also point to the existence of a critical value of $n_{\rm DH}$ that separates eigenstates with and without mirror-glass order in the MBL phase, although finite-size restrictions also pose a challenge to confirming this idea numerically.  Indeed, interactions between singlons are the crucial ingredient that drives the spontaneous $\mathbb Z_2$ mirror symmetry breaking that gives rise to mirror-glass order, and their strength must be compared to that of the local transverse field to determine whether a given singlon remains paramagnetic or participates in the long-range order.  It is important to stress in this case that the putative critical doublon/holon density for mirror-glass order does not imply the presence of delocalized states in the many-body spectrum; all states are localized in the MBL mirror glass phase, and a finite fraction of them participate in the mirror-glass order, as our numerical results in Sec.~\ref{sec:model} indicate.  Any states that do not participate in the long-range order are simply ``paramagnetic" MBL states. 

%%%%%%%%%%%%%%%%%%
%%%%%%%%%%%%%%%%%%
\begin{figure*}[t!]
(a)\includegraphics[width=0.8\columnwidth]{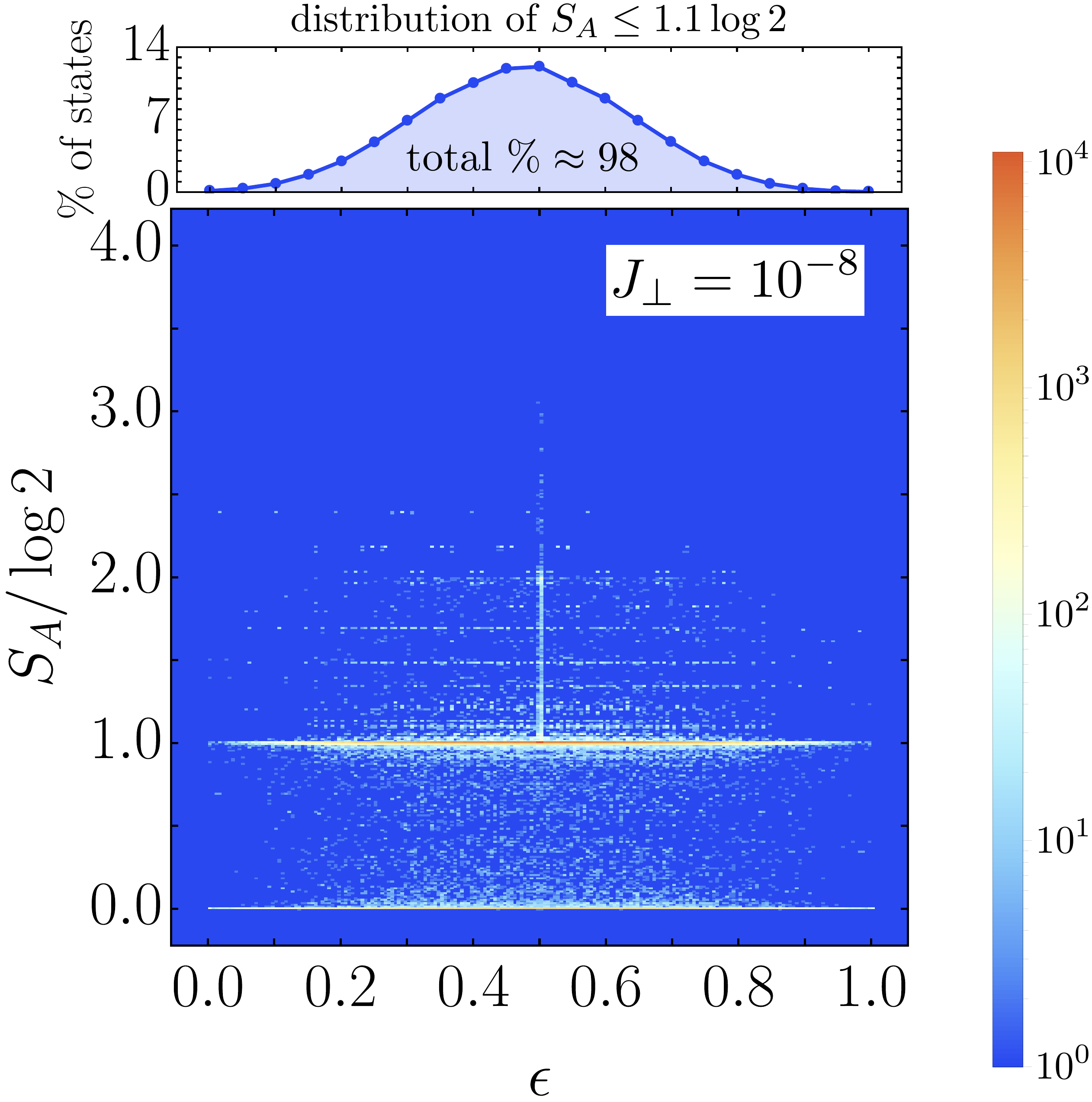}
\hspace{1.5cm}
(b)\includegraphics[width=0.8\columnwidth]{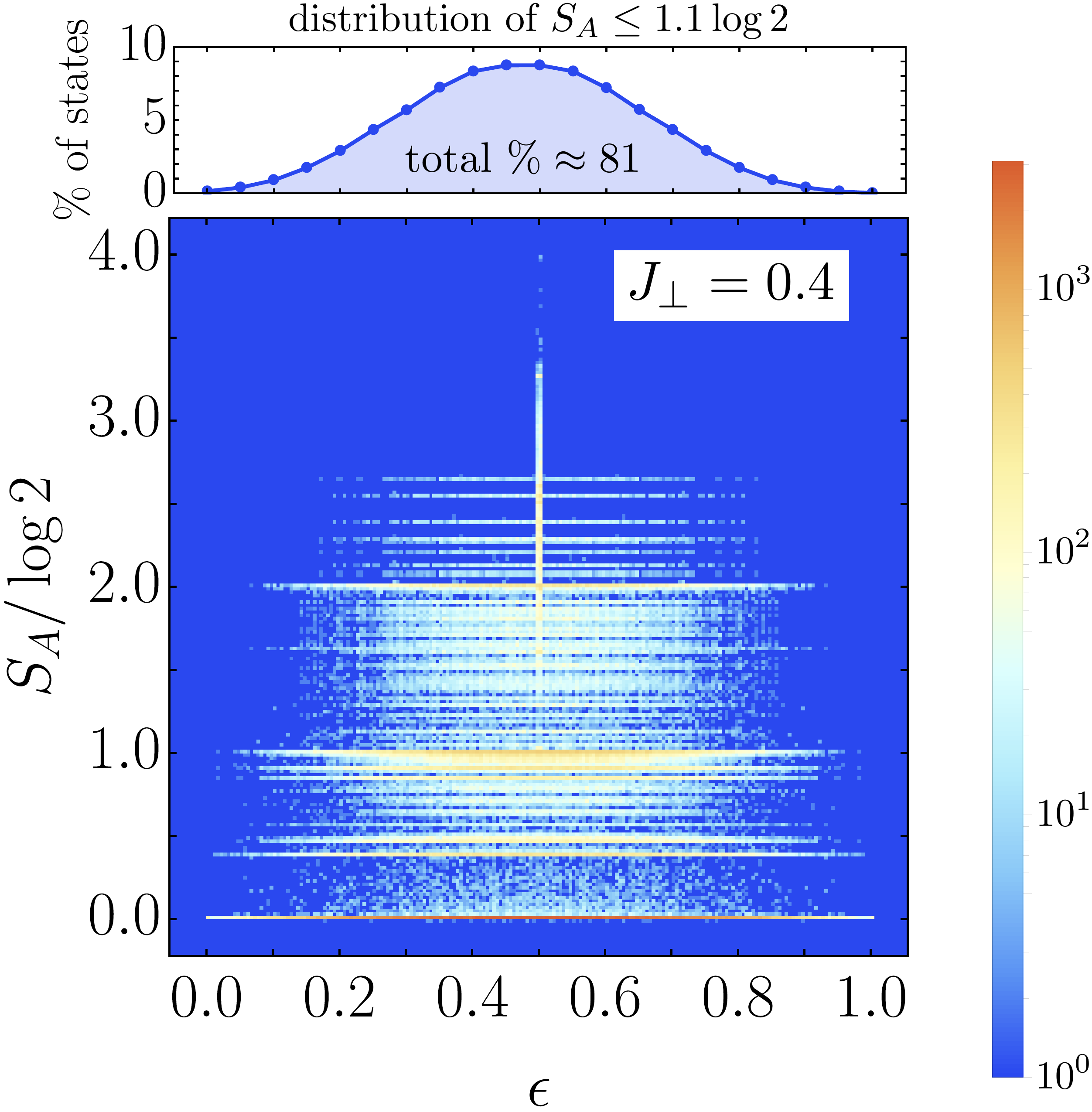}
\caption{Histogram of the bipartite entanglement entropy $S_A$ and  many-body energy density $\epsilon$ accumulated from every eigenstate of 100 disorder realizations.  The color scale indicates the \emph{log} of the number of counts in each bin in order to make outliers more visible. In both panels $J=1,\Delta=0.5, W=1000$ and $L=8$ (16 sites). (a) In the MBL mirror-glass phase, the strong peaks of $S_A$ near the values $0$ and $\log 2$ result from fully-localized paramagnetic states and symmetry-broken cat states, respectively. (b) In the ME phase the distribution of entanglement entropy spreads out to much larger values, indicating a stronger tendency towards thermalization. This is particularly evident for high-entanglement states near $\epsilon\approx 0.5$ which \emph{coexist} with the band of localized states near $S_A\approx 0$. The strong increase of entanglement in the ME phase is \emph{not} accompanied by a many-body mobility edge, as a nonzero fraction of localized states exist and violate ETH for all $\epsilon$ [compare top panels of (a) and (b)].}
\label{fig:7}
\end{figure*}
%%%%%%%%%%%%%%%%%%
%%%%%%%%%%%%%%%%%%

%%%%%%%%%%%%%%%%%%
%%%%%%%%%%%%%%%%%%
\begin{figure*}[t!]
(a)\includegraphics[width=0.8\columnwidth]{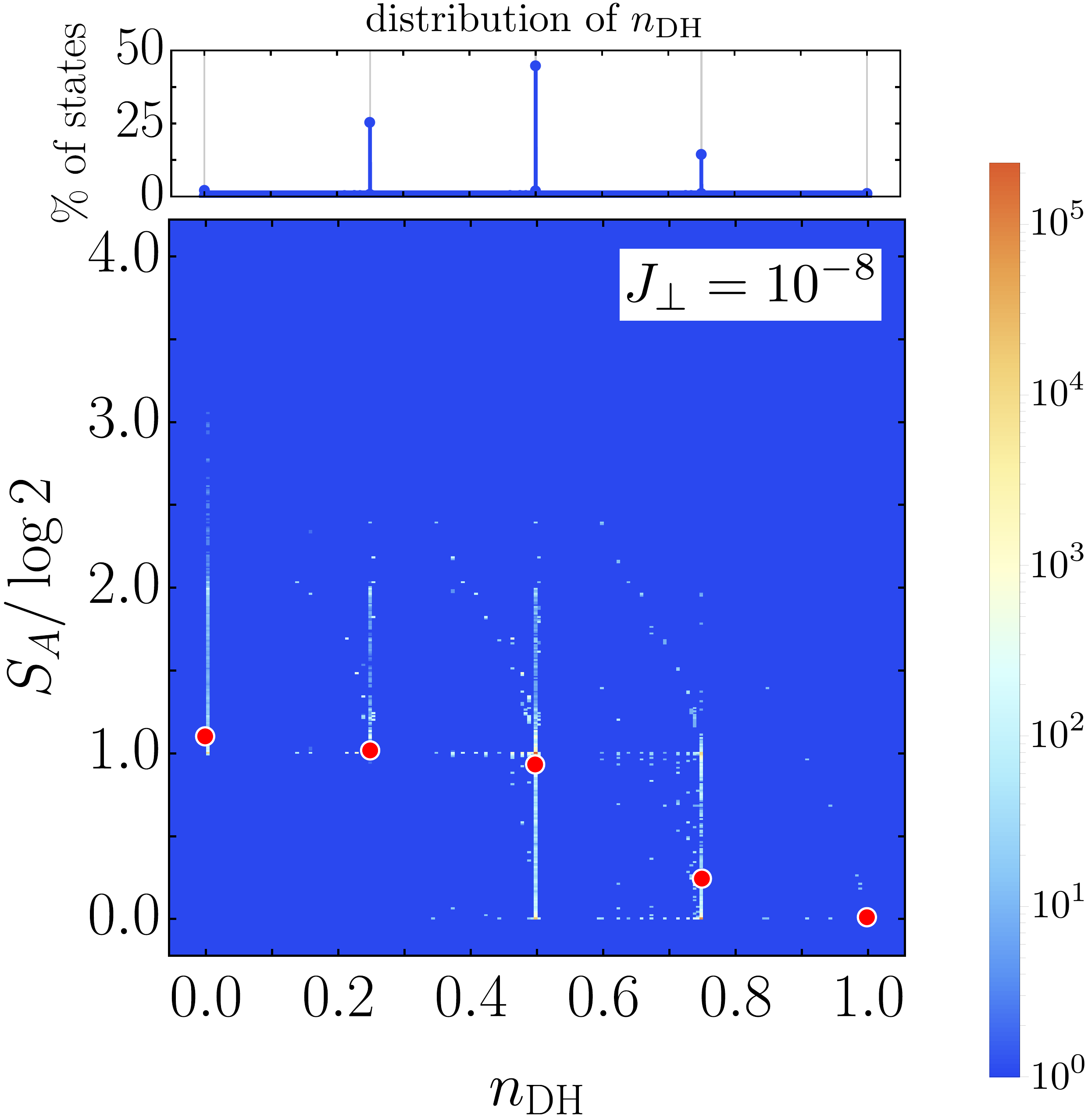}
\hspace{1.5cm}
(b)\includegraphics[width=0.8\columnwidth]{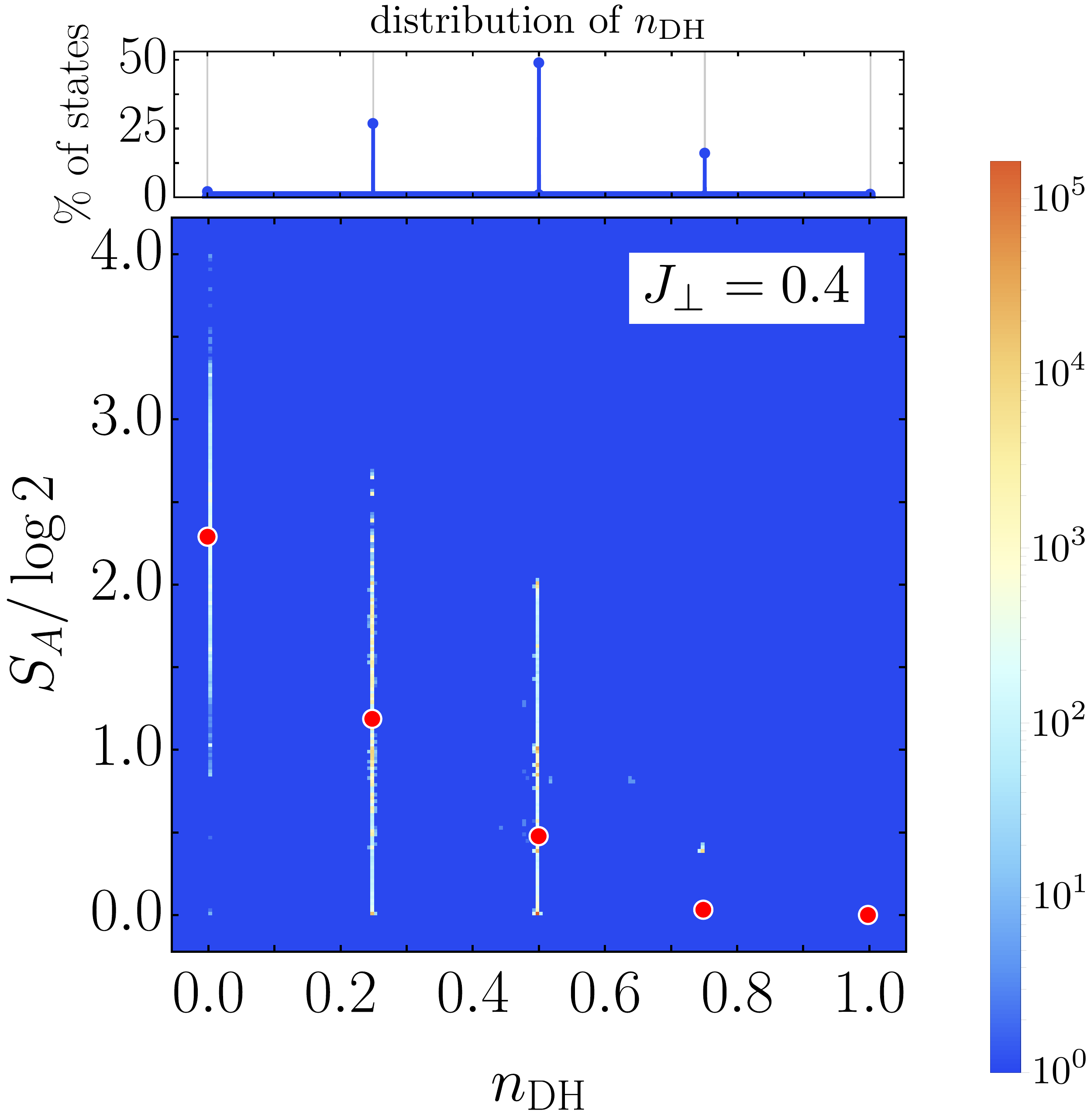}
\caption{Histogram of the bipartite entanglement and doublon/holon density $n_{\rm DH}$ using the same data set and color scale as in Fig.~\ref{fig:7}. The distribution of entanglement organizes into well-separated doublon/holon density bands labeled by \emph{quantized} values of $n_{\rm DH}=2k/L$ ($L=8$ shown) with $k=0,1,\dots,L/2$ [see top panels of (a) and (b)]. (a) In the MBL phase the mean of the distribution of $S_A$ in a given density band (red dots) grows with decreasing $n_{\rm DH}$ but saturates to $\sim \log 2$ for $n_{\rm DH}\lesssim 0.5$.  (b) In the ME phase $S_A$ grows monotonically without saturating as $n_{\rm DH}$ decreases. This indicates that the degree of localization of a given eigenstate is \emph{configuration-controlled}. }
\label{fig:8}
\end{figure*}
%%%%%%%%%%%%%%%%%%
%%%%%%%%%%%%%%%%%%

\subsubsection{Analogy with MBL coupled to a heat bath}

Finally, we note that one can make an analogy between the above stability analysis of the localized and thermalizing limits and the problem of an MBL system coupled to a heat bath~\cite{Nandkishore14,Johri15,Nandkishore15a,Fischer16,Hyatt17}.  In this analogy, one can view the interacting singlons as the ``bath" and the inert doublons and holons as the ``system" of interest.  If $n_{\rm DH} \ll 1/2$, then there are many more singlons in the system than there are doublons and holons. In this case, the many-body density of states of the singlons is nearly continuous and their many-body bandwidth is much larger than that of the doublons and holons. In this case, the system truly resembles the generic case of an MBL system coupled to a heat bath, where it is known on general grounds that the initially localized system will thermalize at infinite time~\cite{Nandkishore14,Johri15,Fischer16}.  However, if $n_{\rm DH} \gg 1/2$, then the ``bath" contains many fewer degrees of freedom than the system and can itself become localized due to their coupling~\cite{Nandkishore15a,Hyatt17}.

The fascinating aspect of this problem in the context of the ME phase is that the emergent parameter $n_{\rm DH}$ essentially \emph{tunes} the ``quality" of the thermal bath as a function of the many-body configuration of an eigenstate. When $n_{\rm DH} \ll 1/2$, the singlon ``bath" has a dense many-body spectrum and can easily mediate thermalization via the VRH process outlined in Sec.~\ref{sec:thermalizing}.  Increasing $n_{\rm DH}$ degrades the quality of the bath until ultimately it is incapable of thermalizing the system, and becomes localized itself via the self-generated random field discussed in Sec.~\ref{sec:localized}.

\section{Hilbert space structure of the mobility emulsion: Absence of a many-body mobility edge}\label{sec:mixed}

The discussion in Sec.~\ref{sec:singlon} indicates that it is the doublon/holon density $n_{\rm DH}$, which is an emergent approximately conserved quantity at strong disorder, that controls whether eigenstates are localized ($n_{\rm DH}>n_{\rm DH}^*$) or thermalizing ($n_{\rm DH}<n_{\rm DH}^*$).  This is in sharp contrast to the case of a putative many-body mobility edge, where localization is controlled by the many-body energy density $\epsilon$, which is not an emergent quantity.  This implies that the ME phase is both sharply distinct from the ETH phase as we have already argued, and from other possible intermediate phases with putative many-body mobility edges. (In particular, unlike in several recently studied incommensurate MBL models in the literature \cite{Luschen18,Modak15,Li15,Li16,Li17}, in the noninteracting limit our system manifests no single particle mobility edge.) In this section, we analyze this claim in further detail and present additional numerical results on the full model \eqref{eq:H} that substantiate it.

To demonstrate the absence of a many-body mobility edge in this model, we investigate the distribution of the bipartite entanglement entropy $S_A$ defined in Eq.~\eqref{eq:sA} as a function of the many-body energy density $\epsilon$.  More concretely, we perform full exact diagonalization of the model \eqref{eq:H} at $L=8$ for 100 realizations of the disorder and record $S_A$ and $\epsilon$ for each eigenstate in each realization.  We choose a large disorder strength, $W=1000$, so that the discrepancy between the degree of localization of the singlons and the doublons and holons is more pronounced.  We plot the results in a two dimensional-histogram for two representative values of $J_{\perp}$ in Fig.~\ref{fig:7}.  Deep in the MBL phase, at $J_{\perp}=10^{-8}$ [Fig.~\ref{fig:7}(a)], the distribution shows that the entanglement entropy clusters around two characteristic values, $0$ and $\log 2$, for any energy density.  This is to be expected, since at such large disorder strengths states where nearly all rungs are occupied by doublons or holons are essentially product states, while states with more singlons, which spontaneously break the mirror symmetry, form many-body cat states with entanglement $\log 2$.

%%%%%%%%%%%%%%%%%%
%%%%%%%%%%%%%%%%%%
\begin{figure}[t!]
\includegraphics[width=\columnwidth]{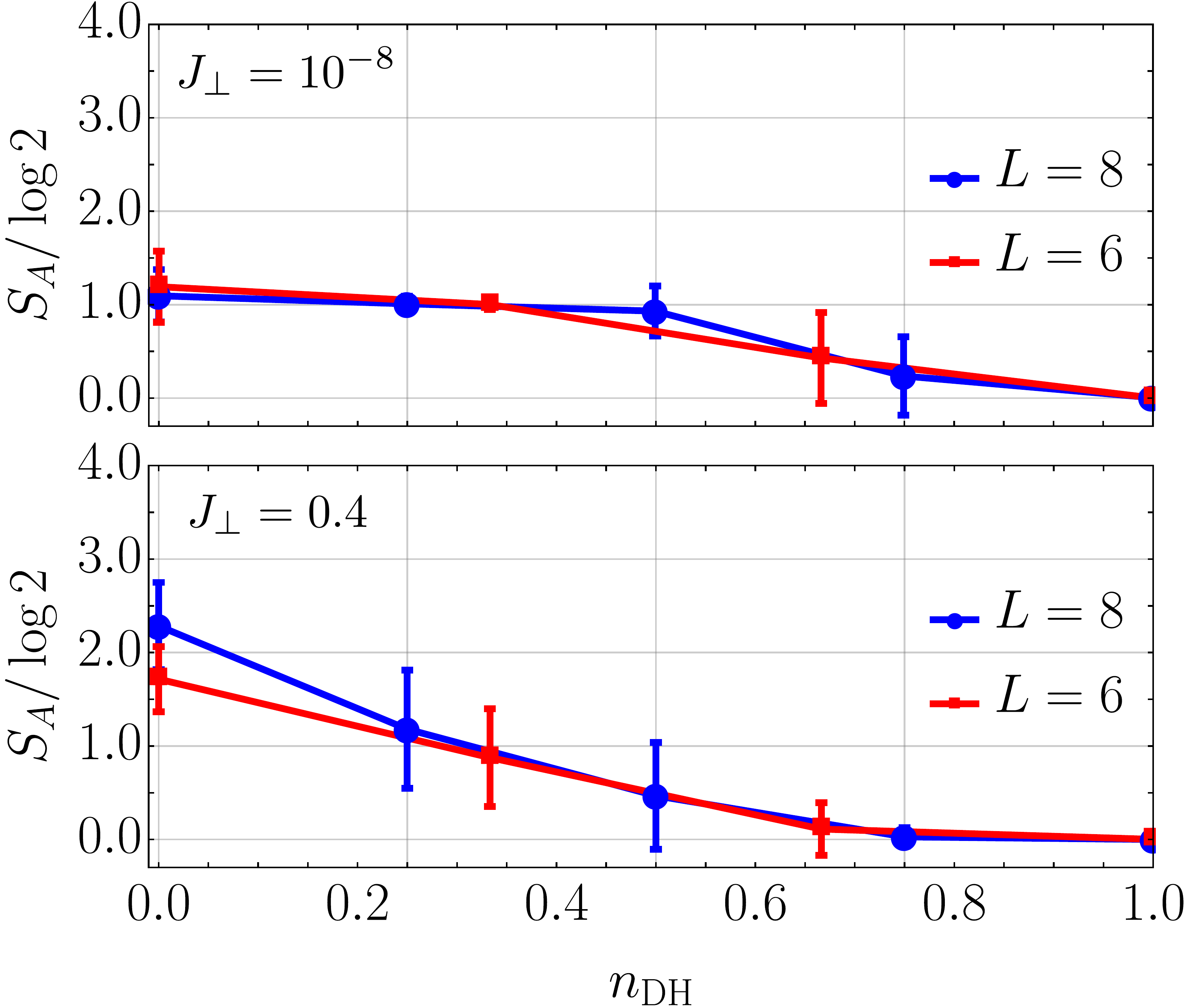}
\caption{Average bipartite entanglement entropy $S_A$ in each doublon/holon density sector at $L=6$ and $8$ (for $L=8$, we use the same data sets as in Figs.~\ref{fig:7}-\ref{fig:8}). Error bars represent one standard deviation of the distribution of $S_A$ (horizontal error bars are smaller than the point size). Top: In the MBL phase $S_A$ saturates to $\log 2$ for $n_{\rm DH}\lesssim 1/2$ and is insensitive to the system size. Bottom: In the ME phase $S_A$ is nearly independent of $n^{\,}_{\rm DH}$ for $n^{\,}_{\rm DH}\geq 2/3$ and increases with decreasing $n^{\,}_{\rm DH}$ for $n^{\,}_{\rm DH}< 2/3$. For $n_{\rm DH}\lesssim 1/2$, $S_A$ clearly increases with $L$, suggesting delocalization below some critical doublon/holon density.}
\label{fig:9}
\end{figure}
%%%%%%%%%%%%%%%%%%
%%%%%%%%%%%%%%%%%%

In contrast, for $J_\perp=0.4$ [Fig.~\ref{fig:7}(b)], deep in the ME phase, the distribution of the entanglement entropy is much broader, encompassing values between $0$ and $\sim 4\log 2$.  Indeed, in going from the MBL to the ME phase a finite fraction of eigenstates are redistributed to entanglement entropies larger than $\sim\log 2$, as one can see by comparing the top panels in Figs.~\ref{fig:7}(a) and \ref{fig:7}(b). Moreover, the largest values of the entanglement entropy increase with system size, as shown in Fig.~\ref{fig:9}.  However, these high-entanglement eigenstates \textit{coexist} with low-entanglement eigenstates at the same energy density, with no obvious demarcation between them as a function of $\epsilon$.  

In Fig.~\ref{fig:8}, we show the same data as in Fig.~\ref{fig:7}, but binned as a function of the doublon/holon density $n_{\rm DH}$ rather than $\epsilon$.
At such strong disorder $n_{\rm DH}$ assumes sharply quantized values, as can be seen in the top panels of Figs.~\ref{fig:8}(a) and \ref{fig:8}(b).  Once the data have been reorganized in this way, a clear trend emerges.  In both the MBL and ME phases, states in the all-doublon/holon sector at $n_{\rm DH}=1$ have nearly zero entanglement, and are essentially product states.  In the MBL phase, the average entanglement entropy in each eigenstate increases as $n_{\rm DH}$ is decreased from $1$, but saturates once it reaches a value near $\log 2$.  In contrast, in the ME phase, the entanglement entropy \textit{does not} saturate as $n_{\rm DH}$ decreases.  Instead, the entanglement entropy increases monotonically until $n_{\rm DH}$ reaches $0$ and the all-singlon limit is achieved.  Thus, in the ME phase there is a clear correlation between the density of singlons and the entanglement entropy: eigenstates containing more singlons have substantially more entanglement on average than those with fewer singlons.

In Fig.~\ref{fig:9}, we compare the data from Figs.~\ref{fig:7} and \ref{fig:8}, obtained at $L=8$, to data obtained at $L=6$ with the same parameters and 2000 disorder realizations.  We plot the means and standard deviations of the entanglement distributions for each system size as functions of $n_{\rm DH}$ in both the MBL and ME phases. (Note that if we had instead plotted the standard \emph{error}, indicating convergence of the \textit{mean} of the distribution as a function of the number of samples, the error bars would be comparable to or smaller than the size of the plot markers.) In the MBL phase, the two entanglement-vs.-doublon/holon density curves are nearly indistinguishable, consistent with the expected area-law scaling.  In the ME phase, the two curves are indistinguishable for $n^{\,}_{\rm DH}\gtrsim 1/2$, and begin to diverge from one another for  $n^{\,}_{\rm DH}\lesssim 1/2$.  The entanglement growth as $n^{\,}_{\rm DH}$ decreases appears to be \emph{faster}, and the final mean value of $S_{A}$ at $n_{\rm DH}=0$ markedly higher, for the larger system.  Furthermore, we observe that $S_A$ appears to be independent of both $L$ and $n^{\,}_{\rm DH}$ for $n^{\,}_{\rm DH}\geq 2/3$, which may indicate that the singlons ``freeze out" above $n^{\,}_{\rm DH}\sim 2/3$, as argued heuristically in Sec.~\ref{subsubsec: Critical doublon/holon density for localization}. However, in order to test the predictions of Sec.~\ref{subsubsec: Critical doublon/holon density for localization} more rigorously, it is necessary to consider larger system sizes.

The data presented in this section serve as an important consistency check on the picture of the ME phase developed in Sec.~\ref{sec:singlon}.  However, by no means do they constitute proof that there is a critical doublon/holon density for localization in the ME phase.  Indeed, finite size scaling at fixed $n_{\rm DH}$ requires access to much larger system sizes than are available to exact diagonalization.  Nevertheless, these data show that there is no many-body mobility edge in the ME phase, but that instead the emergent doublon/holon density controls the (de)localization of eigenstates. Thus, eigenstate properties in ME phase are controlled by the many-body configuration rather than the many-body energy density.

%%%%%%%%%%%%%%%%%%%%
\section{Dynamical signatures of the ME phase}\label{sec:dynamics}
%%%%%%%%%%%%%%%%%%%%

%%%%%%%%%%%%%%%%%%
%%%%%%%%%%%%%%%%%%
\begin{figure}[t!]
\includegraphics[width=0.9\columnwidth]{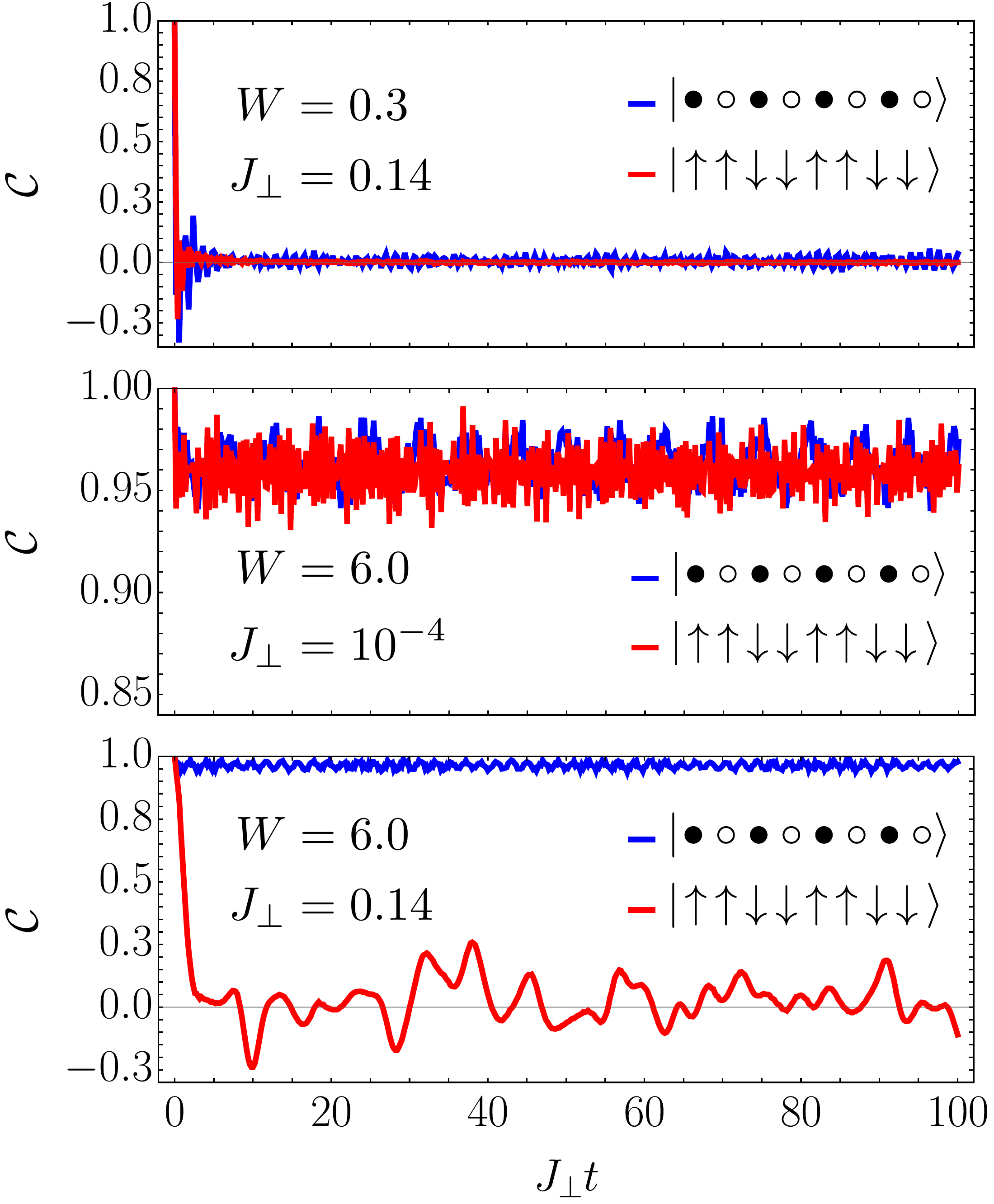}
\caption{Dynamics of the local density autocorrelator~\eqref{eq:C} in the ETH (top), MBL (middle), and ME (bottom) phases at $J=1,\Delta=0.5,$ and $L=8$.  In each case, we compare the dynamics starting from two initial states: the all-doublon state $|\!\bullet\circ\bullet\circ\bullet\circ\bullet\circ\rangle$ and the all-singlon state $|\!\uparrow\,\uparrow\,\downarrow\,\downarrow\,\uparrow\,\uparrow\,\downarrow\,\downarrow\rangle$.
In the ETH and MBL phases, the dynamics of the autocorrelator does not depend on the choice of initial state; in the former case, it quickly decays to zero, whereas in the latter case it remains frozen near its initial value.  In the ME phase, however, the autocorrelator does not decay when the system is initialized in the all-doublon state (similar to its behavior in the MBL phase), while it does when initialized in the all-singlon state (similar to its behavior in the ETH phase). Note that the data for the MBL and ME phases were obtained using the same disorder realization---the only change in going from the middle to the bottom panel is the increase in $J_\perp$.
}
\label{fig:10}
\end{figure}
%%%%%%%%%%%%%%%%%%
%%%%%%%%%%%%%%%%%%

In this section we discuss dynamics in the ME phase. In particular, we consider how the dynamics after a quantum quench depends on the choice of the initial state. We take the initial states to be local density product states, which are most relevant to ongoing experiments studying non-ergodic dynamics. The defining dynamical feature of the mobility emulsion is that particle and energy transport depend strongly on the choice of initial configuration of the system. For initial states with a subcritical doublon/holon density, $n_{\rm DH}<n_{\rm DH}^*$, the system will show thermalizing behavior due to the large fraction of interacting singlons. For initial states with $n_{\rm DH}>n_{\rm DH}^*$ the singlons, doublons, and holons are frozen and remain localized in the initial configuration.  By contrast, in the ETH and MBL phases, a generic product state will instead relax towards thermal equilibrium or remain localized, respectively, regardless of the value of $n_{\rm DH}$ in the initial state.

This strong dependence of the quench dynamics on the choice of initial product state is striking (and to the best of our knowledge, never discussed before in the MBL literature)
 and provides a useful tool in the study of non-ergodic dynamics in that the initial state can be used to \emph{select} the dynamical regime of interest. In this sense, the physics of the ME phase can be qualitatively altered by the appropriate tuning of the initial state! We demonstrate this ability by studying the behavior of several spatially averaged autocorrelation functions that sharply distinguish the ETH, MBL, and ME phases.

We first consider the local density autocorrelation function,
\begin{align}\label{eq:C}
\mathcal C(t)
=
\frac{1}{2L}\!
\sum_{\alpha=1,2}\sum^{L}_{i=1}
\left\langle n_{\alpha,i}(t)\!-\!\frac{1}{2}\right\rangle\!
\left\langle n_{\alpha,i}(0)\!-\!\frac{1}{2}\right\rangle,
\end{align}
where we have used the fact that the initial state is a local density product state, so that the connected part of the autocorrelator vanishes.  $\mathcal C(t)$ probes the localization of individual particles.  Defined such that $\mathcal C(0)=1$, it remains finite as $t\to\infty$ when particles remain confined to their initial positions and tends to zero as $t\to\infty$ when particles are delocalized.  The dynamics of $\mathcal{C}$ at $L=8$ starting from an all-doublon and an all-singlon state are shown in Fig.~\ref{fig:10}.  In the ETH phase, $\mathcal C$ rapidly decays to zero on a timescale of order $1/J$ irrespective of the initial state.  In the MBL phase, $\mathcal C$ remains frozen near its initial value out to arbitrarily late times, again irrespective of the initial state.  However, in the ME phase, $\mathcal C$ remains finite when the system is initialized in the all-doublon state, while it decays on a timescale $1/J_\perp$ when the system is initialized in the all-singlon state.  This provides evidence that the mobility of particles in the ME phase is \emph{configuration-controlled}, in sharp contrast to the ETH and MBL phases, where the late-time dynamics of $\mathcal C$ is independent of the choice of initial state.

Next, we consider autocorrelators that allow one to probe separately the dynamics of the singlons and doublons/holons contained in the initial state. We first define the singlon autocorrelation function,
\begin{equation}\label{eq:imbalance_M}
\mathcal{C}_{\rm S}(t)=\frac{1}{n_{\rm S}L}\sum_{i=1}^{L}  \langle\sigma_i(t)\rangle\langle\sigma_i(0)\rangle,
\end{equation}
where $n_{\rm S}=1-n_{\rm DH}$ is the density of singlons in the initial density product state, so that $\mathcal{C}_{\rm S}(0)=1$. We implicitly assume $n_{\rm S}>0$ when using $\mathcal C_{\rm S}$ and $n_{\rm DH}>0$ when using $\mathcal C_{\rm DH}$ defined below. With this normalization $\mathcal{C}_{\rm S}(t)$ measures the fraction of singlons that remain in their initial configuration under time evolution. If one starts in a configuration with only singlons, $n_{\rm S}=1$, and the singlons are localized (i.e. $\sigma_i=\pm1$), then $\mathcal{C}_{\rm S}$ will remain nonzero as $t\to\infty$. Such dynamical behavior sharply distinguishes the MBL phase from both the ETH and ME phases, where the all-singlon state thermalizes and $\mathcal{C}_{\rm S}$ decays to zero at late times. Such thermalization and decay of $\mathcal{C}_{\rm S}$ is qualitatively similar in both the ETH and ME phases, thus requiring at least one other measure to distinguish the two.

The distinction between the ETH and ME phases can be observed by studying the doublon/holon autocorrelation function, defined as
\begin{equation}\label{eq:imbalance_cdw}
\mathcal{C}_{\rm DH}(t)=\frac{1}{n_{\rm DH}L}\sum_{i=1}^{L} \langle d_i(t)\rangle\langle d_i(0)\rangle.
\end{equation}
Here $\mathcal{C}_{\rm DH}$ measures the fraction of the doublons and holons that remain confined to their initial sites. This measure also determines the charge transport of the system when initialized in a mirror-symmetric charge density wave (CDW) state. In the MBL phase neither doublons nor holons propagate, and in this case $\mathcal{C}_{\rm DH}$ will remain close to its initial value as $t\to \infty$. In the ETH phase $\mathcal{C}_{\rm DH}$ will generically decay to zero due to the spreading of charge. However, in the ME phase the late-time behavior of $\mathcal{C}_{\rm DH}$ is instead determined by the initial configuration and will show either localized or thermalizing behavior depending on the value of $n_{\rm DH}$ in the initial state. For example, in the all-doublon/holon sector, $n_{\rm DH}=1$, the system is fully localized in the ME phase and any initial CDW will survive indefinitely under time evolution (just as in the MBL phase). In the ETH phase, however, a generic CDW will thermalize rapidly. The doublon/holon autocorrelation function $\mathcal{C}_{\rm DH}$ thus sharply distinguishes the ETH phase from both the ME and MBL phases and, in combination with the singlon autocorrelation function $\mathcal{C}_{\rm S}$, allows the three phases to be uniquely determined.

%%%%%%%%%%%%%%%%%%
%%%%%%%%%%%%%%%%%%
\begin{figure}[t!]
\includegraphics[width=\columnwidth]{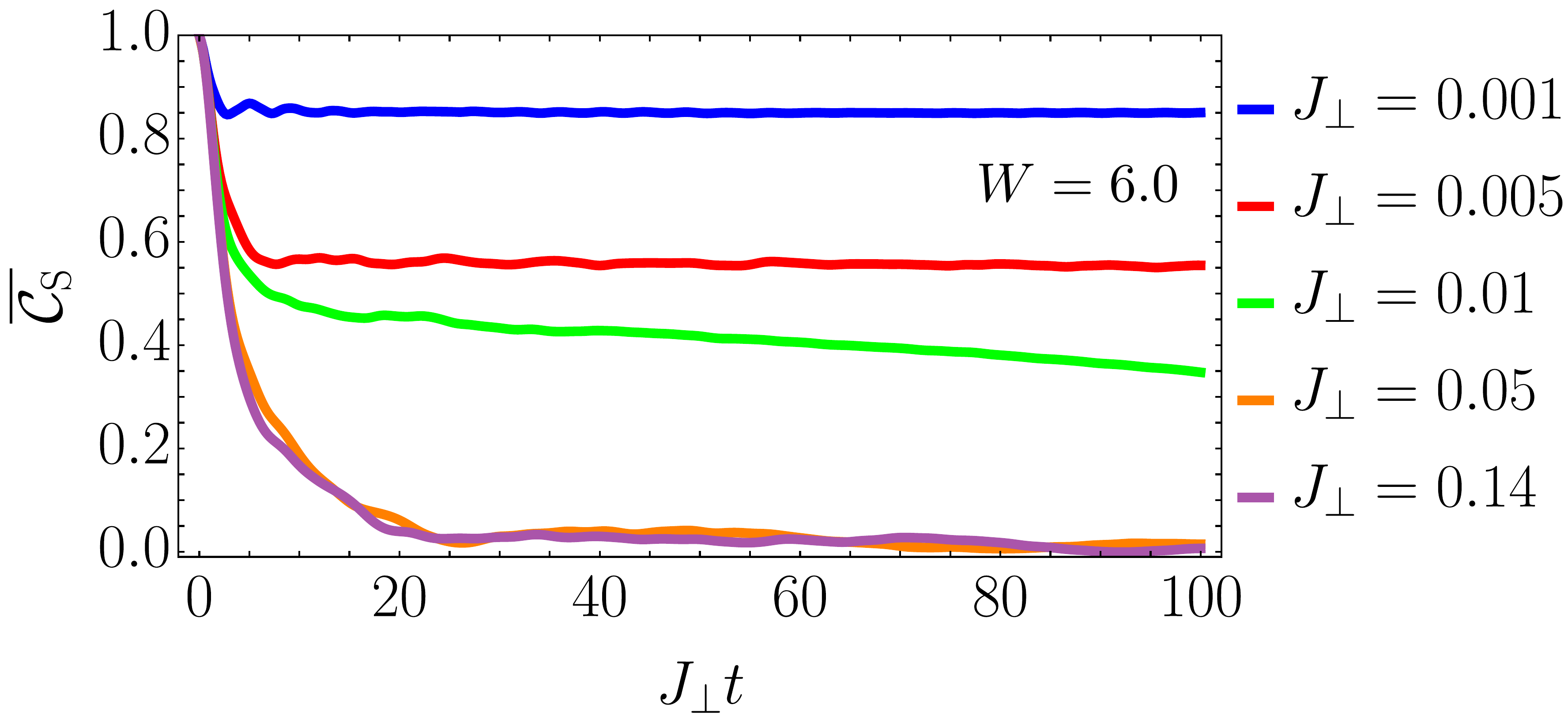}
\caption{Dynamics of the moving average of the imbalance $\mathcal{C}_{\rm S}$ as a function of $J_\perp$ at $J=1,\Delta=0.5,$ and $L=7$.  The data shown were generated using a single disorder realization at $W=6$. The initial state is taken to be the all-singlon configuration $|\!\uparrow\,\uparrow\,\downarrow\,\downarrow\,\uparrow\,\uparrow\,\downarrow\rangle$. In the MBL phase $J_\perp\lesssim 0.03$ the imbalance has a nonzero late time average, while in the ME phase, $J_\perp \gtrsim 0.03$, the late time average vanishes.}
\label{fig:11}
\end{figure}
%%%%%%%%%%%%%%%%%%
%%%%%%%%%%%%%%%%%%

\begin{figure}[t!]
\includegraphics[width=0.95\columnwidth]{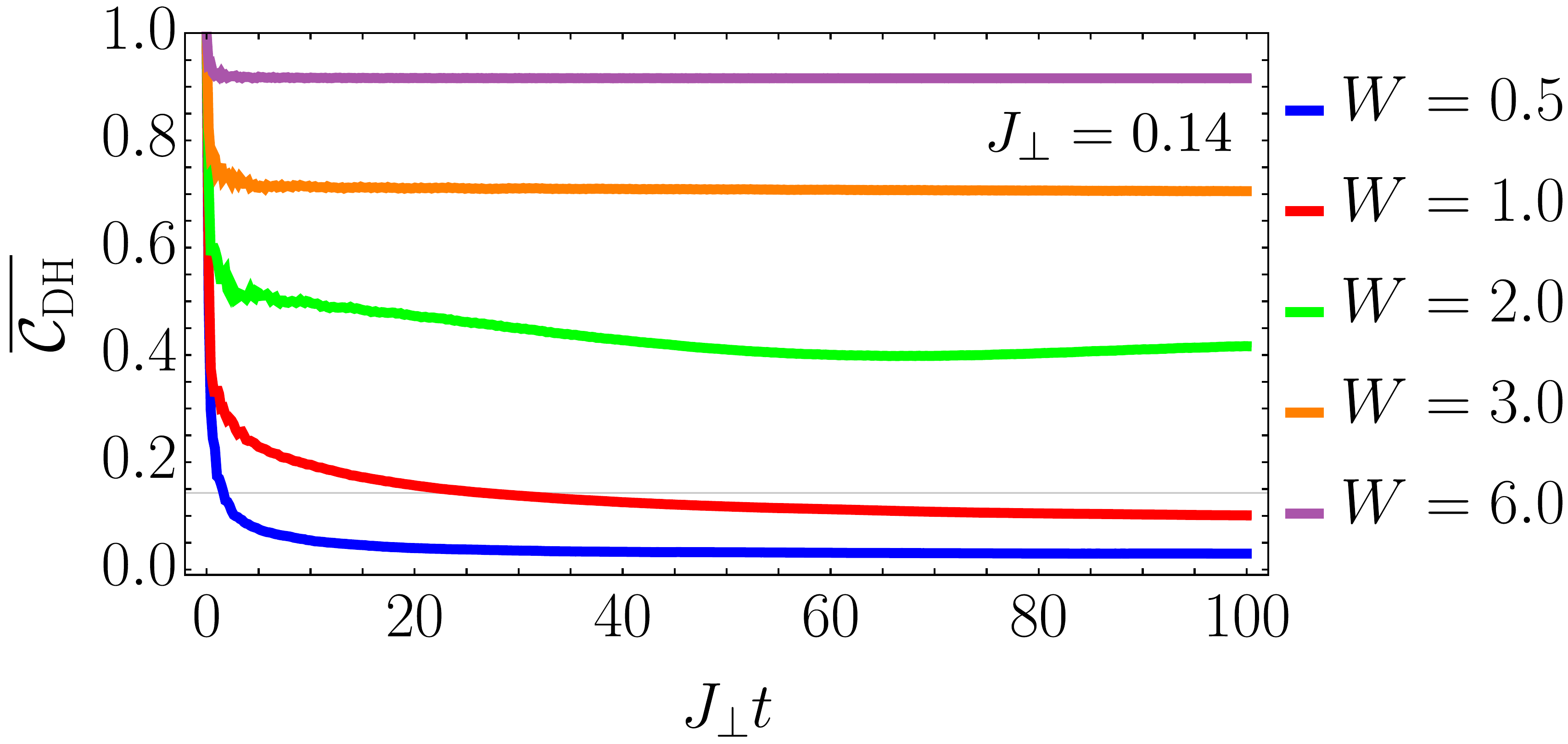}
\caption{
Dynamics of the moving average of the imbalance $\mathcal{C}_{\rm DH}$ as a function of disorder strength at $J=1,\Delta=0.5,J_\perp=0.14,$ and $L=7$. The initial state is taken to be the all-doublon/holon configuration $|\!\bullet\circ\bullet\circ\bullet\circ\ \!\bullet\rangle$. To investigate the dependence of the late-time average of $\mathcal{C}_{\rm DH}$ on the disorder strength, we fix the disorder potential to be the one used in Fig.~\ref{fig:5}, and globally rescale it to change the effective value of $W$. In the ME phase, $W\gtrsim 1.4$, the imbalance has a nonzero late-time average, while in the ETH phase, $W\lesssim 1.4$, the late time average vanishes.  The gray horizontal indicates the value $1/L=0.14\dots$, below which the late-time average should be viewed as indistinguishable from zero.  Note that the parameters and disorder potential used for the purple curve at $W=6$ coincides with those of the purple curve in Fig.~\ref{fig:11}; the only difference between the two curves (besides the quantity being measured) is the initial state used.
}
\label{fig:12}
\end{figure}

In Figs.~\ref{fig:11}--\ref{fig:12} we plot the moving time average of the singlon and doublon/holon autocorrelators, defined for any quantity $\mathcal O(t)$ as
\begin{align}
\overline{\mathcal{O}}(t)=\int_0^t \frac{dt^\prime}{t} \mathcal{O}(t^\prime).
\end{align}
In Fig.~\ref{fig:11} we show the behavior of $\mathcal{C}_{\rm S}$ starting from a fixed all-singlon configuration for various values of $J_\perp$ at large $W$ where only the MBL and ME phases exist. At small transverse hopping, $J_\perp<J_\perp^*$, one sees that the late-time value of $\overline{\mathcal{C}_{\rm S}}(t)$ converges to a nonzero value due to singlon localization that arises in the MBL phase. Beyond a critical value $J_\perp^*\approx 0.03$ the system enters the ME phase and $\mathcal{C}_{\rm S}$ decays to zero at late-times, indicating a thermalizing state in the all-singlon sector, $n_{\rm S}=1$.  This behavior is consistent with the data in Figs.~\ref{fig:8}-\ref{fig:9} that shows the entanglement growing with every singlon added to the system at $n_{\rm DH}=0$.  Near the transition ($J_\perp=0.01$), the timescale on which $\overline{\mathcal{C}_{\rm S}}(t)$ converges becomes much longer than it is when the system is deep in either phase. 

To distinguish the ETH and ME phases, in Fig.~\ref{fig:12} we plot the doublon-holon autocorrelation function starting from a fixed doublon CDW initial state.  The data are taken at a fixed transverse hopping strength where only the ETH and ME phases exist. We fix a random disorder configuration at $W=6.0$ (the same one used in Fig.~\ref{fig:11}) and globally rescale it to vary its strength $W$; this drives the system through the ETH to ME transition. At strong disorder the system is in the ME phase and the CDW is dynamically stable. As one lowers the disorder strength the late-time average of $\mathcal{C}_{\rm DH}$ is reduced until  it vanishes upon reaching the critical point. Below the critical disorder strength  the CDW melts at late times, indicating behavior consistent with the ETH phase.  At $W=2$, near the transition $W^*\approx1.4$ shown in Fig.~\ref{fig:3}, the the timescale on which $\overline{\mathcal{C}_{\rm S}}(t)$ converges becomes longer than it is when the system is deep in either phase, similar to what is observed near the transition in Fig.~\ref{fig:11}.

We emphasize that the dynamical data shown in Figs.~\ref{fig:10}--\ref{fig:12} further exemplify a defining feature of the ME phase, namely that the dynamics of the system is localized when the system is prepared in an all-doublon/holon initial state, and thermalizing when the system is prepared in an all-singlon initial state. By strong contradistinction, in the MBL phase both of the two states studied (as well as generic states) remain localized, while in the ETH phase both states thermalize (see Fig.~\ref{fig:10}). The fact that two distinct diagnostics, namely, $\mathcal C_{\rm S}$ and $\mathcal C_{\rm DH}$, are needed to uniquely characterize the ME phase makes perfect sense since the ME phase shares properties of both ETH and MBL phases in a configuration-dependent (tuned by $n_{\rm DH}$) manner, necessitating two separate correlators to distinguish it from the ETH and MBL phases.  In fact, any intermediate phase in any situation is likely to require two distinct diagnostics to distinguish it from both ETH and MBL phases whereas ETH and MBL phases themselves can be distinguished by one diagnostic.

%%%%%%%%%%%%%%%%%%
%%%%%%%%%%%%%%%%%%
%%%%%%%%%%%%%%%%%%
%%%%%%%%%%%%%%%%%%

We should note that when the singlons are in a \emph{paramagnetic} state, as occurs when singlons are sufficiently dilute, they localize along the $\sigma^x$-direction and in this case $\mathcal C_{\rm S}(t)$ will show oscillations associated with singlons initialized in a $\sigma_i^z$ eigenstate precessing about the local $\sigma^x$ field of strength $\sim J_\perp$ (see Sec.~\ref{sec:singlon}). To see localization in this case, one should first perform a $\pi/2$ rotation about the $\sigma^y$-axis on the singlons in the initial $\sigma^z$-basis product state to obtain a $\sigma^x$-basis product state. [Note that here we are talking about the spin axes of the emergent singlon spin states; the hardcore bosons we consider are spinless.] After evolving the state with the Hamiltonian \eqref{eq:H}, one performs an additional $\pi/2$ rotation before measuring in the $\sigma^z$-basis. In this way, one will effectively measure the dynamics of $\sigma^x_i$ starting from a $\sigma^x_i$ product state, which will remain localized under time evolution for a paramagnetic singlon state. It is interesting to note that the necessary $\pi/2$ rotation of the singlons around the $\sigma^y$-axis can be achieved \emph{in situ} in a system of hardcore bosons.  Starting from a deep lattice, such that hopping is suppressed both within and between the legs, one can first apply the inter-leg hopping Hamiltonian $H_{\perp}$ for a time $\pi/(2 J_\perp)$, which enacts a $\pi/2$ rotation about the $\sigma^x$-axis for the singlons and does nothing to the doublons and holons.  Subsequently, one can turn on a bias of the form $h\sum_i (n_{1,i}-n_{2,i})$ for a time $\pi/(4h)$, which performs a $\pi/2$ rotation about the $\sigma^z$-axis for the singlons and again does nothing to the doublons and holons.

Our arguments in Sec.~\ref{sec:mixed} suggest that the qualitatively distinct dynamical behavior of the ME phase, illustrated in Figs.~\ref{fig:10}--\ref{fig:12}, persists away from the special limits $n_{\rm DH}=0$ and $n_{\rm DH}=1$. However, due to severe finite size restrictions imposed by exact diagonalization, we are not able to see a sharp transition between these two behaviors upon changing $n_{\rm DH}$ in dynamical or eigenstate properties. This is primarily due to the strong quantization of $n_{\rm DH}$ that arises for small systems at half-filling (i.e. $n_{\rm DH}$ cannot be tuned continuously like $J_\perp$ or $W$), c.f.~Fig.~\ref{fig:8}. It would be very interesting to implement another method, e.g. the recently developed time-dependent variational principle (TDVP) \cite{Haegeman11,Haegeman16,Leviatan17,Doggen18}, that can study dynamics in larger systems where the sensitive dependence of the dynamics on the initial value of $n_{\rm DH}$ can be seen explicitly. This would also potentially allow one to observe the drastic separation of charge and energy transport discussed in Sec.~\ref{sec:thermalizing}, which is not visible in the small systems studied here. This issue is also coupled to an accurate determination of the critical $n_{\rm DH}^*$ itself, requiring future studies  using much larger system sizes.

\section{Discussions and Conclusion}\label{sec:conclusion}

\subsection{Summary}

In this paper we have proposed a new non-ergodic phase of matter---the mobility emulsion. It has the peculiar property of neither satisfying ETH nor being fully MBL, and as such is an intermediate phase; indeed, in Sec.~\ref{sec:model}, we presented numerical evidence that it is separated from both phases by a phase transition, and is thus sharply distinct.  It is characterized by
\begin{enumerate}
\item the \emph{coexistence} of localized and thermalizing eigenstates at fixed energy density (and thus manifests no mobility edge),
\item the \emph{emergence} of a parameter, the doublon/holon density $n_{\rm DH}$ in the case of the model~\eqref{eq:H}, that can be used to label eigenstates and that determines whether they exhibit localized or thermalizing behavior, and
\item the ability to \emph{select} the dynamical behavior following a quantum quench by initializing the system in many-body configurations with different values of this emergent parameter.
\end{enumerate}
The ME phase of the model~\eqref{eq:H} provides a striking example of a quantum system with robust and \emph{incomplete} emergent integrability, wherein the number of integrals of motion (here represented by the emergent dressed doublons and holons) is a nonzero, but nonunity, fraction of the total number of degrees of freedom. Thus, it is neither ETH nor MBL, but is a distinct new phase. As discussed in Secs.~\ref{sec:singlon} and \ref{sec:mixed}, this feature has a natural explanation and characterization in terms of many-body configurations labeled by the doublon/holon density $n_{\rm DH}$. Morever, the strong configuration dependence of the quench dynamics examined in Sec.~\ref{sec:dynamics} enables one to selectively access various dynamical regimes by initial state preparation. This suggests a new paradigm in which the choice of initial state can be used as a tool to study and manipulate non-ergodic disordered phases of matter. This paradigm could be implemented, e.g., in systems of cold atoms, trapped ions, or Rydberg atoms, as long as a system with the appropriate hierarchy of energy scales can be engineered.

Quite apart from possible experimental realizations, which should be possible since the system we study is equivalent to two coupled XXZ spin chains, the ME phase is of considerable fundamental interest, explicitly demonstrating that the presence of hopping, disorder, and interaction could lead to sufficient dynamical frustration in a quantum system so as to produce only partial emergent integrability so that the resultant phase, depending on its internal configurations, may, even at infinite temperature, be neither an ergodic metal nor a non-ergodic insulator.  The fact that such a phase could arise without the physics of mobility edge playing any role is indeed extremely intriguing.

\subsection{Discussion and Outlook}

Although various aspects of the ME phase have been studied and characterized numerically and analytically in this work, some important properties remain yet to be fully characterized. Perhaps the most notable is the critical doublon/holon density $n_{\rm DH}^*$ that separates localized from thermalizing states. While we have argued for the existence of a nontrivial critical point, $0<n_{\rm DH}^*<1$, its precise location is difficult to extract using exact diagonalization due to severe finite size restrictions.  We do, however, establish the stability of the localized and thermalizing limits of the ME phase to small changes in $n_{\rm DH}$.

On the other hand, analytical attempts to determine $n_{\rm DH}^*$ are challenging due to the presence of rare region effects, the importance of which is well-appreciated but that are notoriously difficult to address \cite{DeRoeck16,DeRoeck17a,DeRoeck17b,Thiery17a,Thiery17b,Ponte17}. In our setup, rare regions of weak disorder can influence the form of the effective singlon Hamiltonian \eqref{eq:Heff} by locally admixing singlon and doublon states. Such regions are statistically rare for $W\gg J$ (where we find the ME phase---see Fig.~\ref{fig:3}) and are not expected to qualitatively modify the phase diagram as compared to the case of a quasiperiodic potential where rare regions are absent. Even in the latter case, however, rare \emph{configurations} with anomalously high local singlon density could play an important role near the critical doublon/holon density. (The effects of similar rare configurations may have been observed experimentally in Ref.~\cite{Bordia17}.) Consequently, it would be highly desirable to implement a numerical method, such as the TDVP \cite{Haegeman11,Haegeman16,Leviatan17,Doggen18}, that not only captures rare region effects but also allows one to access substantially larger system sizes where the doublon/holon density quantization is less prominent. Such a future work could also help to better pin down the critical density $n_{\rm DH}^*$. Moreover, using such a method one might be able to observe the strong separation of timescales for energy and charge transport, the latter being associated with the slow variable-range hopping of doublons and holons. This physics should be directly visible in a sufficiently large experimental system with on-site density resolution and a long particle lifetime. 

One intriguing aspect of the model studied here is that it resembles the Fermi-Hubbard model in a transverse field. In this analogy, the two legs of the ladder represent the spin-up and spin-down states of spin-1/2 fermions on a single chain. The doublons and holons of the resulting model are those of the usual Fermi-Hubbard model, and the singlon states on a rung correspond to the spin states of a fermion on a singly occupied site. However, because of the form of Eq.~\eqref{eq:H}, there is no $SU(2)$ spin-rotation symmetry, even for $J_\perp=\Delta=0$. This breaking of $SU(2)$ symmetry lifts the obstruction to localization~\cite{Protopopov17,Potter16}, ultimately leading to the mobility emulsion. In the bosonic language, the interaction between particles on the same leg (or spin state) is hardcore, i.e. $U_{\uparrow\uparrow}=U_{\downarrow\downarrow}=\infty$, while the interaction between particles on opposite legs vanishes, $U_{\downarrow\uparrow}=0$. The breaking of $SU(2)$ symmetry thus arises not only from the transverse hopping $J_\perp$, which can be viewed as a homogeneous transverse field in spin space, but also from the spin-anisotropic interactions that further break conservation of spin $U(1)$ symmetry associated with rotations about the transverse-field axis. 

This feature represents a crucial difference from the Fermi-Hubbard model, which always preserves a spin $U(1)$ symmetry in the presence of a uniform field. The similarity to the Fermi-Hubbard model nevertheless provides an additional starting point and guiding principle in the search for and engineering of systems that may harbor the ME phase. There is particular motivation to use such a broken $SU(2)$ Fermi-Hubbard prescription for the laboratory observation of the ME phase since recent experimental work has been able to study the Fermi-Hubbard model with single-site resolution in optical lattices \cite{Mazurenko17,Mitra17,Chiu18}. 

A more promising route to experimentally observing the separation of singlon and doublon dynamics in the ME phase would be to implement  the disordered Bose-Hubbard model on a two-leg ladder in a quantum gas microscope. This model should exhibit qualitatively similar physics when the hardcore constraint is softened by a finite on-site repulsion $U<\infty$; a nearest-neighbor interaction like $\Delta$ can then be generated via fluctuations into and out of intermediate states with double occupancies. A recent experiment by Lukin et al.~\cite{Lukin18} has shown the feasibility of this setup for studying entanglement in a single-leg disordered Bose-Hubbard model, while Kaufman et al.~\cite{Kaufman16} studied the clean Bose-Hubbard model on the two-leg ladder with single-site resolution. By combining these techniques, one could prepare and directly image the nonergodic dynamics of singlons and doublons that can be controlled exclusively by the initial state.

%%%%%%%%%%
%%%%%%%%%%

\acknowledgements{We acknowledge helpful discussions with Vedika Khemani, Markus M\"uller, and Ivan Protopopov. This work is supported in part by the Laboratory for Physical Sciences and Microsoft. T.I. acknowledges a JQI postdoctoral fellowship.}

%%%%%%%%%%
%%%%%%%%%%

\appendix*

\section{Derivation of the optimal doublon variable-range hopping rate}

We consider the variable-range hopping of a doublon embedded in a thermalizing background of singlons.  The coupling between the doublon and singlons is easiest to derive when the inter-leg hopping $J_{\perp}$ is either much smaller or much greater than the nearest-neighbor interaction $\Delta$.  In both cases, we work in the limit $W \gg J,J_\perp,\Delta$, so that the doublons and holons are strongly localized and the effective disorder for the singlons is very small.

\subsection{Domain-wall limit}

We first consider the limit $J_\perp \ll \Delta$. In this limit, the domain walls of the singlon ``spin" configurations are good quasiparticles. In this case, the doublon can annihilate a domain wall with amplitude $J_\perp$ because the former acts essentially as a hard boundary at large disorder $W\gg J$. Denoting a domain wall creation/annihilation operator on bond $i$ by $a^\dagger_i/a_i$, we have 
\begin{equation}\label{eq:coupling}
H_{\rm int}=\frac{J_\perp}{2}\sum_i f^\dagger_i f_i\, (a_i+a^\dagger_i),
\end{equation}
where $f^\dagger_i/f_i$ is a doublon creation/annihilation operator on site $i$. The domain walls form a bath with single-particle spectrum 
\begin{equation}\label{eq:Hdw}
H_{\rm dw}=\sum_k \omega_k\, a^\dagger_k a_k,\indent \omega_k=\Delta+\frac{J_\perp}{2}\cos k,
\end{equation}
where $k$ labels the domain-wall momentum and we neglected the weak disorder-dependent terms. In addition to the domain wall Hamiltonian, we have the localized doublon Hamiltonian written in the basis of (exponentially) localized states $\alpha$ as
\begin{equation}\label{eq:doublon}
H_{\rm dbl}=\sum_{\alpha}\tilde{h}_\alpha\,  f^\dagger_\alpha f_\alpha.
\end{equation}
In the localized basis $H_{\rm int}$ takes the form 
\begin{equation}\label{eq:coupling2}
H_{\rm int}=\frac{J_\perp}{2} \sum_{i,\alpha,\alpha^\prime} \psi^*_{i,\alpha}\psi_{i,\alpha^\prime}\, f^\dagger_{\alpha} f_{\alpha^\prime}\, (a_i+a^\dagger_i),
\end{equation}
where $\psi_{i,\alpha}$ is the wavefunction of a localized state $\alpha$. The coupling \eqref{eq:coupling2} shows that domain wall absorption or emission leads to transitions between distinct localized states whose amplitude decays exponentially (due to the decay of $\psi_{i,\alpha}$) with the separation from the site of absorption $i$. At the same time the typical energy difference between states $\alpha,\alpha^\prime$, $\tilde h_\alpha - \tilde h_{\alpha^\prime}\sim W$, is much larger than the energy of a single domain wall, $\sim \Delta$. To find the optimal transition rate we first find the transition amplitude for a process involving an energy transfer $E<W$.

Starting from some site $i$, we can estimate the distance $r$ required before finding another site with probability $\mathcal O(1)$ within the energy window $h_i\pm E/2$ to be $r(E)\sim W/E$ [see Fig.~\ref{fig:6}(b)]. In order for the doublon to gain or lose energy $E$, it must absorb or emit $N(E)\sim E/\Delta$ domain walls. This requires going to $N$th order perturbation theory in $H_{\rm int}$, leading to an effective doublon hopping amplitude scaling as
\begin{equation}\label{eq:amplitude}
J(E)\sim J e^{-r(E)/\xi}\left(\frac{J_\perp/2}{N(E)\Delta}\right)^{N(E)},
\end{equation}
where $\xi$ is the doublon localization length. At strong disorder $\xi \sim 1/\log(W/J)\ll1$. The factor $N(E)$ in the denominator of Eq.~\eqref{eq:amplitude} stems from the fact that the energy difference grows linearly in each order of perturbation theory, and we used the approximation $n!\simeq n^n$. Optimizing $J(E)$ over $E$ leads to an optimal hopping rate
\begin{subequations}
\label{eq:hoptimal}
\begin{equation}
J^*\sim J e^{-2r^*/\xi},
\end{equation}
where
\begin{equation}
r^*\approx\sqrt{\frac{W\xi}{\Delta}\log\left(\frac{\sqrt{W\Delta/\xi}}{J_\perp/2}\right)}.
\end{equation}
\end{subequations}

%%%%%%%%%%%%%%%%%
\subsection{Paramagnetic Limit}\label{sec:paramagnet}
%%%%%%%%%%%%%%%%%
In the limit $J_\perp\gg \Delta$, the domain walls are no longer good quasiparticles, but the local spin projections onto the $x$-axis, $|\!\rightarrow\rangle_i$ and $|\!\leftarrow\rangle_i$, are. Such ``spin flips" have a single-particle spectrum given by 
\begin{equation}\label{eq:spin}
H_{\rm spin}=\sum_k \varepsilon_k\, s^\dagger_k s_k,\indent \varepsilon_k=J_\perp+\frac{\Delta}{2}\cos k,
\end{equation} 
where $k$ is the spin-flip momentum and $s^\dagger_k/s_k$ are spin-flip quasiparticle creation/annihilation operators. The coupling to the doublon now acquires a scattering form 
\begin{equation}\label{eq:coupling3}
H_{\rm int}=\Delta\sum_{j,k,k^\prime}g_{k,k^\prime}\, e^{ij(k-k^\prime)}\, f^\dagger_j f_j\,  s^\dagger_k s_{k^\prime},
\end{equation}
with dimensionless form factor $g_{k,k^\prime}$ whose precise structure we neglect in the following estimate of the effective doublon hopping amplitude. In each scattering process the doublon can acquire a maximal energy of $\Delta$ from the spin excitations scattered across the Brillouin zone from $k=0$ to $k^\prime=\pi$ or vice versa. The multiparticle scattering process that absorbs energy $E\gg \Delta$ acquires a structure similar to Eq.~\eqref{eq:amplitude} and thus gives rise to an effective doublon hopping amplitude scaling as
\begin{equation}\label{eq:amplitude2}
J(E)=J e^{-r(E)/\xi}\left(\frac{\Delta}{E}\right)^{\frac{E}{\Delta}}.
\end{equation}
Optimizing over energy $E$ leads to the hopping rate
\begin{subequations}
\label{eq:hoptimal2}
\begin{equation}
J^*\sim J e^{-2r^*/\xi},
\end{equation}
where
\begin{equation}
r^*\approx\sqrt{\frac{W\xi}{\Delta}\log\left(\sqrt{\frac{W}{\xi\Delta}}\right)}.
\end{equation}
\end{subequations}
The result Eq.~\eqref{eq:hoptimal2} can be obtained from Eq.~\eqref{eq:hoptimal} upon the substitution $J_\perp/2 \to \Delta$ due to the difference between the interaction Hamiltonians~\eqref{eq:coupling2} and \eqref{eq:coupling3}.

\bibliography{refs_iMBL}

%merlin.mbs apsrev4-1.bst 2010-07-25 4.21a (PWD, AO, DPC) hacked
%Control: key (0)
%Control: author (8) initials jnrlst
%Control: editor formatted (1) identically to author
%Control: production of article title (-1) disabled
%Control: page (0) single
%Control: year (1) truncated
%Control: production of eprint (0) enabled
\begin{thebibliography}{98}%
\makeatletter
\providecommand \@ifxundefined [1]{%
 \@ifx{#1\undefined}
}%
\providecommand \@ifnum [1]{%
 \ifnum #1\expandafter \@firstoftwo
 \else \expandafter \@secondoftwo
 \fi
}%
\providecommand \@ifx [1]{%
 \ifx #1\expandafter \@firstoftwo
 \else \expandafter \@secondoftwo
 \fi
}%
\providecommand \natexlab [1]{#1}%
\providecommand \enquote  [1]{``#1''}%
\providecommand \bibnamefont  [1]{#1}%
\providecommand \bibfnamefont [1]{#1}%
\providecommand \citenamefont [1]{#1}%
\providecommand \href@noop [0]{\@secondoftwo}%
\providecommand \href [0]{\begingroup \@sanitize@url \@href}%
\providecommand \@href[1]{\@@startlink{#1}\@@href}%
\providecommand \@@href[1]{\endgroup#1\@@endlink}%
\providecommand \@sanitize@url [0]{\catcode `\\12\catcode `\$12\catcode
  `\&12\catcode `\#12\catcode `\^12\catcode `\_12\catcode `\%12\relax}%
\providecommand \@@startlink[1]{}%
\providecommand \@@endlink[0]{}%
\providecommand \url  [0]{\begingroup\@sanitize@url \@url }%
\providecommand \@url [1]{\endgroup\@href {#1}{\urlprefix }}%
\providecommand \urlprefix  [0]{URL }%
\providecommand \Eprint [0]{\href }%
\providecommand \doibase [0]{http://dx.doi.org/}%
\providecommand \selectlanguage [0]{\@gobble}%
\providecommand \bibinfo  [0]{\@secondoftwo}%
\providecommand \bibfield  [0]{\@secondoftwo}%
\providecommand \translation [1]{[#1]}%
\providecommand \BibitemOpen [0]{}%
\providecommand \bibitemStop [0]{}%
\providecommand \bibitemNoStop [0]{.\EOS\space}%
\providecommand \EOS [0]{\spacefactor3000\relax}%
\providecommand \BibitemShut  [1]{\csname bibitem#1\endcsname}%
\let\auto@bib@innerbib\@empty
%</preamble>
\bibitem [{\citenamefont {Bernien}\ \emph {et~al.}(2017)\citenamefont
  {Bernien}, \citenamefont {Schwartz}, \citenamefont {Keesling}, \citenamefont
  {Levine}, \citenamefont {Omran}, \citenamefont {Pichler}, \citenamefont
  {Choi}, \citenamefont {Zibrov}, \citenamefont {Endres}, \citenamefont
  {Greiner} \emph {et~al.}}]{Bernien17}%
  \BibitemOpen
  \bibfield  {author} {\bibinfo {author} {\bibfnamefont {H.}~\bibnamefont
  {Bernien}}, \bibinfo {author} {\bibfnamefont {S.}~\bibnamefont {Schwartz}},
  \bibinfo {author} {\bibfnamefont {A.}~\bibnamefont {Keesling}}, \bibinfo
  {author} {\bibfnamefont {H.}~\bibnamefont {Levine}}, \bibinfo {author}
  {\bibfnamefont {A.}~\bibnamefont {Omran}}, \bibinfo {author} {\bibfnamefont
  {H.}~\bibnamefont {Pichler}}, \bibinfo {author} {\bibfnamefont
  {S.}~\bibnamefont {Choi}}, \bibinfo {author} {\bibfnamefont {A.~S.}\
  \bibnamefont {Zibrov}}, \bibinfo {author} {\bibfnamefont {M.}~\bibnamefont
  {Endres}}, \bibinfo {author} {\bibfnamefont {M.}~\bibnamefont {Greiner}},
  \emph {et~al.},\ }\href {\doibase 10.1038/nature24622} {\bibfield  {journal}
  {\bibinfo  {journal} {Nature}\ }\textbf {\bibinfo {volume} {551}},\ \bibinfo
  {pages} {579} (\bibinfo {year} {2017})}\BibitemShut {NoStop}%
\bibitem [{\citenamefont {Guardado-Sanchez}\ \emph {et~al.}(2018)\citenamefont
  {Guardado-Sanchez}, \citenamefont {Brown}, \citenamefont {Mitra},
  \citenamefont {Devakul}, \citenamefont {Huse}, \citenamefont {Schau\ss{}},\
  and\ \citenamefont {Bakr}}]{Guardado-Sanchez17}%
  \BibitemOpen
  \bibfield  {author} {\bibinfo {author} {\bibfnamefont {E.}~\bibnamefont
  {Guardado-Sanchez}}, \bibinfo {author} {\bibfnamefont {P.~T.}\ \bibnamefont
  {Brown}}, \bibinfo {author} {\bibfnamefont {D.}~\bibnamefont {Mitra}},
  \bibinfo {author} {\bibfnamefont {T.}~\bibnamefont {Devakul}}, \bibinfo
  {author} {\bibfnamefont {D.~A.}\ \bibnamefont {Huse}}, \bibinfo {author}
  {\bibfnamefont {P.}~\bibnamefont {Schau\ss{}}}, \ and\ \bibinfo {author}
  {\bibfnamefont {W.~S.}\ \bibnamefont {Bakr}},\ }\href {\doibase
  10.1103/PhysRevX.8.021069} {\bibfield  {journal} {\bibinfo  {journal} {Phys.
  Rev. X}\ }\textbf {\bibinfo {volume} {8}},\ \bibinfo {pages} {021069}
  (\bibinfo {year} {2018})}\BibitemShut {NoStop}%
\bibitem [{\citenamefont {Lienhard}\ \emph {et~al.}(2018)\citenamefont
  {Lienhard}, \citenamefont {de~L\'es\'eleuc}, \citenamefont {Barredo},
  \citenamefont {Lahaye}, \citenamefont {Browaeys}, \citenamefont {Schuler},
  \citenamefont {Henry},\ and\ \citenamefont {L\"auchli}}]{Lienhard17}%
  \BibitemOpen
  \bibfield  {author} {\bibinfo {author} {\bibfnamefont {V.}~\bibnamefont
  {Lienhard}}, \bibinfo {author} {\bibfnamefont {S.}~\bibnamefont
  {de~L\'es\'eleuc}}, \bibinfo {author} {\bibfnamefont {D.}~\bibnamefont
  {Barredo}}, \bibinfo {author} {\bibfnamefont {T.}~\bibnamefont {Lahaye}},
  \bibinfo {author} {\bibfnamefont {A.}~\bibnamefont {Browaeys}}, \bibinfo
  {author} {\bibfnamefont {M.}~\bibnamefont {Schuler}}, \bibinfo {author}
  {\bibfnamefont {L.-P.}\ \bibnamefont {Henry}}, \ and\ \bibinfo {author}
  {\bibfnamefont {A.~M.}\ \bibnamefont {L\"auchli}},\ }\href {\doibase
  10.1103/PhysRevX.8.021070} {\bibfield  {journal} {\bibinfo  {journal} {Phys.
  Rev. X}\ }\textbf {\bibinfo {volume} {8}},\ \bibinfo {pages} {021070}
  (\bibinfo {year} {2018})}\BibitemShut {NoStop}%
\bibitem [{\citenamefont {Schreiber}\ \emph {et~al.}(2015)\citenamefont
  {Schreiber}, \citenamefont {Hodgman}, \citenamefont {Bordia}, \citenamefont
  {L{\"u}schen}, \citenamefont {Fischer}, \citenamefont {Vosk}, \citenamefont
  {Altman}, \citenamefont {Schneider},\ and\ \citenamefont
  {Bloch}}]{Schreiber15}%
  \BibitemOpen
  \bibfield  {author} {\bibinfo {author} {\bibfnamefont {M.}~\bibnamefont
  {Schreiber}}, \bibinfo {author} {\bibfnamefont {S.~S.}\ \bibnamefont
  {Hodgman}}, \bibinfo {author} {\bibfnamefont {P.}~\bibnamefont {Bordia}},
  \bibinfo {author} {\bibfnamefont {H.~P.}\ \bibnamefont {L{\"u}schen}},
  \bibinfo {author} {\bibfnamefont {M.~H.}\ \bibnamefont {Fischer}}, \bibinfo
  {author} {\bibfnamefont {R.}~\bibnamefont {Vosk}}, \bibinfo {author}
  {\bibfnamefont {E.}~\bibnamefont {Altman}}, \bibinfo {author} {\bibfnamefont
  {U.}~\bibnamefont {Schneider}}, \ and\ \bibinfo {author} {\bibfnamefont
  {I.}~\bibnamefont {Bloch}},\ }\href {\doibase 10.1126/science.aaa7432}
  {\bibfield  {journal} {\bibinfo  {journal} {Science}\ }\textbf {\bibinfo
  {volume} {349}},\ \bibinfo {pages} {842} (\bibinfo {year}
  {2015})}\BibitemShut {NoStop}%
\bibitem [{\citenamefont {Kondov}\ \emph {et~al.}(2015)\citenamefont {Kondov},
  \citenamefont {McGehee}, \citenamefont {Xu},\ and\ \citenamefont
  {DeMarco}}]{Kondov15}%
  \BibitemOpen
  \bibfield  {author} {\bibinfo {author} {\bibfnamefont {S.~S.}\ \bibnamefont
  {Kondov}}, \bibinfo {author} {\bibfnamefont {W.~R.}\ \bibnamefont {McGehee}},
  \bibinfo {author} {\bibfnamefont {W.}~\bibnamefont {Xu}}, \ and\ \bibinfo
  {author} {\bibfnamefont {B.}~\bibnamefont {DeMarco}},\ }\href {\doibase
  10.1103/PhysRevLett.114.083002} {\bibfield  {journal} {\bibinfo  {journal}
  {Phys. Rev. Lett.}\ }\textbf {\bibinfo {volume} {114}},\ \bibinfo {pages}
  {083002} (\bibinfo {year} {2015})}\BibitemShut {NoStop}%
\bibitem [{\citenamefont {Bordia}\ \emph {et~al.}(2016)\citenamefont {Bordia},
  \citenamefont {L\"uschen}, \citenamefont {Hodgman}, \citenamefont
  {Schreiber}, \citenamefont {Bloch},\ and\ \citenamefont
  {Schneider}}]{Bordia16}%
  \BibitemOpen
  \bibfield  {author} {\bibinfo {author} {\bibfnamefont {P.}~\bibnamefont
  {Bordia}}, \bibinfo {author} {\bibfnamefont {H.~P.}\ \bibnamefont
  {L\"uschen}}, \bibinfo {author} {\bibfnamefont {S.~S.}\ \bibnamefont
  {Hodgman}}, \bibinfo {author} {\bibfnamefont {M.}~\bibnamefont {Schreiber}},
  \bibinfo {author} {\bibfnamefont {I.}~\bibnamefont {Bloch}}, \ and\ \bibinfo
  {author} {\bibfnamefont {U.}~\bibnamefont {Schneider}},\ }\href {\doibase
  10.1103/PhysRevLett.116.140401} {\bibfield  {journal} {\bibinfo  {journal}
  {Phys. Rev. Lett.}\ }\textbf {\bibinfo {volume} {116}},\ \bibinfo {pages}
  {140401} (\bibinfo {year} {2016})}\BibitemShut {NoStop}%
\bibitem [{\citenamefont {Kaufman}\ \emph {et~al.}(2016)\citenamefont
  {Kaufman}, \citenamefont {Tai}, \citenamefont {Lukin}, \citenamefont
  {Rispoli}, \citenamefont {Schittko}, \citenamefont {Preiss},\ and\
  \citenamefont {Greiner}}]{Kaufman16}%
  \BibitemOpen
  \bibfield  {author} {\bibinfo {author} {\bibfnamefont {A.~M.}\ \bibnamefont
  {Kaufman}}, \bibinfo {author} {\bibfnamefont {M.~E.}\ \bibnamefont {Tai}},
  \bibinfo {author} {\bibfnamefont {A.}~\bibnamefont {Lukin}}, \bibinfo
  {author} {\bibfnamefont {M.}~\bibnamefont {Rispoli}}, \bibinfo {author}
  {\bibfnamefont {R.}~\bibnamefont {Schittko}}, \bibinfo {author}
  {\bibfnamefont {P.~M.}\ \bibnamefont {Preiss}}, \ and\ \bibinfo {author}
  {\bibfnamefont {M.}~\bibnamefont {Greiner}},\ }\href {\doibase
  10.1126/science.aaf6725} {\bibfield  {journal} {\bibinfo  {journal}
  {Science}\ }\textbf {\bibinfo {volume} {353}},\ \bibinfo {pages} {794}
  (\bibinfo {year} {2016})}\BibitemShut {NoStop}%
\bibitem [{\citenamefont {Choi}\ \emph {et~al.}(2016)\citenamefont {Choi},
  \citenamefont {Hild}, \citenamefont {Zeiher}, \citenamefont {Schau{\ss}},
  \citenamefont {Rubio-Abadal}, \citenamefont {Yefsah}, \citenamefont
  {Khemani}, \citenamefont {Huse}, \citenamefont {Bloch},\ and\ \citenamefont
  {Gross}}]{Choi16}%
  \BibitemOpen
  \bibfield  {author} {\bibinfo {author} {\bibfnamefont {J.-y.}\ \bibnamefont
  {Choi}}, \bibinfo {author} {\bibfnamefont {S.}~\bibnamefont {Hild}}, \bibinfo
  {author} {\bibfnamefont {J.}~\bibnamefont {Zeiher}}, \bibinfo {author}
  {\bibfnamefont {P.}~\bibnamefont {Schau{\ss}}}, \bibinfo {author}
  {\bibfnamefont {A.}~\bibnamefont {Rubio-Abadal}}, \bibinfo {author}
  {\bibfnamefont {T.}~\bibnamefont {Yefsah}}, \bibinfo {author} {\bibfnamefont
  {V.}~\bibnamefont {Khemani}}, \bibinfo {author} {\bibfnamefont {D.~A.}\
  \bibnamefont {Huse}}, \bibinfo {author} {\bibfnamefont {I.}~\bibnamefont
  {Bloch}}, \ and\ \bibinfo {author} {\bibfnamefont {C.}~\bibnamefont
  {Gross}},\ }\href {\doibase 10.1126/science.aaf8834} {\bibfield  {journal}
  {\bibinfo  {journal} {Science}\ }\textbf {\bibinfo {volume} {352}},\ \bibinfo
  {pages} {1547} (\bibinfo {year} {2016})}\BibitemShut {NoStop}%
\bibitem [{\citenamefont {Bordia}\ \emph {et~al.}(2017)\citenamefont {Bordia},
  \citenamefont {L\"uschen}, \citenamefont {Scherg}, \citenamefont
  {Gopalakrishnan}, \citenamefont {Knap}, \citenamefont {Schneider},\ and\
  \citenamefont {Bloch}}]{Bordia17}%
  \BibitemOpen
  \bibfield  {author} {\bibinfo {author} {\bibfnamefont {P.}~\bibnamefont
  {Bordia}}, \bibinfo {author} {\bibfnamefont {H.}~\bibnamefont {L\"uschen}},
  \bibinfo {author} {\bibfnamefont {S.}~\bibnamefont {Scherg}}, \bibinfo
  {author} {\bibfnamefont {S.}~\bibnamefont {Gopalakrishnan}}, \bibinfo
  {author} {\bibfnamefont {M.}~\bibnamefont {Knap}}, \bibinfo {author}
  {\bibfnamefont {U.}~\bibnamefont {Schneider}}, \ and\ \bibinfo {author}
  {\bibfnamefont {I.}~\bibnamefont {Bloch}},\ }\href {\doibase
  10.1103/PhysRevX.7.041047} {\bibfield  {journal} {\bibinfo  {journal} {Phys.
  Rev. X}\ }\textbf {\bibinfo {volume} {7}},\ \bibinfo {pages} {041047}
  (\bibinfo {year} {2017})}\BibitemShut {NoStop}%
\bibitem [{\citenamefont {L{\"u}schen}\ \emph {et~al.}(2017)\citenamefont
  {L{\"u}schen}, \citenamefont {Bordia}, \citenamefont {Scherg}, \citenamefont
  {Alet}, \citenamefont {Altman}, \citenamefont {Schneider},\ and\
  \citenamefont {Bloch}}]{Luschen17}%
  \BibitemOpen
  \bibfield  {author} {\bibinfo {author} {\bibfnamefont {H.~P.}\ \bibnamefont
  {L{\"u}schen}}, \bibinfo {author} {\bibfnamefont {P.}~\bibnamefont {Bordia}},
  \bibinfo {author} {\bibfnamefont {S.}~\bibnamefont {Scherg}}, \bibinfo
  {author} {\bibfnamefont {F.}~\bibnamefont {Alet}}, \bibinfo {author}
  {\bibfnamefont {E.}~\bibnamefont {Altman}}, \bibinfo {author} {\bibfnamefont
  {U.}~\bibnamefont {Schneider}}, \ and\ \bibinfo {author} {\bibfnamefont
  {I.}~\bibnamefont {Bloch}},\ }\href {\doibase 10.1103/PhysRevLett.119.260401}
  {\bibfield  {journal} {\bibinfo  {journal} {Phys. Rev. Lett.}\ }\textbf
  {\bibinfo {volume} {119}},\ \bibinfo {pages} {260401} (\bibinfo {year}
  {2017})}\BibitemShut {NoStop}%
\bibitem [{\citenamefont {L{\"u}schen}\ \emph {et~al.}(2018)\citenamefont
  {L{\"u}schen}, \citenamefont {Scherg}, \citenamefont {Kohlert}, \citenamefont
  {Schreiber}, \citenamefont {Bordia}, \citenamefont {Li}, \citenamefont
  {Das~Sarma},\ and\ \citenamefont {Bloch}}]{Luschen18}%
  \BibitemOpen
  \bibfield  {author} {\bibinfo {author} {\bibfnamefont {H.~P.}\ \bibnamefont
  {L{\"u}schen}}, \bibinfo {author} {\bibfnamefont {S.}~\bibnamefont {Scherg}},
  \bibinfo {author} {\bibfnamefont {T.}~\bibnamefont {Kohlert}}, \bibinfo
  {author} {\bibfnamefont {M.}~\bibnamefont {Schreiber}}, \bibinfo {author}
  {\bibfnamefont {P.}~\bibnamefont {Bordia}}, \bibinfo {author} {\bibfnamefont
  {X.}~\bibnamefont {Li}}, \bibinfo {author} {\bibfnamefont {S.}~\bibnamefont
  {Das~Sarma}}, \ and\ \bibinfo {author} {\bibfnamefont {I.}~\bibnamefont
  {Bloch}},\ }\href {\doibase 10.1103/PhysRevLett.120.160404} {\bibfield
  {journal} {\bibinfo  {journal} {Phys. Rev. Lett.}\ }\textbf {\bibinfo
  {volume} {120}},\ \bibinfo {pages} {160404} (\bibinfo {year}
  {2018})}\BibitemShut {NoStop}%
\bibitem [{\citenamefont {{Lukin}}\ \emph {et~al.}(2018)\citenamefont
  {{Lukin}}, \citenamefont {{Rispoli}}, \citenamefont {{Schittko}},
  \citenamefont {{Tai}}, \citenamefont {{Kaufman}}, \citenamefont {{Choi}},
  \citenamefont {{Khemani}}, \citenamefont {{L{\'e}onard}},\ and\ \citenamefont
  {{Greiner}}}]{Lukin18}%
  \BibitemOpen
  \bibfield  {author} {\bibinfo {author} {\bibfnamefont {A.}~\bibnamefont
  {{Lukin}}}, \bibinfo {author} {\bibfnamefont {M.}~\bibnamefont {{Rispoli}}},
  \bibinfo {author} {\bibfnamefont {R.}~\bibnamefont {{Schittko}}}, \bibinfo
  {author} {\bibfnamefont {M.~E.}\ \bibnamefont {{Tai}}}, \bibinfo {author}
  {\bibfnamefont {A.~M.}\ \bibnamefont {{Kaufman}}}, \bibinfo {author}
  {\bibfnamefont {S.}~\bibnamefont {{Choi}}}, \bibinfo {author} {\bibfnamefont
  {V.}~\bibnamefont {{Khemani}}}, \bibinfo {author} {\bibfnamefont
  {J.}~\bibnamefont {{L{\'e}onard}}}, \ and\ \bibinfo {author} {\bibfnamefont
  {M.}~\bibnamefont {{Greiner}}},\ }\href@noop {} {\bibfield  {journal}
  {\bibinfo  {journal} {ArXiv e-prints}\ } (\bibinfo {year} {2018})},\ \Eprint
  {http://arxiv.org/abs/1805.09819} {arXiv:1805.09819 [cond-mat.quant-gas]}
  \BibitemShut {NoStop}%
\bibitem [{\citenamefont {Smith}\ \emph {et~al.}(2016)\citenamefont {Smith},
  \citenamefont {Lee}, \citenamefont {Richerme}, \citenamefont {Neyenhuis},
  \citenamefont {Hess}, \citenamefont {Hauke}, \citenamefont {Huse},\ and\
  \citenamefont {Monroe}}]{Smith16}%
  \BibitemOpen
  \bibfield  {author} {\bibinfo {author} {\bibfnamefont {J.}~\bibnamefont
  {Smith}}, \bibinfo {author} {\bibfnamefont {A.}~\bibnamefont {Lee}}, \bibinfo
  {author} {\bibfnamefont {P.}~\bibnamefont {Richerme}}, \bibinfo {author}
  {\bibfnamefont {B.}~\bibnamefont {Neyenhuis}}, \bibinfo {author}
  {\bibfnamefont {P.~W.}\ \bibnamefont {Hess}}, \bibinfo {author}
  {\bibfnamefont {P.}~\bibnamefont {Hauke}}, \bibinfo {author} {\bibfnamefont
  {D.~A.}\ \bibnamefont {Huse}}, \ and\ \bibinfo {author} {\bibfnamefont
  {C.}~\bibnamefont {Monroe}},\ }\href {\doibase 10.1038/NPHYS3783} {\bibfield
  {journal} {\bibinfo  {journal} {Nat. Phys.}\ }\textbf {\bibinfo {volume}
  {12}},\ \bibinfo {pages} {907} (\bibinfo {year} {2016})}\BibitemShut
  {NoStop}%
\bibitem [{\citenamefont {Zhang}\ \emph {et~al.}(2017)\citenamefont {Zhang},
  \citenamefont {Hess}, \citenamefont {Kyprianidis}, \citenamefont {Becker},
  \citenamefont {Lee}, \citenamefont {Smith}, \citenamefont {Pagano},
  \citenamefont {Potirniche}, \citenamefont {Potter}, \citenamefont
  {Vishwanath}, \citenamefont {Yao},\ and\ \citenamefont {Monroe}}]{Zhang17}%
  \BibitemOpen
  \bibfield  {author} {\bibinfo {author} {\bibfnamefont {J.}~\bibnamefont
  {Zhang}}, \bibinfo {author} {\bibfnamefont {P.~W.}\ \bibnamefont {Hess}},
  \bibinfo {author} {\bibfnamefont {A.}~\bibnamefont {Kyprianidis}}, \bibinfo
  {author} {\bibfnamefont {P.}~\bibnamefont {Becker}}, \bibinfo {author}
  {\bibfnamefont {A.}~\bibnamefont {Lee}}, \bibinfo {author} {\bibfnamefont
  {J.}~\bibnamefont {Smith}}, \bibinfo {author} {\bibfnamefont
  {G.}~\bibnamefont {Pagano}}, \bibinfo {author} {\bibfnamefont {I.-D.}\
  \bibnamefont {Potirniche}}, \bibinfo {author} {\bibfnamefont {A.~C.}\
  \bibnamefont {Potter}}, \bibinfo {author} {\bibfnamefont {A.}~\bibnamefont
  {Vishwanath}}, \bibinfo {author} {\bibfnamefont {N.~Y.}\ \bibnamefont {Yao}},
  \ and\ \bibinfo {author} {\bibfnamefont {C.}~\bibnamefont {Monroe}},\ }\href
  {\doibase 10.1038/nature21413} {\bibfield  {journal} {\bibinfo  {journal}
  {Nature}\ }\textbf {\bibinfo {volume} {543}},\ \bibinfo {pages} {217}
  (\bibinfo {year} {2017})}\BibitemShut {NoStop}%
\bibitem [{\citenamefont {Choi}\ \emph {et~al.}(2017)\citenamefont {Choi},
  \citenamefont {Choi}, \citenamefont {Landig}, \citenamefont {Kucsko},
  \citenamefont {Zhou}, \citenamefont {Isoya}, \citenamefont {Jelezko},
  \citenamefont {Onoda}, \citenamefont {Sumiya}, \citenamefont {Khemani},
  \citenamefont {von Keyserlingk}, \citenamefont {Yao}, \citenamefont
  {Demler},\ and\ \citenamefont {Lukin}}]{Choi17}%
  \BibitemOpen
  \bibfield  {author} {\bibinfo {author} {\bibfnamefont {S.}~\bibnamefont
  {Choi}}, \bibinfo {author} {\bibfnamefont {J.}~\bibnamefont {Choi}}, \bibinfo
  {author} {\bibfnamefont {R.}~\bibnamefont {Landig}}, \bibinfo {author}
  {\bibfnamefont {G.}~\bibnamefont {Kucsko}}, \bibinfo {author} {\bibfnamefont
  {H.}~\bibnamefont {Zhou}}, \bibinfo {author} {\bibfnamefont {J.}~\bibnamefont
  {Isoya}}, \bibinfo {author} {\bibfnamefont {F.}~\bibnamefont {Jelezko}},
  \bibinfo {author} {\bibfnamefont {S.}~\bibnamefont {Onoda}}, \bibinfo
  {author} {\bibfnamefont {H.}~\bibnamefont {Sumiya}}, \bibinfo {author}
  {\bibfnamefont {V.}~\bibnamefont {Khemani}}, \bibinfo {author} {\bibfnamefont
  {C.}~\bibnamefont {von Keyserlingk}}, \bibinfo {author} {\bibfnamefont
  {N.~Y.}\ \bibnamefont {Yao}}, \bibinfo {author} {\bibfnamefont
  {E.}~\bibnamefont {Demler}}, \ and\ \bibinfo {author} {\bibfnamefont {M.~D.}\
  \bibnamefont {Lukin}},\ }\href {\doibase 10.1038/nature21426} {\bibfield
  {journal} {\bibinfo  {journal} {Nature}\ }\textbf {\bibinfo {volume} {543}},\
  \bibinfo {pages} {221} (\bibinfo {year} {2017})}\BibitemShut {NoStop}%
\bibitem [{Fer()}]{Fermi55}%
  \BibitemOpen
  \href@noop {} {}\bibinfo {howpublished} {E.~Fermi, J.~Pasta and S.~Ulam,
  “Studies of Nonlinear Problems.”, Los Alamos report LA-1940
  (1955).}\BibitemShut {Stop}%
\bibitem [{\citenamefont {Zabusky}\ and\ \citenamefont
  {Kruskal}(1965)}]{Zabusky65}%
  \BibitemOpen
  \bibfield  {author} {\bibinfo {author} {\bibfnamefont {N.~J.}\ \bibnamefont
  {Zabusky}}\ and\ \bibinfo {author} {\bibfnamefont {M.~D.}\ \bibnamefont
  {Kruskal}},\ }\href {\doibase 10.1103/PhysRevLett.15.240} {\bibfield
  {journal} {\bibinfo  {journal} {Phys. Rev. Lett.}\ }\textbf {\bibinfo
  {volume} {15}},\ \bibinfo {pages} {240} (\bibinfo {year} {1965})}\BibitemShut
  {NoStop}%
\bibitem [{Gal()}]{Gallavotti08}%
  \BibitemOpen
  \href@noop {} {}\bibinfo {howpublished} {G.~Gallavotti (Editor), {\it The
  Fermi-Pasta-Ulam Problem: A Status Report} (Springer-Verlag, Berlin,
  2008).}\BibitemShut {Stop}%
\bibitem [{\citenamefont {Caux}\ and\ \citenamefont {Mossel}(2011)}]{Caux11}%
  \BibitemOpen
  \bibfield  {author} {\bibinfo {author} {\bibfnamefont {J.-S.}\ \bibnamefont
  {Caux}}\ and\ \bibinfo {author} {\bibfnamefont {J.}~\bibnamefont {Mossel}},\
  }\href {http://stacks.iop.org/1742-5468/2011/i=02/a=P02023} {\bibfield
  {journal} {\bibinfo  {journal} {J. Stat. Mech. Theory Exp.}\ }\textbf
  {\bibinfo {volume} {2011}},\ \bibinfo {pages} {P02023} (\bibinfo {year}
  {2011})}\BibitemShut {NoStop}%
\bibitem [{\citenamefont {Brandino}\ \emph {et~al.}(2015)\citenamefont
  {Brandino}, \citenamefont {Caux},\ and\ \citenamefont {Konik}}]{Brandino15}%
  \BibitemOpen
  \bibfield  {author} {\bibinfo {author} {\bibfnamefont {G.~P.}\ \bibnamefont
  {Brandino}}, \bibinfo {author} {\bibfnamefont {J.-S.}\ \bibnamefont {Caux}},
  \ and\ \bibinfo {author} {\bibfnamefont {R.~M.}\ \bibnamefont {Konik}},\
  }\href {\doibase 10.1103/PhysRevX.5.041043} {\bibfield  {journal} {\bibinfo
  {journal} {Phys. Rev. X}\ }\textbf {\bibinfo {volume} {5}},\ \bibinfo {pages}
  {041043} (\bibinfo {year} {2015})}\BibitemShut {NoStop}%
\bibitem [{\citenamefont {Kinoshita}\ \emph {et~al.}(2006)\citenamefont
  {Kinoshita}, \citenamefont {Wenger},\ and\ \citenamefont
  {Weiss}}]{Kinoshita06}%
  \BibitemOpen
  \bibfield  {author} {\bibinfo {author} {\bibfnamefont {T.}~\bibnamefont
  {Kinoshita}}, \bibinfo {author} {\bibfnamefont {T.}~\bibnamefont {Wenger}}, \
  and\ \bibinfo {author} {\bibfnamefont {D.~S.}\ \bibnamefont {Weiss}},\ }\href
  {\doibase 10.1038/nature04693} {\bibfield  {journal} {\bibinfo  {journal}
  {Nature}\ }\textbf {\bibinfo {volume} {440}},\ \bibinfo {pages} {900}
  (\bibinfo {year} {2006})}\BibitemShut {NoStop}%
\bibitem [{\citenamefont {Gornyi}\ \emph {et~al.}(2005)\citenamefont {Gornyi},
  \citenamefont {Mirlin},\ and\ \citenamefont {Polyakov}}]{Gornyi05}%
  \BibitemOpen
  \bibfield  {author} {\bibinfo {author} {\bibfnamefont {I.~V.}\ \bibnamefont
  {Gornyi}}, \bibinfo {author} {\bibfnamefont {A.~D.}\ \bibnamefont {Mirlin}},
  \ and\ \bibinfo {author} {\bibfnamefont {D.~G.}\ \bibnamefont {Polyakov}},\
  }\href {\doibase 10.1103/PhysRevLett.95.206603} {\bibfield  {journal}
  {\bibinfo  {journal} {Phys. Rev. Lett.}\ }\textbf {\bibinfo {volume} {95}},\
  \bibinfo {pages} {206603} (\bibinfo {year} {2005})}\BibitemShut {NoStop}%
\bibitem [{\citenamefont {Basko}\ \emph {et~al.}(2006)\citenamefont {Basko},
  \citenamefont {Aleiner},\ and\ \citenamefont {Altshuler}}]{Basko06}%
  \BibitemOpen
  \bibfield  {author} {\bibinfo {author} {\bibfnamefont {D.}~\bibnamefont
  {Basko}}, \bibinfo {author} {\bibfnamefont {I.}~\bibnamefont {Aleiner}}, \
  and\ \bibinfo {author} {\bibfnamefont {B.}~\bibnamefont {Altshuler}},\ }\href
  {\doibase https://doi.org/10.1016/j.aop.2005.11.014} {\bibfield  {journal}
  {\bibinfo  {journal} {Ann. Phys. (N.Y.)}\ }\textbf {\bibinfo {volume}
  {321}},\ \bibinfo {pages} {1126 } (\bibinfo {year} {2006})}\BibitemShut
  {NoStop}%
\bibitem [{\citenamefont {Oganesyan}\ and\ \citenamefont
  {Huse}(2007)}]{Oganesyan07}%
  \BibitemOpen
  \bibfield  {author} {\bibinfo {author} {\bibfnamefont {V.}~\bibnamefont
  {Oganesyan}}\ and\ \bibinfo {author} {\bibfnamefont {D.~A.}\ \bibnamefont
  {Huse}},\ }\href {\doibase 10.1103/PhysRevB.75.155111} {\bibfield  {journal}
  {\bibinfo  {journal} {Phys. Rev. B}\ }\textbf {\bibinfo {volume} {75}},\
  \bibinfo {pages} {155111} (\bibinfo {year} {2007})}\BibitemShut {NoStop}%
\bibitem [{\citenamefont {\ifmmode \check{Z}\else
  \v{Z}\fi{}nidari\ifmmode~\check{c}\else \v{c}\fi{}}\ \emph
  {et~al.}(2008)\citenamefont {\ifmmode \check{Z}\else
  \v{Z}\fi{}nidari\ifmmode~\check{c}\else \v{c}\fi{}}, \citenamefont {Prosen},\
  and\ \citenamefont {Prelov\ifmmode~\check{s}\else
  \v{s}\fi{}ek}}]{Znidaric08}%
  \BibitemOpen
  \bibfield  {author} {\bibinfo {author} {\bibfnamefont {M.}~\bibnamefont
  {\ifmmode \check{Z}\else \v{Z}\fi{}nidari\ifmmode~\check{c}\else
  \v{c}\fi{}}}, \bibinfo {author} {\bibfnamefont {T.}~\bibnamefont {Prosen}}, \
  and\ \bibinfo {author} {\bibfnamefont {P.}~\bibnamefont
  {Prelov\ifmmode~\check{s}\else \v{s}\fi{}ek}},\ }\href {\doibase
  10.1103/PhysRevB.77.064426} {\bibfield  {journal} {\bibinfo  {journal} {Phys.
  Rev. B}\ }\textbf {\bibinfo {volume} {77}},\ \bibinfo {pages} {064426}
  (\bibinfo {year} {2008})}\BibitemShut {NoStop}%
\bibitem [{\citenamefont {Vosk}\ \emph {et~al.}(2015)\citenamefont {Vosk},
  \citenamefont {Huse},\ and\ \citenamefont {Altman}}]{Vosk10}%
  \BibitemOpen
  \bibfield  {author} {\bibinfo {author} {\bibfnamefont {R.}~\bibnamefont
  {Vosk}}, \bibinfo {author} {\bibfnamefont {D.~A.}\ \bibnamefont {Huse}}, \
  and\ \bibinfo {author} {\bibfnamefont {E.}~\bibnamefont {Altman}},\ }\href
  {\doibase 10.1103/PhysRevX.5.031032} {\bibfield  {journal} {\bibinfo
  {journal} {Phys. Rev. X}\ }\textbf {\bibinfo {volume} {5}},\ \bibinfo {pages}
  {031032} (\bibinfo {year} {2015})}\BibitemShut {NoStop}%
\bibitem [{\citenamefont {Pal}\ and\ \citenamefont {Huse}(2010)}]{Pal10}%
  \BibitemOpen
  \bibfield  {author} {\bibinfo {author} {\bibfnamefont {A.}~\bibnamefont
  {Pal}}\ and\ \bibinfo {author} {\bibfnamefont {D.~A.}\ \bibnamefont {Huse}},\
  }\href {\doibase 10.1103/PhysRevB.82.174411} {\bibfield  {journal} {\bibinfo
  {journal} {Phys. Rev. B}\ }\textbf {\bibinfo {volume} {82}},\ \bibinfo
  {pages} {174411} (\bibinfo {year} {2010})}\BibitemShut {NoStop}%
\bibitem [{\citenamefont {Berkelbach}\ and\ \citenamefont
  {Reichman}(2010)}]{Berkelbach10}%
  \BibitemOpen
  \bibfield  {author} {\bibinfo {author} {\bibfnamefont {T.~C.}\ \bibnamefont
  {Berkelbach}}\ and\ \bibinfo {author} {\bibfnamefont {D.~R.}\ \bibnamefont
  {Reichman}},\ }\href {\doibase 10.1103/PhysRevB.81.224429} {\bibfield
  {journal} {\bibinfo  {journal} {Phys. Rev. B}\ }\textbf {\bibinfo {volume}
  {81}},\ \bibinfo {pages} {224429} (\bibinfo {year} {2010})}\BibitemShut
  {NoStop}%
\bibitem [{\citenamefont {Bardarson}\ \emph {et~al.}(2012)\citenamefont
  {Bardarson}, \citenamefont {Pollmann},\ and\ \citenamefont
  {Moore}}]{Bardarson12}%
  \BibitemOpen
  \bibfield  {author} {\bibinfo {author} {\bibfnamefont {J.~H.}\ \bibnamefont
  {Bardarson}}, \bibinfo {author} {\bibfnamefont {F.}~\bibnamefont {Pollmann}},
  \ and\ \bibinfo {author} {\bibfnamefont {J.~E.}\ \bibnamefont {Moore}},\
  }\href {\doibase 10.1103/PhysRevLett.109.017202} {\bibfield  {journal}
  {\bibinfo  {journal} {Phys. Rev. Lett.}\ }\textbf {\bibinfo {volume} {109}},\
  \bibinfo {pages} {017202} (\bibinfo {year} {2012})}\BibitemShut {NoStop}%
\bibitem [{\citenamefont {Serbyn}\ \emph {et~al.}(2013)\citenamefont {Serbyn},
  \citenamefont {Papi\ifmmode~\acute{c}\else \'{c}\fi{}},\ and\ \citenamefont
  {Abanin}}]{Serbyn13}%
  \BibitemOpen
  \bibfield  {author} {\bibinfo {author} {\bibfnamefont {M.}~\bibnamefont
  {Serbyn}}, \bibinfo {author} {\bibfnamefont {Z.}~\bibnamefont
  {Papi\ifmmode~\acute{c}\else \'{c}\fi{}}}, \ and\ \bibinfo {author}
  {\bibfnamefont {D.~A.}\ \bibnamefont {Abanin}},\ }\href {\doibase
  10.1103/PhysRevLett.111.127201} {\bibfield  {journal} {\bibinfo  {journal}
  {Phys. Rev. Lett.}\ }\textbf {\bibinfo {volume} {111}},\ \bibinfo {pages}
  {127201} (\bibinfo {year} {2013})}\BibitemShut {NoStop}%
\bibitem [{\citenamefont {Huse}\ \emph {et~al.}(2014)\citenamefont {Huse},
  \citenamefont {Nandkishore},\ and\ \citenamefont {Oganesyan}}]{Huse14}%
  \BibitemOpen
  \bibfield  {author} {\bibinfo {author} {\bibfnamefont {D.~A.}\ \bibnamefont
  {Huse}}, \bibinfo {author} {\bibfnamefont {R.}~\bibnamefont {Nandkishore}}, \
  and\ \bibinfo {author} {\bibfnamefont {V.}~\bibnamefont {Oganesyan}},\ }\href
  {\doibase 10.1103/PhysRevB.90.174202} {\bibfield  {journal} {\bibinfo
  {journal} {Phys. Rev. B}\ }\textbf {\bibinfo {volume} {90}},\ \bibinfo
  {pages} {174202} (\bibinfo {year} {2014})}\BibitemShut {NoStop}%
\bibitem [{\citenamefont {Bar~Lev}\ and\ \citenamefont
  {Reichman}(2014)}]{Barlev14}%
  \BibitemOpen
  \bibfield  {author} {\bibinfo {author} {\bibfnamefont {Y.}~\bibnamefont
  {Bar~Lev}}\ and\ \bibinfo {author} {\bibfnamefont {D.~R.}\ \bibnamefont
  {Reichman}},\ }\href {\doibase 10.1103/PhysRevB.89.220201} {\bibfield
  {journal} {\bibinfo  {journal} {Phys. Rev. B}\ }\textbf {\bibinfo {volume}
  {89}},\ \bibinfo {pages} {220201} (\bibinfo {year} {2014})}\BibitemShut
  {NoStop}%
\bibitem [{\citenamefont {Serbyn}\ \emph {et~al.}(2014)\citenamefont {Serbyn},
  \citenamefont {Papi\ifmmode~\acute{c}\else \'{c}\fi{}},\ and\ \citenamefont
  {Abanin}}]{Serbyn14}%
  \BibitemOpen
  \bibfield  {author} {\bibinfo {author} {\bibfnamefont {M.}~\bibnamefont
  {Serbyn}}, \bibinfo {author} {\bibfnamefont {Z.}~\bibnamefont
  {Papi\ifmmode~\acute{c}\else \'{c}\fi{}}}, \ and\ \bibinfo {author}
  {\bibfnamefont {D.~A.}\ \bibnamefont {Abanin}},\ }\href {\doibase
  10.1103/PhysRevB.90.174302} {\bibfield  {journal} {\bibinfo  {journal} {Phys.
  Rev. B}\ }\textbf {\bibinfo {volume} {90}},\ \bibinfo {pages} {174302}
  (\bibinfo {year} {2014})}\BibitemShut {NoStop}%
\bibitem [{\citenamefont {Serbyn}\ \emph {et~al.}(2015)\citenamefont {Serbyn},
  \citenamefont {Papi\ifmmode~\acute{c}\else \'{c}\fi{}},\ and\ \citenamefont
  {Abanin}}]{Serbyn15}%
  \BibitemOpen
  \bibfield  {author} {\bibinfo {author} {\bibfnamefont {M.}~\bibnamefont
  {Serbyn}}, \bibinfo {author} {\bibfnamefont {Z.}~\bibnamefont
  {Papi\ifmmode~\acute{c}\else \'{c}\fi{}}}, \ and\ \bibinfo {author}
  {\bibfnamefont {D.~A.}\ \bibnamefont {Abanin}},\ }\href {\doibase
  10.1103/PhysRevX.5.041047} {\bibfield  {journal} {\bibinfo  {journal} {Phys.
  Rev. X}\ }\textbf {\bibinfo {volume} {5}},\ \bibinfo {pages} {041047}
  (\bibinfo {year} {2015})}\BibitemShut {NoStop}%
\bibitem [{\citenamefont {Chandran}\ \emph {et~al.}(2015)\citenamefont
  {Chandran}, \citenamefont {Kim}, \citenamefont {Vidal},\ and\ \citenamefont
  {Abanin}}]{Chandran15a}%
  \BibitemOpen
  \bibfield  {author} {\bibinfo {author} {\bibfnamefont {A.}~\bibnamefont
  {Chandran}}, \bibinfo {author} {\bibfnamefont {I.~H.}\ \bibnamefont {Kim}},
  \bibinfo {author} {\bibfnamefont {G.}~\bibnamefont {Vidal}}, \ and\ \bibinfo
  {author} {\bibfnamefont {D.~A.}\ \bibnamefont {Abanin}},\ }\href {\doibase
  10.1103/PhysRevB.91.085425} {\bibfield  {journal} {\bibinfo  {journal} {Phys.
  Rev. B}\ }\textbf {\bibinfo {volume} {91}},\ \bibinfo {pages} {085425}
  (\bibinfo {year} {2015})}\BibitemShut {NoStop}%
\bibitem [{\citenamefont {Luitz}\ \emph {et~al.}(2015)\citenamefont {Luitz},
  \citenamefont {Laflorencie},\ and\ \citenamefont {Alet}}]{Luitz15}%
  \BibitemOpen
  \bibfield  {author} {\bibinfo {author} {\bibfnamefont {D.~J.}\ \bibnamefont
  {Luitz}}, \bibinfo {author} {\bibfnamefont {N.}~\bibnamefont {Laflorencie}},
  \ and\ \bibinfo {author} {\bibfnamefont {F.}~\bibnamefont {Alet}},\ }\href
  {\doibase 10.1103/PhysRevB.91.081103} {\bibfield  {journal} {\bibinfo
  {journal} {Phys. Rev. B}\ }\textbf {\bibinfo {volume} {91}},\ \bibinfo
  {pages} {081103} (\bibinfo {year} {2015})}\BibitemShut {NoStop}%
\bibitem [{\citenamefont {Potter}\ \emph {et~al.}(2015)\citenamefont {Potter},
  \citenamefont {Vasseur},\ and\ \citenamefont {Parameswaran}}]{Potter15}%
  \BibitemOpen
  \bibfield  {author} {\bibinfo {author} {\bibfnamefont {A.~C.}\ \bibnamefont
  {Potter}}, \bibinfo {author} {\bibfnamefont {R.}~\bibnamefont {Vasseur}}, \
  and\ \bibinfo {author} {\bibfnamefont {S.~A.}\ \bibnamefont {Parameswaran}},\
  }\href {\doibase 10.1103/PhysRevX.5.031033} {\bibfield  {journal} {\bibinfo
  {journal} {Phys. Rev. X}\ }\textbf {\bibinfo {volume} {5}},\ \bibinfo {pages}
  {031033} (\bibinfo {year} {2015})}\BibitemShut {NoStop}%
\bibitem [{\citenamefont {Agarwal}\ \emph {et~al.}(2015)\citenamefont
  {Agarwal}, \citenamefont {Gopalakrishnan}, \citenamefont {Knap},
  \citenamefont {M{\"u}ller},\ and\ \citenamefont {Demler}}]{Agarwal15}%
  \BibitemOpen
  \bibfield  {author} {\bibinfo {author} {\bibfnamefont {K.}~\bibnamefont
  {Agarwal}}, \bibinfo {author} {\bibfnamefont {S.}~\bibnamefont
  {Gopalakrishnan}}, \bibinfo {author} {\bibfnamefont {M.}~\bibnamefont
  {Knap}}, \bibinfo {author} {\bibfnamefont {M.}~\bibnamefont {M{\"u}ller}}, \
  and\ \bibinfo {author} {\bibfnamefont {E.}~\bibnamefont {Demler}},\ }\href
  {\doibase 10.1103/PhysRevLett.114.160401} {\bibfield  {journal} {\bibinfo
  {journal} {Phys. Rev. Lett.}\ }\textbf {\bibinfo {volume} {114}},\ \bibinfo
  {pages} {160401} (\bibinfo {year} {2015})}\BibitemShut {NoStop}%
\bibitem [{\citenamefont {Ros}\ \emph {et~al.}(2015)\citenamefont {Ros},
  \citenamefont {M{\"u}ller},\ and\ \citenamefont {Scardicchio}}]{Ros15}%
  \BibitemOpen
  \bibfield  {author} {\bibinfo {author} {\bibfnamefont {V.}~\bibnamefont
  {Ros}}, \bibinfo {author} {\bibfnamefont {M.}~\bibnamefont {M{\"u}ller}}, \
  and\ \bibinfo {author} {\bibfnamefont {A.}~\bibnamefont {Scardicchio}},\
  }\href {\doibase 10.1016/j.nuclphysb.2014.12.014} {\bibfield  {journal}
  {\bibinfo  {journal} {Nucl. Phys. B}\ }\textbf {\bibinfo {volume} {891}},\
  \bibinfo {pages} {420 } (\bibinfo {year} {2015})}\BibitemShut {NoStop}%
\bibitem [{\citenamefont {Gopalakrishnan}\ \emph {et~al.}(2015)\citenamefont
  {Gopalakrishnan}, \citenamefont {M{\"u}ller}, \citenamefont {Khemani},
  \citenamefont {Knap}, \citenamefont {Demler},\ and\ \citenamefont
  {Huse}}]{Gopalakrishnan15}%
  \BibitemOpen
  \bibfield  {author} {\bibinfo {author} {\bibfnamefont {S.}~\bibnamefont
  {Gopalakrishnan}}, \bibinfo {author} {\bibfnamefont {M.}~\bibnamefont
  {M{\"u}ller}}, \bibinfo {author} {\bibfnamefont {V.}~\bibnamefont {Khemani}},
  \bibinfo {author} {\bibfnamefont {M.}~\bibnamefont {Knap}}, \bibinfo {author}
  {\bibfnamefont {E.}~\bibnamefont {Demler}}, \ and\ \bibinfo {author}
  {\bibfnamefont {D.~A.}\ \bibnamefont {Huse}},\ }\href {\doibase
  10.1103/PhysRevB.92.104202} {\bibfield  {journal} {\bibinfo  {journal} {Phys.
  Rev. B}\ }\textbf {\bibinfo {volume} {92}},\ \bibinfo {pages} {104202}
  (\bibinfo {year} {2015})}\BibitemShut {NoStop}%
\bibitem [{\citenamefont {\ifmmode \check{Z}\else
  \v{Z}\fi{}nidari\ifmmode~\check{c}\else \v{c}\fi{}}\ \emph
  {et~al.}(2016)\citenamefont {\ifmmode \check{Z}\else
  \v{Z}\fi{}nidari\ifmmode~\check{c}\else \v{c}\fi{}}, \citenamefont
  {Scardicchio},\ and\ \citenamefont {Varma}}]{Znidaric16}%
  \BibitemOpen
  \bibfield  {author} {\bibinfo {author} {\bibfnamefont {M.}~\bibnamefont
  {\ifmmode \check{Z}\else \v{Z}\fi{}nidari\ifmmode~\check{c}\else
  \v{c}\fi{}}}, \bibinfo {author} {\bibfnamefont {A.}~\bibnamefont
  {Scardicchio}}, \ and\ \bibinfo {author} {\bibfnamefont {V.~K.}\ \bibnamefont
  {Varma}},\ }\href {\doibase 10.1103/PhysRevLett.117.040601} {\bibfield
  {journal} {\bibinfo  {journal} {Phys. Rev. Lett.}\ }\textbf {\bibinfo
  {volume} {117}},\ \bibinfo {pages} {040601} (\bibinfo {year}
  {2016})}\BibitemShut {NoStop}%
\bibitem [{\citenamefont {Rademaker}\ and\ \citenamefont
  {Ortu\~no}(2016)}]{Rademaker16}%
  \BibitemOpen
  \bibfield  {author} {\bibinfo {author} {\bibfnamefont {L.}~\bibnamefont
  {Rademaker}}\ and\ \bibinfo {author} {\bibfnamefont {M.}~\bibnamefont
  {Ortu\~no}},\ }\href {\doibase 10.1103/PhysRevLett.116.010404} {\bibfield
  {journal} {\bibinfo  {journal} {Phys. Rev. Lett.}\ }\textbf {\bibinfo
  {volume} {116}},\ \bibinfo {pages} {010404} (\bibinfo {year}
  {2016})}\BibitemShut {NoStop}%
\bibitem [{\citenamefont {Imbrie}(2016{\natexlab{a}})}]{Imbrie16a}%
  \BibitemOpen
  \bibfield  {author} {\bibinfo {author} {\bibfnamefont {J.~Z.}\ \bibnamefont
  {Imbrie}},\ }\href {\doibase 10.1103/PhysRevLett.117.027201} {\bibfield
  {journal} {\bibinfo  {journal} {Phys. Rev. Lett.}\ }\textbf {\bibinfo
  {volume} {117}},\ \bibinfo {pages} {027201} (\bibinfo {year}
  {2016}{\natexlab{a}})}\BibitemShut {NoStop}%
\bibitem [{\citenamefont {Imbrie}(2016{\natexlab{b}})}]{Imbrie16b}%
  \BibitemOpen
  \bibfield  {author} {\bibinfo {author} {\bibfnamefont {J.~Z.}\ \bibnamefont
  {Imbrie}},\ }\href {\doibase 10.1007/s10955-016-1508-x} {\bibfield  {journal}
  {\bibinfo  {journal} {J.~Stat.~Phys}\ }\textbf {\bibinfo {volume} {163}},\
  \bibinfo {pages} {998} (\bibinfo {year} {2016}{\natexlab{b}})}\BibitemShut
  {NoStop}%
\bibitem [{\citenamefont {O'Brien}\ \emph {et~al.}(2016)\citenamefont
  {O'Brien}, \citenamefont {Abanin}, \citenamefont {Vidal},\ and\ \citenamefont
  {Papi\ifmmode~\acute{c}\else \'{c}\fi{}}}]{OBrien16}%
  \BibitemOpen
  \bibfield  {author} {\bibinfo {author} {\bibfnamefont {T.~E.}\ \bibnamefont
  {O'Brien}}, \bibinfo {author} {\bibfnamefont {D.~A.}\ \bibnamefont {Abanin}},
  \bibinfo {author} {\bibfnamefont {G.}~\bibnamefont {Vidal}}, \ and\ \bibinfo
  {author} {\bibfnamefont {Z.}~\bibnamefont {Papi\ifmmode~\acute{c}\else
  \'{c}\fi{}}},\ }\href {\doibase 10.1103/PhysRevB.94.144208} {\bibfield
  {journal} {\bibinfo  {journal} {Phys. Rev. B}\ }\textbf {\bibinfo {volume}
  {94}},\ \bibinfo {pages} {144208} (\bibinfo {year} {2016})}\BibitemShut
  {NoStop}%
\bibitem [{\citenamefont {Khemani}\ \emph {et~al.}(2017)\citenamefont
  {Khemani}, \citenamefont {Lim}, \citenamefont {Sheng},\ and\ \citenamefont
  {Huse}}]{Khemani17}%
  \BibitemOpen
  \bibfield  {author} {\bibinfo {author} {\bibfnamefont {V.}~\bibnamefont
  {Khemani}}, \bibinfo {author} {\bibfnamefont {S.~P.}\ \bibnamefont {Lim}},
  \bibinfo {author} {\bibfnamefont {D.~N.}\ \bibnamefont {Sheng}}, \ and\
  \bibinfo {author} {\bibfnamefont {D.~A.}\ \bibnamefont {Huse}},\ }\href
  {\doibase 10.1103/PhysRevX.7.021013} {\bibfield  {journal} {\bibinfo
  {journal} {Phys. Rev. X}\ }\textbf {\bibinfo {volume} {7}},\ \bibinfo {pages}
  {021013} (\bibinfo {year} {2017})}\BibitemShut {NoStop}%
\bibitem [{\citenamefont {Yang}\ \emph {et~al.}(2017)\citenamefont {Yang},
  \citenamefont {Hamma}, \citenamefont {Giampaolo}, \citenamefont {Mucciolo},\
  and\ \citenamefont {Chamon}}]{Yang17}%
  \BibitemOpen
  \bibfield  {author} {\bibinfo {author} {\bibfnamefont {Z.-C.}\ \bibnamefont
  {Yang}}, \bibinfo {author} {\bibfnamefont {A.}~\bibnamefont {Hamma}},
  \bibinfo {author} {\bibfnamefont {S.~M.}\ \bibnamefont {Giampaolo}}, \bibinfo
  {author} {\bibfnamefont {E.~R.}\ \bibnamefont {Mucciolo}}, \ and\ \bibinfo
  {author} {\bibfnamefont {C.}~\bibnamefont {Chamon}},\ }\href {\doibase
  10.1103/PhysRevB.96.020408} {\bibfield  {journal} {\bibinfo  {journal} {Phys.
  Rev. B}\ }\textbf {\bibinfo {volume} {96}},\ \bibinfo {pages} {020408}
  (\bibinfo {year} {2017})}\BibitemShut {NoStop}%
\bibitem [{\citenamefont {Li}\ \emph {et~al.}(2015)\citenamefont {Li},
  \citenamefont {Ganeshan}, \citenamefont {Pixley},\ and\ \citenamefont
  {Das~Sarma}}]{Li15}%
  \BibitemOpen
  \bibfield  {author} {\bibinfo {author} {\bibfnamefont {X.}~\bibnamefont
  {Li}}, \bibinfo {author} {\bibfnamefont {S.}~\bibnamefont {Ganeshan}},
  \bibinfo {author} {\bibfnamefont {J.~H.}\ \bibnamefont {Pixley}}, \ and\
  \bibinfo {author} {\bibfnamefont {S.}~\bibnamefont {Das~Sarma}},\ }\href
  {\doibase 10.1103/PhysRevLett.115.186601} {\bibfield  {journal} {\bibinfo
  {journal} {Phys. Rev. Lett.}\ }\textbf {\bibinfo {volume} {115}},\ \bibinfo
  {pages} {186601} (\bibinfo {year} {2015})}\BibitemShut {NoStop}%
\bibitem [{\citenamefont {Li}\ \emph {et~al.}(2017)\citenamefont {Li},
  \citenamefont {Li},\ and\ \citenamefont {Das~Sarma}}]{Li17}%
  \BibitemOpen
  \bibfield  {author} {\bibinfo {author} {\bibfnamefont {X.}~\bibnamefont
  {Li}}, \bibinfo {author} {\bibfnamefont {X.}~\bibnamefont {Li}}, \ and\
  \bibinfo {author} {\bibfnamefont {S.}~\bibnamefont {Das~Sarma}},\ }\href
  {\doibase 10.1103/PhysRevB.96.085119} {\bibfield  {journal} {\bibinfo
  {journal} {Phys. Rev. B}\ }\textbf {\bibinfo {volume} {96}},\ \bibinfo
  {pages} {085119} (\bibinfo {year} {2017})}\BibitemShut {NoStop}%
\bibitem [{\citenamefont {Gornyi}\ \emph {et~al.}(2017)\citenamefont {Gornyi},
  \citenamefont {Mirlin}, \citenamefont {M{\"u}ller},\ and\ \citenamefont
  {Polyakov}}]{Gornyi17}%
  \BibitemOpen
  \bibfield  {author} {\bibinfo {author} {\bibfnamefont {I.~V.}\ \bibnamefont
  {Gornyi}}, \bibinfo {author} {\bibfnamefont {A.~D.}\ \bibnamefont {Mirlin}},
  \bibinfo {author} {\bibfnamefont {M.}~\bibnamefont {M{\"u}ller}}, \ and\
  \bibinfo {author} {\bibfnamefont {D.~G.}\ \bibnamefont {Polyakov}},\ }\href
  {\doibase 10.1002/andp.201600365} {\bibfield  {journal} {\bibinfo  {journal}
  {Ann. Phys. (Berl.)}\ }\textbf {\bibinfo {volume} {529}},\ \bibinfo {pages}
  {1600365} (\bibinfo {year} {2017})}\BibitemShut {NoStop}%
\bibitem [{\citenamefont {Huse}\ \emph {et~al.}(2013)\citenamefont {Huse},
  \citenamefont {Nandkishore}, \citenamefont {Oganesyan}, \citenamefont {Pal},\
  and\ \citenamefont {Sondhi}}]{Huse13}%
  \BibitemOpen
  \bibfield  {author} {\bibinfo {author} {\bibfnamefont {D.~A.}\ \bibnamefont
  {Huse}}, \bibinfo {author} {\bibfnamefont {R.}~\bibnamefont {Nandkishore}},
  \bibinfo {author} {\bibfnamefont {V.}~\bibnamefont {Oganesyan}}, \bibinfo
  {author} {\bibfnamefont {A.}~\bibnamefont {Pal}}, \ and\ \bibinfo {author}
  {\bibfnamefont {S.~L.}\ \bibnamefont {Sondhi}},\ }\href {\doibase
  10.1103/PhysRevB.88.014206} {\bibfield  {journal} {\bibinfo  {journal} {Phys.
  Rev. B}\ }\textbf {\bibinfo {volume} {88}},\ \bibinfo {pages} {014206}
  (\bibinfo {year} {2013})}\BibitemShut {NoStop}%
\bibitem [{Bau()}]{Bauer13}%
  \BibitemOpen
  \href@noop {} {}\bibinfo {howpublished} {B.~Bauer and C.~Nayak
  \href{http://iopscience.iop.org/article/10.1088/1742-5468/2013/09/P09005/meta}{J.~Stat.~Mech.
  (2013) P09005}.}\BibitemShut {Stop}%
\bibitem [{\citenamefont {Pekker}\ \emph {et~al.}(2014)\citenamefont {Pekker},
  \citenamefont {Refael}, \citenamefont {Altman}, \citenamefont {Demler},\ and\
  \citenamefont {Oganesyan}}]{Pekker14}%
  \BibitemOpen
  \bibfield  {author} {\bibinfo {author} {\bibfnamefont {D.}~\bibnamefont
  {Pekker}}, \bibinfo {author} {\bibfnamefont {G.}~\bibnamefont {Refael}},
  \bibinfo {author} {\bibfnamefont {E.}~\bibnamefont {Altman}}, \bibinfo
  {author} {\bibfnamefont {E.}~\bibnamefont {Demler}}, \ and\ \bibinfo {author}
  {\bibfnamefont {V.}~\bibnamefont {Oganesyan}},\ }\href {\doibase
  10.1103/PhysRevX.4.011052} {\bibfield  {journal} {\bibinfo  {journal} {Phys.
  Rev. X}\ }\textbf {\bibinfo {volume} {4}},\ \bibinfo {pages} {011052}
  (\bibinfo {year} {2014})}\BibitemShut {NoStop}%
\bibitem [{\citenamefont {Kj\"all}\ \emph {et~al.}(2014)\citenamefont
  {Kj\"all}, \citenamefont {Bardarson},\ and\ \citenamefont
  {Pollmann}}]{Kjall14}%
  \BibitemOpen
  \bibfield  {author} {\bibinfo {author} {\bibfnamefont {J.~A.}\ \bibnamefont
  {Kj\"all}}, \bibinfo {author} {\bibfnamefont {J.~H.}\ \bibnamefont
  {Bardarson}}, \ and\ \bibinfo {author} {\bibfnamefont {F.}~\bibnamefont
  {Pollmann}},\ }\href {\doibase 10.1103/PhysRevLett.113.107204} {\bibfield
  {journal} {\bibinfo  {journal} {Phys. Rev. Lett.}\ }\textbf {\bibinfo
  {volume} {113}},\ \bibinfo {pages} {107204} (\bibinfo {year}
  {2014})}\BibitemShut {NoStop}%
\bibitem [{\citenamefont {Chandran}\ \emph {et~al.}(2014)\citenamefont
  {Chandran}, \citenamefont {Khemani}, \citenamefont {Laumann},\ and\
  \citenamefont {Sondhi}}]{Chandran14}%
  \BibitemOpen
  \bibfield  {author} {\bibinfo {author} {\bibfnamefont {A.}~\bibnamefont
  {Chandran}}, \bibinfo {author} {\bibfnamefont {V.}~\bibnamefont {Khemani}},
  \bibinfo {author} {\bibfnamefont {C.~R.}\ \bibnamefont {Laumann}}, \ and\
  \bibinfo {author} {\bibfnamefont {S.~L.}\ \bibnamefont {Sondhi}},\ }\href
  {\doibase 10.1103/PhysRevB.89.144201} {\bibfield  {journal} {\bibinfo
  {journal} {Phys. Rev. B}\ }\textbf {\bibinfo {volume} {89}},\ \bibinfo
  {pages} {144201} (\bibinfo {year} {2014})}\BibitemShut {NoStop}%
\bibitem [{Sla()}]{Slagle15}%
  \BibitemOpen
  \href@noop {} {}\bibinfo {howpublished} {K.~Slagle, Z.~Bi, Y.-Z.~You, and
  C.~Xu, \href{https://arxiv.org/abs/1505.05147}{arXiv:1505.05147
  (unpublished)}.}\BibitemShut {Stop}%
\bibitem [{Bah()}]{Bahri15}%
  \BibitemOpen
  \href@noop {} {}\bibinfo {howpublished} {Y.~Bahri, R.~Vosk, E.~Altman, and
  A.~Vishwanath, \href{https://www.nature.com/articles/ncomms8341}{Nat. Commun.
  \textbf{6}, 7341 (2015).}}\BibitemShut {Stop}%
\bibitem [{\citenamefont {Potter}\ and\ \citenamefont
  {Vasseur}(2016)}]{Potter16}%
  \BibitemOpen
  \bibfield  {author} {\bibinfo {author} {\bibfnamefont {A.~C.}\ \bibnamefont
  {Potter}}\ and\ \bibinfo {author} {\bibfnamefont {R.}~\bibnamefont
  {Vasseur}},\ }\href {\doibase 10.1103/PhysRevB.94.224206} {\bibfield
  {journal} {\bibinfo  {journal} {Phys. Rev. B}\ }\textbf {\bibinfo {volume}
  {94}},\ \bibinfo {pages} {224206} (\bibinfo {year} {2016})}\BibitemShut
  {NoStop}%
\bibitem [{\citenamefont {Vasseur}\ \emph {et~al.}(2016)\citenamefont
  {Vasseur}, \citenamefont {Friedman}, \citenamefont {Parameswaran},\ and\
  \citenamefont {Potter}}]{Vasseur16}%
  \BibitemOpen
  \bibfield  {author} {\bibinfo {author} {\bibfnamefont {R.}~\bibnamefont
  {Vasseur}}, \bibinfo {author} {\bibfnamefont {A.~J.}\ \bibnamefont
  {Friedman}}, \bibinfo {author} {\bibfnamefont {S.~A.}\ \bibnamefont
  {Parameswaran}}, \ and\ \bibinfo {author} {\bibfnamefont {A.~C.}\
  \bibnamefont {Potter}},\ }\href {\doibase 10.1103/PhysRevB.93.134207}
  {\bibfield  {journal} {\bibinfo  {journal} {Phys. Rev. B}\ }\textbf {\bibinfo
  {volume} {93}},\ \bibinfo {pages} {134207} (\bibinfo {year}
  {2016})}\BibitemShut {NoStop}%
\bibitem [{Fri()}]{Friedman17}%
  \BibitemOpen
  \href@noop {} {}\bibinfo {howpublished} {A.~J.~Friedman, R.~Vasseur, A.~C.
  Potter, and S.~A.~Parameswaran,
  \href{https://arxiv.org/abs/1706.00022}{arXiv:1706.00022
  (unpublished)}.}\BibitemShut {Stop}%
\bibitem [{\citenamefont {Parameswaran}\ \emph {et~al.}(2017)\citenamefont
  {Parameswaran}, \citenamefont {Potter},\ and\ \citenamefont
  {Vasseur}}]{Parameswaran17}%
  \BibitemOpen
  \bibfield  {author} {\bibinfo {author} {\bibfnamefont {S.~A.}\ \bibnamefont
  {Parameswaran}}, \bibinfo {author} {\bibfnamefont {A.~C.}\ \bibnamefont
  {Potter}}, \ and\ \bibinfo {author} {\bibfnamefont {R.}~\bibnamefont
  {Vasseur}},\ }\href
  {https://onlinelibrary.wiley.com/doi/abs/10.1002/andp.201600302} {\bibfield
  {journal} {\bibinfo  {journal} {Ann. Phys. (Berl.)}\ }\textbf {\bibinfo
  {volume} {529}} (\bibinfo {year} {2017})}\BibitemShut {NoStop}%
\bibitem [{\citenamefont {Chandran}\ and\ \citenamefont
  {Laumann}(2017)}]{Chandran17}%
  \BibitemOpen
  \bibfield  {author} {\bibinfo {author} {\bibfnamefont {A.}~\bibnamefont
  {Chandran}}\ and\ \bibinfo {author} {\bibfnamefont {C.~R.}\ \bibnamefont
  {Laumann}},\ }\href {\doibase 10.1103/PhysRevX.7.031061} {\bibfield
  {journal} {\bibinfo  {journal} {Phys. Rev. X}\ }\textbf {\bibinfo {volume}
  {7}},\ \bibinfo {pages} {031061} (\bibinfo {year} {2017})}\BibitemShut
  {NoStop}%
\bibitem [{\citenamefont {Prakash}\ \emph {et~al.}(2017)\citenamefont
  {Prakash}, \citenamefont {Ganeshan}, \citenamefont {Fidkowski},\ and\
  \citenamefont {Wei}}]{Prakash17}%
  \BibitemOpen
  \bibfield  {author} {\bibinfo {author} {\bibfnamefont {A.}~\bibnamefont
  {Prakash}}, \bibinfo {author} {\bibfnamefont {S.}~\bibnamefont {Ganeshan}},
  \bibinfo {author} {\bibfnamefont {L.}~\bibnamefont {Fidkowski}}, \ and\
  \bibinfo {author} {\bibfnamefont {T.-C.}\ \bibnamefont {Wei}},\ }\href
  {\doibase 10.1103/PhysRevB.96.165136} {\bibfield  {journal} {\bibinfo
  {journal} {Phys. Rev. B}\ }\textbf {\bibinfo {volume} {96}},\ \bibinfo
  {pages} {165136} (\bibinfo {year} {2017})}\BibitemShut {NoStop}%
\bibitem [{\citenamefont {Deutsch}(1991)}]{Deutsch91}%
  \BibitemOpen
  \bibfield  {author} {\bibinfo {author} {\bibfnamefont {J.~M.}\ \bibnamefont
  {Deutsch}},\ }\href {\doibase 10.1103/PhysRevA.43.2046} {\bibfield  {journal}
  {\bibinfo  {journal} {Phys. Rev. A}\ }\textbf {\bibinfo {volume} {43}},\
  \bibinfo {pages} {2046} (\bibinfo {year} {1991})}\BibitemShut {NoStop}%
\bibitem [{\citenamefont {Page}(1993)}]{Page93}%
  \BibitemOpen
  \bibfield  {author} {\bibinfo {author} {\bibfnamefont {D.~N.}\ \bibnamefont
  {Page}},\ }\href {\doibase 10.1103/PhysRevLett.71.1291} {\bibfield  {journal}
  {\bibinfo  {journal} {Phys. Rev. Lett.}\ }\textbf {\bibinfo {volume} {71}},\
  \bibinfo {pages} {1291} (\bibinfo {year} {1993})}\BibitemShut {NoStop}%
\bibitem [{\citenamefont {Srednicki}(1994)}]{Srednicki94}%
  \BibitemOpen
  \bibfield  {author} {\bibinfo {author} {\bibfnamefont {M.}~\bibnamefont
  {Srednicki}},\ }\href {\doibase 10.1103/PhysRevE.50.888} {\bibfield
  {journal} {\bibinfo  {journal} {Phys. Rev. E}\ }\textbf {\bibinfo {volume}
  {50}},\ \bibinfo {pages} {888} (\bibinfo {year} {1994})}\BibitemShut
  {NoStop}%
\bibitem [{\citenamefont {D'Alessio}\ \emph {et~al.}(2016)\citenamefont
  {D'Alessio}, \citenamefont {Kafri}, \citenamefont {Polkovnikov},\ and\
  \citenamefont {Rigol}}]{D'Alessio16}%
  \BibitemOpen
  \bibfield  {author} {\bibinfo {author} {\bibfnamefont {L.}~\bibnamefont
  {D'Alessio}}, \bibinfo {author} {\bibfnamefont {Y.}~\bibnamefont {Kafri}},
  \bibinfo {author} {\bibfnamefont {A.}~\bibnamefont {Polkovnikov}}, \ and\
  \bibinfo {author} {\bibfnamefont {M.}~\bibnamefont {Rigol}},\ }\href
  {\doibase 10.1080/00018732.2016.1198134} {\bibfield  {journal} {\bibinfo
  {journal} {Adv.~Phys.}\ }\textbf {\bibinfo {volume} {65}},\ \bibinfo {pages}
  {239} (\bibinfo {year} {2016})}\BibitemShut {NoStop}%
\bibitem [{\citenamefont {Mori}\ \emph {et~al.}(2018)\citenamefont {Mori},
  \citenamefont {Ikeda}, \citenamefont {Kaminishi},\ and\ \citenamefont
  {Ueda}}]{Mori17}%
  \BibitemOpen
  \bibfield  {author} {\bibinfo {author} {\bibfnamefont {T.}~\bibnamefont
  {Mori}}, \bibinfo {author} {\bibfnamefont {T.~N.}\ \bibnamefont {Ikeda}},
  \bibinfo {author} {\bibfnamefont {E.}~\bibnamefont {Kaminishi}}, \ and\
  \bibinfo {author} {\bibfnamefont {M.}~\bibnamefont {Ueda}},\ }\href {\doibase
  10.1088/1361-6455/aabcdf} {\bibfield  {journal} {\bibinfo  {journal} {J.
  Phys. B}\ }\textbf {\bibinfo {volume} {51}},\ \bibinfo {pages} {112001}
  (\bibinfo {year} {2018})}\BibitemShut {NoStop}%
\bibitem [{\citenamefont {Modak}\ and\ \citenamefont
  {Mukerjee}(2015)}]{Modak15}%
  \BibitemOpen
  \bibfield  {author} {\bibinfo {author} {\bibfnamefont {R.}~\bibnamefont
  {Modak}}\ and\ \bibinfo {author} {\bibfnamefont {S.}~\bibnamefont
  {Mukerjee}},\ }\href {\doibase 10.1103/PhysRevLett.115.230401} {\bibfield
  {journal} {\bibinfo  {journal} {Phys. Rev. Lett.}\ }\textbf {\bibinfo
  {volume} {115}},\ \bibinfo {pages} {230401} (\bibinfo {year}
  {2015})}\BibitemShut {NoStop}%
\bibitem [{\citenamefont {Li}\ \emph {et~al.}(2016)\citenamefont {Li},
  \citenamefont {Pixley}, \citenamefont {Deng}, \citenamefont {Ganeshan},\ and\
  \citenamefont {Das~Sarma}}]{Li16}%
  \BibitemOpen
  \bibfield  {author} {\bibinfo {author} {\bibfnamefont {X.}~\bibnamefont
  {Li}}, \bibinfo {author} {\bibfnamefont {J.~H.}\ \bibnamefont {Pixley}},
  \bibinfo {author} {\bibfnamefont {D.-L.}\ \bibnamefont {Deng}}, \bibinfo
  {author} {\bibfnamefont {S.}~\bibnamefont {Ganeshan}}, \ and\ \bibinfo
  {author} {\bibfnamefont {S.}~\bibnamefont {Das~Sarma}},\ }\href {\doibase
  10.1103/PhysRevB.93.184204} {\bibfield  {journal} {\bibinfo  {journal} {Phys.
  Rev. B}\ }\textbf {\bibinfo {volume} {93}},\ \bibinfo {pages} {184204}
  (\bibinfo {year} {2016})}\BibitemShut {NoStop}%
\bibitem [{\citenamefont {De~Roeck}\ \emph {et~al.}(2016)\citenamefont
  {De~Roeck}, \citenamefont {Huveneers}, \citenamefont {M{\"u}ller},\ and\
  \citenamefont {Schiulaz}}]{DeRoeck16}%
  \BibitemOpen
  \bibfield  {author} {\bibinfo {author} {\bibfnamefont {W.}~\bibnamefont
  {De~Roeck}}, \bibinfo {author} {\bibfnamefont {F.}~\bibnamefont {Huveneers}},
  \bibinfo {author} {\bibfnamefont {M.}~\bibnamefont {M{\"u}ller}}, \ and\
  \bibinfo {author} {\bibfnamefont {M.}~\bibnamefont {Schiulaz}},\ }\href
  {\doibase 10.1103/PhysRevB.93.014203} {\bibfield  {journal} {\bibinfo
  {journal} {Phys. Rev. B}\ }\textbf {\bibinfo {volume} {93}},\ \bibinfo
  {pages} {014203} (\bibinfo {year} {2016})}\BibitemShut {NoStop}%
\bibitem [{\citenamefont {Bar~Lev}\ \emph {et~al.}(2016)\citenamefont
  {Bar~Lev}, \citenamefont {Reichman},\ and\ \citenamefont {Sagi}}]{BarLev16}%
  \BibitemOpen
  \bibfield  {author} {\bibinfo {author} {\bibfnamefont {Y.}~\bibnamefont
  {Bar~Lev}}, \bibinfo {author} {\bibfnamefont {D.~R.}\ \bibnamefont
  {Reichman}}, \ and\ \bibinfo {author} {\bibfnamefont {Y.}~\bibnamefont
  {Sagi}},\ }\href {\doibase 10.1103/PhysRevB.94.201116} {\bibfield  {journal}
  {\bibinfo  {journal} {Phys. Rev. B}\ }\textbf {\bibinfo {volume} {94}},\
  \bibinfo {pages} {201116} (\bibinfo {year} {2016})}\BibitemShut {NoStop}%
\bibitem [{\citenamefont {Prelov\ifmmode~\check{s}\else \v{s}\fi{}ek}\ \emph
  {et~al.}(2016)\citenamefont {Prelov\ifmmode~\check{s}\else \v{s}\fi{}ek},
  \citenamefont {Bari\ifmmode \check{s}\else \v{s}\fi{}i\ifmmode~\acute{c}\else
  \'{c}\fi{}},\ and\ \citenamefont {\ifmmode \check{Z}\else
  \v{Z}\fi{}nidari\ifmmode~\check{c}\else \v{c}\fi{}}}]{Prelovsek16}%
  \BibitemOpen
  \bibfield  {author} {\bibinfo {author} {\bibfnamefont {P.}~\bibnamefont
  {Prelov\ifmmode~\check{s}\else \v{s}\fi{}ek}}, \bibinfo {author}
  {\bibfnamefont {O.~S.}\ \bibnamefont {Bari\ifmmode \check{s}\else
  \v{s}\fi{}i\ifmmode~\acute{c}\else \'{c}\fi{}}}, \ and\ \bibinfo {author}
  {\bibfnamefont {M.}~\bibnamefont {\ifmmode \check{Z}\else
  \v{Z}\fi{}nidari\ifmmode~\check{c}\else \v{c}\fi{}}},\ }\href {\doibase
  10.1103/PhysRevB.94.241104} {\bibfield  {journal} {\bibinfo  {journal} {Phys.
  Rev. B}\ }\textbf {\bibinfo {volume} {94}},\ \bibinfo {pages} {241104}
  (\bibinfo {year} {2016})}\BibitemShut {NoStop}%
\bibitem [{\citenamefont {Protopopov}\ \emph {et~al.}(2017)\citenamefont
  {Protopopov}, \citenamefont {Ho},\ and\ \citenamefont
  {Abanin}}]{Protopopov17}%
  \BibitemOpen
  \bibfield  {author} {\bibinfo {author} {\bibfnamefont {I.~V.}\ \bibnamefont
  {Protopopov}}, \bibinfo {author} {\bibfnamefont {W.~W.}\ \bibnamefont {Ho}},
  \ and\ \bibinfo {author} {\bibfnamefont {D.~A.}\ \bibnamefont {Abanin}},\
  }\href {\doibase 10.1103/PhysRevB.96.041122} {\bibfield  {journal} {\bibinfo
  {journal} {Phys. Rev. B}\ }\textbf {\bibinfo {volume} {96}},\ \bibinfo
  {pages} {041122} (\bibinfo {year} {2017})}\BibitemShut {NoStop}%
\bibitem [{\citenamefont {Kozarzewski}\ \emph {et~al.}(2018)\citenamefont
  {Kozarzewski}, \citenamefont {Prelov\ifmmode~\check{s}\else \v{s}\fi{}ek},\
  and\ \citenamefont {Mierzejewski}}]{Kozarzewski18}%
  \BibitemOpen
  \bibfield  {author} {\bibinfo {author} {\bibfnamefont {M.}~\bibnamefont
  {Kozarzewski}}, \bibinfo {author} {\bibfnamefont {P.}~\bibnamefont
  {Prelov\ifmmode~\check{s}\else \v{s}\fi{}ek}}, \ and\ \bibinfo {author}
  {\bibfnamefont {M.}~\bibnamefont {Mierzejewski}},\ }\href {\doibase
  10.1103/PhysRevLett.120.246602} {\bibfield  {journal} {\bibinfo  {journal}
  {Phys. Rev. Lett.}\ }\textbf {\bibinfo {volume} {120}},\ \bibinfo {pages}
  {246602} (\bibinfo {year} {2018})}\BibitemShut {NoStop}%
\bibitem [{\citenamefont {{Iadecola}}\ and\ \citenamefont
  {{Schecter}}(2018)}]{Iadecola18}%
  \BibitemOpen
  \bibfield  {author} {\bibinfo {author} {\bibfnamefont {T.}~\bibnamefont
  {{Iadecola}}}\ and\ \bibinfo {author} {\bibfnamefont {M.}~\bibnamefont
  {{Schecter}}},\ }\href@noop {} {\bibfield  {journal} {\bibinfo  {journal}
  {ArXiv e-prints}\ } (\bibinfo {year} {2018})},\ \Eprint
  {http://arxiv.org/abs/1805.05360} {arXiv:1805.05360 [cond-mat.dis-nn]}
  \BibitemShut {NoStop}%
\bibitem [{\citenamefont {Landau}\ and\ \citenamefont
  {Lifshitz}(1980)}]{LandauStatMPhys1980}%
  \BibitemOpen
  \bibfield  {author} {\bibinfo {author} {\bibfnamefont {L.~D.}\ \bibnamefont
  {Landau}}\ and\ \bibinfo {author} {\bibfnamefont {E.~M.}\ \bibnamefont
  {Lifshitz}},\ }\href {\doibase 10.1016/C2009-0-24487-4} {\emph {\bibinfo
  {title} {Statistical Physics (Third Edition)}}}\ (\bibinfo  {publisher}
  {Butterworth-Heinemann},\ \bibinfo {address} {Oxford},\ \bibinfo {year}
  {1980})\BibitemShut {NoStop}%
\bibitem [{\citenamefont {Mermin}\ and\ \citenamefont
  {Wagner}(1966)}]{Mermin66}%
  \BibitemOpen
  \bibfield  {author} {\bibinfo {author} {\bibfnamefont {N.~D.}\ \bibnamefont
  {Mermin}}\ and\ \bibinfo {author} {\bibfnamefont {H.}~\bibnamefont
  {Wagner}},\ }\href {\doibase 10.1103/PhysRevLett.17.1133} {\bibfield
  {journal} {\bibinfo  {journal} {Phys. Rev. Lett.}\ }\textbf {\bibinfo
  {volume} {17}},\ \bibinfo {pages} {1133} (\bibinfo {year}
  {1966})}\BibitemShut {NoStop}%
\bibitem [{\citenamefont {Hohenberg}(1967)}]{Hohenberg67}%
  \BibitemOpen
  \bibfield  {author} {\bibinfo {author} {\bibfnamefont {P.~C.}\ \bibnamefont
  {Hohenberg}},\ }\href {\doibase 10.1103/PhysRev.158.383} {\bibfield
  {journal} {\bibinfo  {journal} {Phys. Rev.}\ }\textbf {\bibinfo {volume}
  {158}},\ \bibinfo {pages} {383} (\bibinfo {year} {1967})}\BibitemShut
  {NoStop}%
\bibitem [{\citenamefont {Schrieffer}\ and\ \citenamefont
  {Wolff}(1966)}]{Schrieffer66}%
  \BibitemOpen
  \bibfield  {author} {\bibinfo {author} {\bibfnamefont {J.~R.}\ \bibnamefont
  {Schrieffer}}\ and\ \bibinfo {author} {\bibfnamefont {P.~A.}\ \bibnamefont
  {Wolff}},\ }\href {\doibase 10.1103/PhysRev.149.491} {\bibfield  {journal}
  {\bibinfo  {journal} {Phys. Rev.}\ }\textbf {\bibinfo {volume} {149}},\
  \bibinfo {pages} {491} (\bibinfo {year} {1966})}\BibitemShut {NoStop}%
\bibitem [{\citenamefont {Mott}(1969)}]{Mott69}%
  \BibitemOpen
  \bibfield  {author} {\bibinfo {author} {\bibfnamefont {N.~F.}\ \bibnamefont
  {Mott}},\ }\href {\doibase 10.1080/14786436908216338} {\bibfield  {journal}
  {\bibinfo  {journal} {Philos. Mag.}\ }\textbf {\bibinfo {volume} {19}},\
  \bibinfo {pages} {835} (\bibinfo {year} {1969})}\BibitemShut {NoStop}%
\bibitem [{\citenamefont {Nandkishore}\ \emph {et~al.}(2014)\citenamefont
  {Nandkishore}, \citenamefont {Gopalakrishnan},\ and\ \citenamefont
  {Huse}}]{Nandkishore14}%
  \BibitemOpen
  \bibfield  {author} {\bibinfo {author} {\bibfnamefont {R.}~\bibnamefont
  {Nandkishore}}, \bibinfo {author} {\bibfnamefont {S.}~\bibnamefont
  {Gopalakrishnan}}, \ and\ \bibinfo {author} {\bibfnamefont {D.~A.}\
  \bibnamefont {Huse}},\ }\href {\doibase 10.1103/PhysRevB.90.064203}
  {\bibfield  {journal} {\bibinfo  {journal} {Phys. Rev. B}\ }\textbf {\bibinfo
  {volume} {90}},\ \bibinfo {pages} {064203} (\bibinfo {year}
  {2014})}\BibitemShut {NoStop}%
\bibitem [{\citenamefont {Johri}\ \emph {et~al.}(2015)\citenamefont {Johri},
  \citenamefont {Nandkishore},\ and\ \citenamefont {Bhatt}}]{Johri15}%
  \BibitemOpen
  \bibfield  {author} {\bibinfo {author} {\bibfnamefont {S.}~\bibnamefont
  {Johri}}, \bibinfo {author} {\bibfnamefont {R.}~\bibnamefont {Nandkishore}},
  \ and\ \bibinfo {author} {\bibfnamefont {R.~N.}\ \bibnamefont {Bhatt}},\
  }\href {\doibase 10.1103/PhysRevLett.114.117401} {\bibfield  {journal}
  {\bibinfo  {journal} {Phys. Rev. Lett.}\ }\textbf {\bibinfo {volume} {114}},\
  \bibinfo {pages} {117401} (\bibinfo {year} {2015})}\BibitemShut {NoStop}%
\bibitem [{\citenamefont {Nandkishore}(2015)}]{Nandkishore15a}%
  \BibitemOpen
  \bibfield  {author} {\bibinfo {author} {\bibfnamefont {R.}~\bibnamefont
  {Nandkishore}},\ }\href {\doibase 10.1103/PhysRevB.92.245141} {\bibfield
  {journal} {\bibinfo  {journal} {Phys. Rev. B}\ }\textbf {\bibinfo {volume}
  {92}},\ \bibinfo {pages} {245141} (\bibinfo {year} {2015})}\BibitemShut
  {NoStop}%
\bibitem [{\citenamefont {Fischer}\ \emph {et~al.}(2016)\citenamefont
  {Fischer}, \citenamefont {Maksymenko},\ and\ \citenamefont
  {Altman}}]{Fischer16}%
  \BibitemOpen
  \bibfield  {author} {\bibinfo {author} {\bibfnamefont {M.~H.}\ \bibnamefont
  {Fischer}}, \bibinfo {author} {\bibfnamefont {M.}~\bibnamefont {Maksymenko}},
  \ and\ \bibinfo {author} {\bibfnamefont {E.}~\bibnamefont {Altman}},\ }\href
  {\doibase 10.1103/PhysRevLett.116.160401} {\bibfield  {journal} {\bibinfo
  {journal} {Phys. Rev. Lett.}\ }\textbf {\bibinfo {volume} {116}},\ \bibinfo
  {pages} {160401} (\bibinfo {year} {2016})}\BibitemShut {NoStop}%
\bibitem [{\citenamefont {Hyatt}\ \emph {et~al.}(2017)\citenamefont {Hyatt},
  \citenamefont {Garrison}, \citenamefont {Potter},\ and\ \citenamefont
  {Bauer}}]{Hyatt17}%
  \BibitemOpen
  \bibfield  {author} {\bibinfo {author} {\bibfnamefont {K.}~\bibnamefont
  {Hyatt}}, \bibinfo {author} {\bibfnamefont {J.~R.}\ \bibnamefont {Garrison}},
  \bibinfo {author} {\bibfnamefont {A.~C.}\ \bibnamefont {Potter}}, \ and\
  \bibinfo {author} {\bibfnamefont {B.}~\bibnamefont {Bauer}},\ }\href
  {\doibase 10.1103/PhysRevB.95.035132} {\bibfield  {journal} {\bibinfo
  {journal} {Phys. Rev. B}\ }\textbf {\bibinfo {volume} {95}},\ \bibinfo
  {pages} {035132} (\bibinfo {year} {2017})}\BibitemShut {NoStop}%
\bibitem [{\citenamefont {Haegeman}\ \emph {et~al.}(2011)\citenamefont
  {Haegeman}, \citenamefont {Cirac}, \citenamefont {Osborne}, \citenamefont
  {Pi\ifmmode~\check{z}\else \v{z}\fi{}orn}, \citenamefont {Verschelde},\ and\
  \citenamefont {Verstraete}}]{Haegeman11}%
  \BibitemOpen
  \bibfield  {author} {\bibinfo {author} {\bibfnamefont {J.}~\bibnamefont
  {Haegeman}}, \bibinfo {author} {\bibfnamefont {J.~I.}\ \bibnamefont {Cirac}},
  \bibinfo {author} {\bibfnamefont {T.~J.}\ \bibnamefont {Osborne}}, \bibinfo
  {author} {\bibfnamefont {I.}~\bibnamefont {Pi\ifmmode~\check{z}\else
  \v{z}\fi{}orn}}, \bibinfo {author} {\bibfnamefont {H.}~\bibnamefont
  {Verschelde}}, \ and\ \bibinfo {author} {\bibfnamefont {F.}~\bibnamefont
  {Verstraete}},\ }\href {\doibase 10.1103/PhysRevLett.107.070601} {\bibfield
  {journal} {\bibinfo  {journal} {Phys. Rev. Lett.}\ }\textbf {\bibinfo
  {volume} {107}},\ \bibinfo {pages} {070601} (\bibinfo {year}
  {2011})}\BibitemShut {NoStop}%
\bibitem [{\citenamefont {Haegeman}\ \emph {et~al.}(2016)\citenamefont
  {Haegeman}, \citenamefont {Lubich}, \citenamefont {Oseledets}, \citenamefont
  {Vandereycken},\ and\ \citenamefont {Verstraete}}]{Haegeman16}%
  \BibitemOpen
  \bibfield  {author} {\bibinfo {author} {\bibfnamefont {J.}~\bibnamefont
  {Haegeman}}, \bibinfo {author} {\bibfnamefont {C.}~\bibnamefont {Lubich}},
  \bibinfo {author} {\bibfnamefont {I.}~\bibnamefont {Oseledets}}, \bibinfo
  {author} {\bibfnamefont {B.}~\bibnamefont {Vandereycken}}, \ and\ \bibinfo
  {author} {\bibfnamefont {F.}~\bibnamefont {Verstraete}},\ }\href {\doibase
  10.1103/PhysRevB.94.165116} {\bibfield  {journal} {\bibinfo  {journal} {Phys.
  Rev. B}\ }\textbf {\bibinfo {volume} {94}},\ \bibinfo {pages} {165116}
  (\bibinfo {year} {2016})}\BibitemShut {NoStop}%
\bibitem [{\citenamefont {{Leviatan}}\ \emph {et~al.}(2017)\citenamefont
  {{Leviatan}}, \citenamefont {{Pollmann}}, \citenamefont {{Bardarson}},
  \citenamefont {{Huse}},\ and\ \citenamefont {{Altman}}}]{Leviatan17}%
  \BibitemOpen
  \bibfield  {author} {\bibinfo {author} {\bibfnamefont {E.}~\bibnamefont
  {{Leviatan}}}, \bibinfo {author} {\bibfnamefont {F.}~\bibnamefont
  {{Pollmann}}}, \bibinfo {author} {\bibfnamefont {J.~H.}\ \bibnamefont
  {{Bardarson}}}, \bibinfo {author} {\bibfnamefont {D.~A.}\ \bibnamefont
  {{Huse}}}, \ and\ \bibinfo {author} {\bibfnamefont {E.}~\bibnamefont
  {{Altman}}},\ }\href@noop {} {\bibfield  {journal} {\bibinfo  {journal}
  {ArXiv e-prints}\ } (\bibinfo {year} {2017})},\ \Eprint
  {http://arxiv.org/abs/1702.08894} {arXiv:1702.08894 [cond-mat.stat-mech]}
  \BibitemShut {NoStop}%
\bibitem [{\citenamefont {{Doggen}}\ \emph {et~al.}(2018)\citenamefont
  {{Doggen}}, \citenamefont {{Schindler}}, \citenamefont {{Tikhonov}},
  \citenamefont {{Mirlin}}, \citenamefont {{Neupert}}, \citenamefont
  {{Polyakov}},\ and\ \citenamefont {{Gornyi}}}]{Doggen18}%
  \BibitemOpen
  \bibfield  {author} {\bibinfo {author} {\bibfnamefont {E.~V.~H.}\
  \bibnamefont {{Doggen}}}, \bibinfo {author} {\bibfnamefont {F.}~\bibnamefont
  {{Schindler}}}, \bibinfo {author} {\bibfnamefont {K.~S.}\ \bibnamefont
  {{Tikhonov}}}, \bibinfo {author} {\bibfnamefont {A.~D.}\ \bibnamefont
  {{Mirlin}}}, \bibinfo {author} {\bibfnamefont {T.}~\bibnamefont {{Neupert}}},
  \bibinfo {author} {\bibfnamefont {D.~G.}\ \bibnamefont {{Polyakov}}}, \ and\
  \bibinfo {author} {\bibfnamefont {I.~V.}\ \bibnamefont {{Gornyi}}},\
  }\href@noop {} {\bibfield  {journal} {\bibinfo  {journal} {ArXiv e-prints}\ }
  (\bibinfo {year} {2018})},\ \Eprint {http://arxiv.org/abs/1807.05051}
  {arXiv:1807.05051 [cond-mat.dis-nn]} \BibitemShut {NoStop}%
\bibitem [{\citenamefont {De~Roeck}\ and\ \citenamefont
  {Huveneers}(2017)}]{DeRoeck17a}%
  \BibitemOpen
  \bibfield  {author} {\bibinfo {author} {\bibfnamefont {W.}~\bibnamefont
  {De~Roeck}}\ and\ \bibinfo {author} {\bibfnamefont {F.}~\bibnamefont
  {Huveneers}},\ }\href {\doibase 10.1103/PhysRevB.95.155129} {\bibfield
  {journal} {\bibinfo  {journal} {Phys. Rev. B}\ }\textbf {\bibinfo {volume}
  {95}},\ \bibinfo {pages} {155129} (\bibinfo {year} {2017})}\BibitemShut
  {NoStop}%
\bibitem [{\citenamefont {De~Roeck}\ and\ \citenamefont
  {Imbrie}(2017)}]{DeRoeck17b}%
  \BibitemOpen
  \bibfield  {author} {\bibinfo {author} {\bibfnamefont {W.}~\bibnamefont
  {De~Roeck}}\ and\ \bibinfo {author} {\bibfnamefont {J.~Z.}\ \bibnamefont
  {Imbrie}},\ }\href
  {http://rsta.royalsocietypublishing.org/content/375/2108/20160422} {\bibfield
   {journal} {\bibinfo  {journal} {Philos. Trans. Royal Soc. A}\ }\textbf
  {\bibinfo {volume} {375}} (\bibinfo {year} {2017})}\BibitemShut {NoStop}%
\bibitem [{\citenamefont {{Thiery}}\ \emph
  {et~al.}(2017{\natexlab{a}})\citenamefont {{Thiery}}, \citenamefont
  {{Huveneers}}, \citenamefont {{M{\"u}ller}},\ and\ \citenamefont {{De
  Roeck}}}]{Thiery17a}%
  \BibitemOpen
  \bibfield  {author} {\bibinfo {author} {\bibfnamefont {T.}~\bibnamefont
  {{Thiery}}}, \bibinfo {author} {\bibfnamefont {F.}~\bibnamefont
  {{Huveneers}}}, \bibinfo {author} {\bibfnamefont {M.}~\bibnamefont
  {{M{\"u}ller}}}, \ and\ \bibinfo {author} {\bibfnamefont {W.}~\bibnamefont
  {{De Roeck}}},\ }\href@noop {} {\bibfield  {journal} {\bibinfo  {journal}
  {ArXiv e-prints}\ } (\bibinfo {year} {2017}{\natexlab{a}})},\ \Eprint
  {http://arxiv.org/abs/1706.09338} {arXiv:1706.09338 [cond-mat.dis-nn]}
  \BibitemShut {NoStop}%
\bibitem [{\citenamefont {{Thiery}}\ \emph
  {et~al.}(2017{\natexlab{b}})\citenamefont {{Thiery}}, \citenamefont
  {{M{\"u}ller}},\ and\ \citenamefont {{De Roeck}}}]{Thiery17b}%
  \BibitemOpen
  \bibfield  {author} {\bibinfo {author} {\bibfnamefont {T.}~\bibnamefont
  {{Thiery}}}, \bibinfo {author} {\bibfnamefont {M.}~\bibnamefont
  {{M{\"u}ller}}}, \ and\ \bibinfo {author} {\bibfnamefont {W.}~\bibnamefont
  {{De Roeck}}},\ }\href@noop {} {\bibfield  {journal} {\bibinfo  {journal}
  {ArXiv e-prints}\ } (\bibinfo {year} {2017}{\natexlab{b}})},\ \Eprint
  {http://arxiv.org/abs/1711.09880} {arXiv:1711.09880 [cond-mat.stat-mech]}
  \BibitemShut {NoStop}%
\bibitem [{\citenamefont {Ponte}\ \emph {et~al.}(2017)\citenamefont {Ponte},
  \citenamefont {Laumann}, \citenamefont {Huse},\ and\ \citenamefont
  {Chandran}}]{Ponte17}%
  \BibitemOpen
  \bibfield  {author} {\bibinfo {author} {\bibfnamefont {P.}~\bibnamefont
  {Ponte}}, \bibinfo {author} {\bibfnamefont {C.~R.}\ \bibnamefont {Laumann}},
  \bibinfo {author} {\bibfnamefont {D.~A.}\ \bibnamefont {Huse}}, \ and\
  \bibinfo {author} {\bibfnamefont {A.}~\bibnamefont {Chandran}},\ }\href
  {http://rsta.royalsocietypublishing.org/content/375/2108/20160428} {\bibfield
   {journal} {\bibinfo  {journal} {Philos. Trans. Royal Soc. A}\ }\textbf
  {\bibinfo {volume} {375}} (\bibinfo {year} {2017})}\BibitemShut {NoStop}%
\bibitem [{\citenamefont {Mazurenko}\ \emph {et~al.}(2017)\citenamefont
  {Mazurenko}, \citenamefont {Chiu}, \citenamefont {Ji}, \citenamefont
  {Parsons}, \citenamefont {Kan\'{a}sz-Nagy}, \citenamefont {Schmidt},
  \citenamefont {Grusdt}, \citenamefont {Demler}, \citenamefont {Greif},\ and\
  \citenamefont {Greiner}}]{Mazurenko17}%
  \BibitemOpen
  \bibfield  {author} {\bibinfo {author} {\bibfnamefont {A.}~\bibnamefont
  {Mazurenko}}, \bibinfo {author} {\bibfnamefont {C.~S.}\ \bibnamefont {Chiu}},
  \bibinfo {author} {\bibfnamefont {G.}~\bibnamefont {Ji}}, \bibinfo {author}
  {\bibfnamefont {M.~F.}\ \bibnamefont {Parsons}}, \bibinfo {author}
  {\bibfnamefont {M.}~\bibnamefont {Kan\'{a}sz-Nagy}}, \bibinfo {author}
  {\bibfnamefont {R.}~\bibnamefont {Schmidt}}, \bibinfo {author} {\bibfnamefont
  {F.}~\bibnamefont {Grusdt}}, \bibinfo {author} {\bibfnamefont
  {E.}~\bibnamefont {Demler}}, \bibinfo {author} {\bibfnamefont
  {D.}~\bibnamefont {Greif}}, \ and\ \bibinfo {author} {\bibfnamefont
  {M.}~\bibnamefont {Greiner}},\ }\href {\doibase 10.1038/nature22362}
  {\bibfield  {journal} {\bibinfo  {journal} {Nature}\ }\textbf {\bibinfo
  {volume} {545}},\ \bibinfo {pages} {462} (\bibinfo {year}
  {2017})}\BibitemShut {NoStop}%
\bibitem [{\citenamefont {Mitra}\ \emph {et~al.}(2017)\citenamefont {Mitra},
  \citenamefont {Brown}, \citenamefont {Guardado-Sanchez}, \citenamefont
  {Kondov}, \citenamefont {Devakul}, \citenamefont {Huse}, \citenamefont
  {Schau{\ss}},\ and\ \citenamefont {Bakr}}]{Mitra17}%
  \BibitemOpen
  \bibfield  {author} {\bibinfo {author} {\bibfnamefont {D.}~\bibnamefont
  {Mitra}}, \bibinfo {author} {\bibfnamefont {P.~T.}\ \bibnamefont {Brown}},
  \bibinfo {author} {\bibfnamefont {E.}~\bibnamefont {Guardado-Sanchez}},
  \bibinfo {author} {\bibfnamefont {S.~S.}\ \bibnamefont {Kondov}}, \bibinfo
  {author} {\bibfnamefont {T.}~\bibnamefont {Devakul}}, \bibinfo {author}
  {\bibfnamefont {D.~A.}\ \bibnamefont {Huse}}, \bibinfo {author}
  {\bibfnamefont {P.}~\bibnamefont {Schau{\ss}}}, \ and\ \bibinfo {author}
  {\bibfnamefont {W.~S.}\ \bibnamefont {Bakr}},\ }\href
  {http://dx.doi.org/10.1038/nphys4297} {\bibfield  {journal} {\bibinfo
  {journal} {Nature Physics}\ }\textbf {\bibinfo {volume} {14}},\ \bibinfo
  {pages} {173} (\bibinfo {year} {2017})}\BibitemShut {NoStop}%
\bibitem [{\citenamefont {Chiu}\ \emph {et~al.}(2018)\citenamefont {Chiu},
  \citenamefont {Ji}, \citenamefont {Mazurenko}, \citenamefont {Greif},\ and\
  \citenamefont {Greiner}}]{Chiu18}%
  \BibitemOpen
  \bibfield  {author} {\bibinfo {author} {\bibfnamefont {C.~S.}\ \bibnamefont
  {Chiu}}, \bibinfo {author} {\bibfnamefont {G.}~\bibnamefont {Ji}}, \bibinfo
  {author} {\bibfnamefont {A.}~\bibnamefont {Mazurenko}}, \bibinfo {author}
  {\bibfnamefont {D.}~\bibnamefont {Greif}}, \ and\ \bibinfo {author}
  {\bibfnamefont {M.}~\bibnamefont {Greiner}},\ }\href {\doibase
  10.1103/PhysRevLett.120.243201} {\bibfield  {journal} {\bibinfo  {journal}
  {Phys. Rev. Lett.}\ }\textbf {\bibinfo {volume} {120}},\ \bibinfo {pages}
  {243201} (\bibinfo {year} {2018})}\BibitemShut {NoStop}%
\end{thebibliography}%

\end{document}